\documentclass[aps,letterpaper,twocolumn,prx,floatfix]{revtex4-1}

\usepackage{amsmath}
\usepackage{amsfonts}
\usepackage{amssymb}
\usepackage[scanall]{psfrag}
\usepackage{psfrag}
\usepackage{graphicx}
\usepackage{color}
\usepackage{amsthm}
\usepackage{thmtools, thm-restate}

\begin{document}
\newcommand{\of}[1]{\left( #1 \right)}
\newcommand{\sqof}[1]{\left[ #1 \right]}
\newcommand{\abs}[1]{\left| #1 \right|}
\newcommand{\avg}[1]{\left< #1 \right>}
\newcommand{\cuof}[1]{\left \{ #1 \right \} }
\newcommand{\bra}[1]{\left < #1 \right | }
\newcommand{\ket}[1]{\left | #1 \right > }
\newcommand{\pil}{\frac{\pi}{L}}
\newcommand{\bx}{\mathbf{x}}
\newcommand{\by}{\mathbf{y}}
\newcommand{\bk}{\mathbf{k}}
\newcommand{\bp}{\mathbf{p}}
\newcommand{\bl}{\mathbf{l}}
\newcommand{\bq}{\mathbf{q}}
\newcommand{\bs}{\mathbf{s}}
\newcommand{\psibar}{\overline{\psi}}
\newcommand{\svec}{\overrightarrow{\sigma}}
\newcommand{\dvec}{\overrightarrow{\partial}}
\newcommand{\bA}{\mathbf{A}}
\newcommand{\bdelta}{\mathbf{\delta}}
\newcommand{\bK}{\mathbf{K}}
\newcommand{\bQ}{\mathbf{Q}}
\newcommand{\bG}{\mathbf{G}}
\newcommand{\bw}{\mathbf{w}}
\newcommand{\bL}{\mathbf{L}}
\newcommand{\ohat}{\widehat{O}}
\newcommand{\up}{\uparrow}
\newcommand{\down}{\downarrow}
\newcommand{\MM}{\mathcal{M}}
\newcommand{\tX}{\tilde{X}}
\newcommand{\tY}{\tilde{Y}}
\newcommand{\tZ}{\tilde{Z}}
\newcommand{\tOm}{\tilde{\Omega}}
\newcommand{\barA}{\bar{\alpha}}

\newtheorem{conjecture}{Conjecture}

\author{Eliot Kapit}
\affiliation{Department of Physics, Colorado School of Mines, 1523 Illinois St, Golden CO 80401}
\author{Vadim Oganesyan}
\affiliation{The Graduate Center, City University of New York, New York, NY}
\affiliation{Department of Physics, College of Staten Island, City University of New York, Staten Island, NY}

\title{Noise-tolerant quantum speedups in quantum annealing without fine tuning}

\begin{abstract}

Quantum annealing is a powerful alternative model of quantum computing, which can succeed in the presence of environmental noise even without error correction. However, despite great effort, no conclusive demonstration of a quantum speedup (relative to state of the art classical algorithms) has been shown for these systems, and rigorous theoretical proofs of a quantum advantage generally rely on exponential precision in at least some aspects of the system, an unphysical resource guaranteed to be scrambled by experimental uncertainties and random noise. In this work, we propose a new variant of quantum annealing, called RFQA, which can maintain a scalable quantum speedup in the face of noise and modest control precision. Specifically, we consider a modification of flux qubit-based quantum annealing which includes low-frequency oscillations in the directions of the transverse field terms as the system evolves. We show that this method produces a quantum speedup for finding ground states in the Grover problem and quantum random energy model, and thus should be widely applicable to other hard optimization problems which can be formulated as quantum spin glasses. Further, we explore three realistic noise channels and show that the speedup from RFQA is resilient to $1/f$-like local potential fluctuations and local heating from interaction with a sufficiently low temperature bath. Another noise channel, bath-assisted quantum cooling transitions, actually accelerates the algorithm and may outweigh the negative effects of the others. We also detail how RFQA may be implemented experimentally with current technology.

\end{abstract}

\maketitle

\section{Introduction}

The possibility of fault tolerant digital quantum computing, where arbitrary quantum algorithms can be executed with noisy qubits and gates given polynomial overhead for error correction, provides much of the promise of quantum computing in the long term. Fault tolerance is guaranteed by the celebrated threshold theorem \cite{aharonov1999fault,knill1998resilient} and a zoo of topological error correction codes \cite{terhal2015}, though the realistic overhead for truly error-free quantum computing is formidable, requiring likely hundreds or even thousands of physical qubits per logical qubit \cite{fowlersurface}. As of this writing, such devices are thought to be at least a decade away. It is hoped that some low-depth quantum algorithms, such as the Variational Quantum Eigensolver \cite{peruzzo2014variational,mcclean2016theory} and Quantum Approximate Optimization Algorithm \cite{farhi2014quantum}, can provide a quantum speedup for realistic problems given a low but finite error rate, but it remains to be seen if this is the case. 

Further, while other approaches to quantum computing, such as quantum annealing \cite{finnila1994quantum,kadowakinishimori1998,das2008colloquium,johnson2011quantum,boixo2014evidence,albashlidar2017}, exist and are able to tolerate noise to some degree, rigorous evidence of a quantum speedup in any realistic implementation of these systems remains elusive (see \cite{albash2018demonstration} and references therein for an extensive discussion). Though quantum annealing has been shown to be formally equivalent to the gate model \cite{PhysRevLett.99.070502,aharonov2008adiabatic}, and algorithms with a provable quantum speedup, such as the adiabatic formulation of Grover's search problem \cite{grover1997,zalka1999,rolandcerf2002,yoderguang2014,dalzellyoder2017,jiang2017near}, exist, these proofs generally break down in the presence of noise and finite control precision. And while quantum annealers do not \textit{need} error correction as obviously as the gate model does, their analog nature makes full quantum error correction unworkable \cite{sarovaryoung2013,youngsarovar2013}. That said, a more modest scheme called Quantum Annealing Correction has shown empirical benefits \cite{pudenzalbash2014,pudenzalbash2015,vincialbash2015,vinci2016nested}. A variant of quantum annealing for which a noise-tolerant quantum speedup is supported by rigorous theoretical evidence would thus be extremely valuable.

% Second, ``appropriate realistic bounds on problem size and runtime," respects that there are some problems where even the best possible quantum algorithm quickly becomes impossible to execute given a large enough problem size. For example, a hypothetical quantum system designed to solve the Grover problem for $N$ spins which, due to noise effects, is expected to lose its quantum speedup beyond $N=300$, could very well still be considered fault welcoming or tolerant, since even an optimal quantum algorithm for $N=300$ would take $\sim 2^{150}$ steps and no (mortal) user could be expected to wait that long! We thus interpret these bounds to be the largest practical problem sizes that a user might conceivably submit and runtimes that a user might conceivably be willing to wait.

% It is not our goal here to prove conclusively that fault welcoming quantum computing is a near term prospect, though we will later argue that two somewhat unrealistic classes of baths for quantum annealing (strictly zero temperature, and ultrafast thermalization, both with all control noise absent) clearly would make the system fault welcoming. 

To address this need, we propose a variant of quantum annealing with coherent oscillations of the transverse fields, which we call RFQA\footnote{The acronym RFQA has the dual meaning of random field quantum annealing and radio frequency quantum annealing.}. Our scheme is capable of dramatically accelerating the search for ground states of frustrated spin glasses, an NP-hard classical problem \cite{lucas2014ising}. It does so by accelerating %accelerating 
collective many-qubit spin rearrangements via a novel mechanism, an exponential proliferation of weak multi-photon resonances generated by applying an extensive number of low-frequency oscillating fields to analog quantum annealing. This results in a reduced difficulty exponent $\Upsilon=\log {\rm TTS }\of{N}/N$, where ${\rm TTS}$ is the time to solution and N, number of qubits.
 We specifically apply it to the Grover problem and closely related quantum random energy model (QREM) \cite{farhigoldstone2008,baldwinlaumann2016,baldwinlaumann2017,baldwin2018quantum,faoro2018non,smelyanskiy2019intermittency}, and show that our method produces a quantum speedup with only inverse polynomial precision in all control parameters. Since it relies on real-time dynamics and approximate resonance conditions, it cannot be efficiently simulated by classical machines. This is in contrast to more basic forms of quantum annealing, which can often (though not always) be efficiently simulated in quantum Monte Carlo \cite{isakov2016understanding,andriyash2017can,jiang2017scaling,jiang2017path,king2019scaling}. We also consider how this system responds to the well-studied noise model of superconducting flux qubits \cite{harris2010experimental,bylandergustavsson2011,yangustavsson2013,yangustavsson2015,webersamach2017}. This model consists of two primary noise channels: $1/f$-like local potential fluctuations, and energy exchange with a finite (low) temperature bath via local spin couplings. In doing so, we show that the resulting quantum speedup can survive against phase noise and local heating, though both of these effects do degrade performance. Further, we show that bath-assisted %phase 
 transitions (including cooling after diabatically missing a phase transition) are also exponentially enhanced by the oscillating fields. This channel may thus dominate the negative influence of phase noise and heating, and possibly even find the solution more quickly than if the bath were absent. And while the Grover problem and QREM have no realistic analog implementation, the QREM in particular is phenomenologically similar to many other hard spin glass problems (including NP-complete ones). Thus, methods which accelerate cooling and thermalization in it will be widely applicable to more realistic problems. 

% \textit{Fault Tolerant (User's view)}: An extensible quantum computing system which, for a given class of problems and appropriate realistic bounds on problem size and runtime, displays a quantum speedup over the best known classical algorithms, and when \textit{all} realistic uncontrollable noise sources are considered, is capable of maintaining this quantum speedup given a polynomial overhead cost (in runtime, quantum hardware, and classical control and processing hardware) relative to the case where all uncontrollable noise sources are absent.

%

% 

% In general, we would expect that nearly any \textit{useful} quantum computing device would be need to be fault tolerant from a user's point of view. But we can go beyond this. The results of this work suggest the new possibility:

% \textit{Fault Welcoming}: An extensible quantum computing system which, for a given class of problems and appropriate realistic bounds on problem size and runtime, displays a quantum speedup over the best known classical algorithms, and when \textit{all} realistic uncontrollable noise sources are considered, finds the solution more quickly than it would if all uncontrollable noise sources were absent.

This paper is organized as follows. First, we qualitatively describe the basic mechanism of RFQA in Sec. \ref{basicsec}, and why it provides a quantum speedup. We proceed to apply it in isolated many-body problems in Sec. \ref{oraclesec}, starting with the Grover problem, where the matrix elements and resulting quantum speedup can be calculated analytically. On phenomenological grounds we expect, and later numerically demonstrate, that RFQA in the QREM displays nearly identical scaling. We then present numerical results for RFQA in the Grover problem and QREM, and show that our scheme provides a quantum speedup in excellent agreement with analytical predictions. Further, RFQA achieves this quantum speedup in regimes where  previously studied (non-oscillatory) algorithms fail. We then consider open noisy systems in Sec. \ref{noisysec}, first in the form of single qubit $1/f$ noise, and then cooling via interaction with a cold spin bath \cite{prokof2000theory}, which we model though weak local coupling to a single auxiliary degree of freedom. 
%Finally, we combine all of our previous results to argue in Sec. \ref{fwsec} that RFQA should maintain its quantum speedup in realistic limits. In fact, due to the positive influence of the cold bath it may even outperform its closed system counterpart, a condition we call Fault Welcoming. We then offer concluding remarks.
Finally, we collect and summarize our results in Sec. \ref{fwsec} emphasizing in particular their implication for the theory of error correction and speed limits in open quantum annealers. We also propose and sharpen a definition of ``fault welcoming''
as refered to quantum annealers (and other platforms more generally) where introduction of the environment leads to improved performance as compared to that of the closed quantum system.

\section{Basic mechanism of RFQA}
\label{basicsec}

\begin{figure*}
\includegraphics[width=6in]{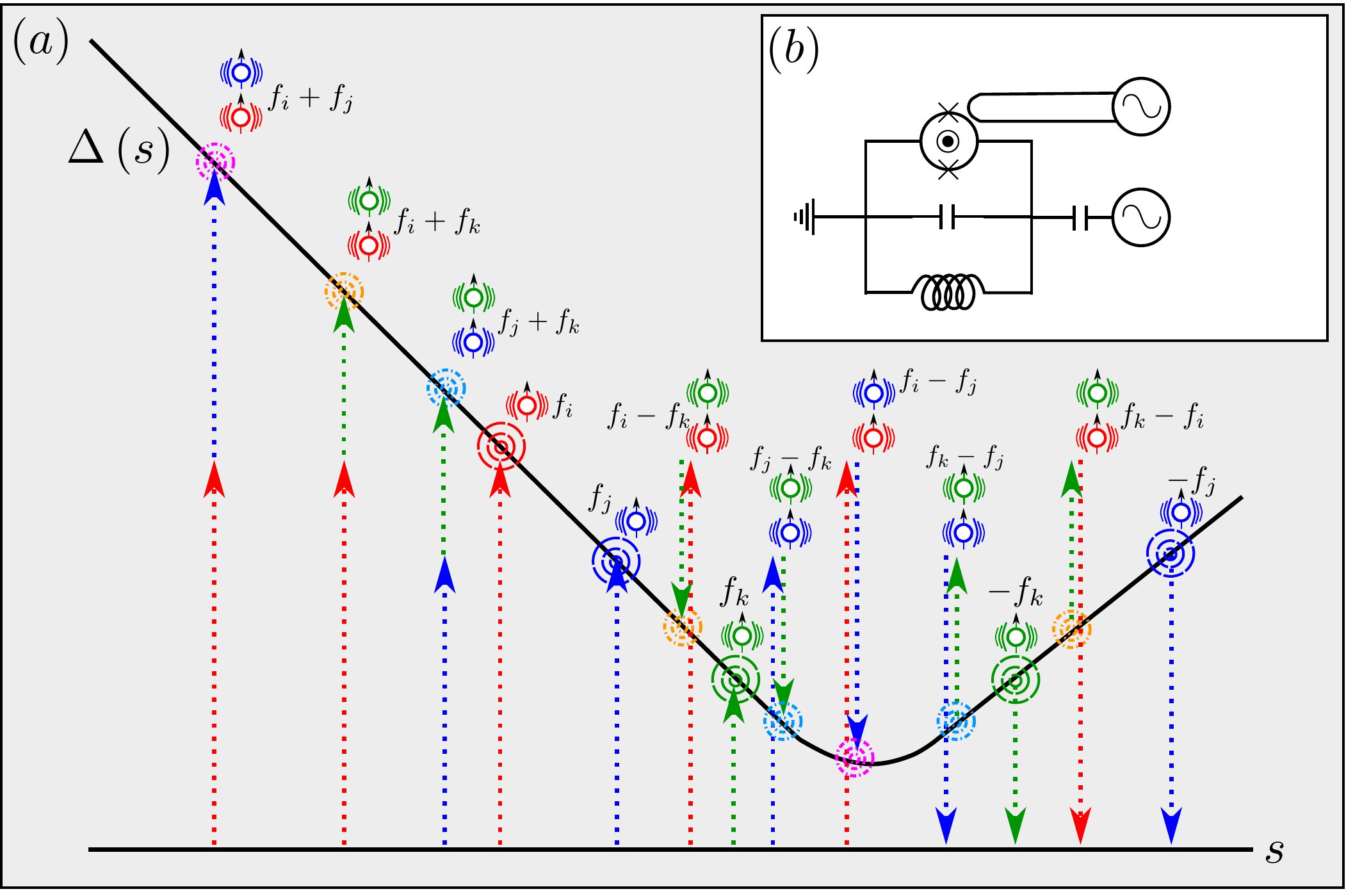}
\caption{Schematic picture for RFQA. In (a), we sketch the energy gap between competing ground states of a quantum many-body quantum system as a function of tuning parameter $s$, with a phase transition at some $s_c$, with three transverse field terms at spins $\cuof{i,j,k}$ which oscillate with frequencies $\cuof{f_i,f_j,f_k}$. Higher excited states are not shown and assumed to be at energies well above the scale of our sketch. Each of these spins produces a virtual avoided crossing whenever the gap $\Delta \of{s}$ crosses $\pm f_{i/j/k}$ (red, blue and green arrows), causing the competing ground states to mix more quickly than they would with oscillating fields absent. Further, two-photon processes produce additional resonant transitions whenever $\Delta \of{s}$ crosses the sum or difference of any pair of frequencies (combinations of colored arrows), and though we do not show them explicitly, three-photon processes produce an additional set of of transitions at the sum of any three frequencies (positive or negative). If an extensive number of oscillating sources are added this produces an exponential proliferation of weak transitions, leading to a quantum speedup. In (b) and also Appendix \ref{fluxqubitsec} we sketch a simple hardware implementation in superconducting flux qubits \cite{harris2010experimental}. An oscillatory flux applied to the qubit's SQUID loop adjusts the tunnel barrier between the two circulating current ground states, and thus oscillates the magnitude of the transverse field. Similarly, an oscillatory voltage adjusts the complex phase acquired when tunneling between minima, and consequently oscillates the effective \textit{direction} of the transverse field in the $x-y$ plane. Note that constant voltages in these qubits can be eliminated by a gauge transformation and thus have no physical consequence to the operation of the quantum computer, but time-varying voltages cannot be gauged away, with important implications for the system's dynamics.}\label{RFQAschematic}
\end{figure*}

We will introduce the RFQA method in stages, beginning with a qualitative introduction to the method and its speedup mechanism. The bulk of this work is then devoted to quantitative analysis for a specific variant of the method, in the Grover/QREM class of problems. In these structureless problems, rigorous classical speed limits are obvious and thus, a quantum speedup is easy to establish. But first, to motivate what follows, we offer a conjecture about the scaling relationship between few-body matrix elements and the minimum gap between competing ground states \textit{near} the phase transition. We argue that this conjecture will be true for a vast array of problems. And while the truth of this conjecture is not required for RFQA to succeed, assuming it makes the resulting speedup mechanism much easier to understand.

\subsection{Transition matrix element scaling conjecture}\label{mscalesec}

The transition matrix element scaling conjecture, which we call MSCALE, can thus be stated as follows. 
\begin{conjecture}
MSCALE: Consider a class of problem Hamiltonians $\cuof{H_P}$ and a driver Hamiltonian $H_D$ (which may be more complex than a simple transverse field), defined over connected graphs of $N$ spins. The total system Hamiltonian is controlled by one or more tuning parameters $\cuof{s}$. We assume that these problems are disordered quantum spin glasses, and that they exhibit one or more ground state  crossings (first order quantum phase transitions) when the $\cuof{s}$ are tuned. The minimum gap $\Delta_{min}$ is typically (though not always) exponentially small in $N$ at first order transitions. Let us imagine that we are in the vicinity of such a transition, such that the energy difference between the competing ground states $\ket{G_0}$ and $\ket{G_1}$ is $O \of{1}$ or inverse polynomially small, but large compared to $\Delta_{min}$. Now consider a class of $k$-body spin operators $\cuof{O_{i}^{(k)}}$, with $k \ll N$, which are not necessarily spatially local. Then, as $N$ becomes large, in almost all cases and ignoring subleading corrections
\begin{eqnarray}\label{MSCALE}
\avg{ \abs{\bra{G_0} O_i^{(k)} \ket{G_1} }^2 / \Delta_{min}^2 } = C_O^{(k)},
\end{eqnarray}
where the average is taken over different spin locations for the same type of operator (e.g. averaging over position $i$ in $\sigma_i^y$, positions $i$ and $j$ in $\sigma_i^z \sigma_j^x$, etc.) and $N$-spin problem Hamiltonians drawn from the class $\cuof{H_P}$, $\Delta_{min}$ is the minimum gap (as a function of $N$) for a given transition, and $C_O^{(k)}$ is a constant which depends on the type of operator, the number of involved spins $k$, and the class of problem Hamiltonians, but \textit{not} on $N$.
\end{conjecture}
In other words, MSCALE\footnote{Based on numerous conversations with other researchers in the field of quantum optimization, we feel that a loose version of this conjecture is tacitly assumed to varying degrees in the thinking of most scholars working on these problems. This is particularly true in studies which consider to how a many-body quantum spin glass interacts with a bath. However, we have been unable to find it stated in print, so we do so here.} states that the transition rate between competing ground states through few-body operators in quantum spin glasses has the same overall scaling with system size as the $N$-body tunnelling rate itself, i.e. $\Delta_{min}$. We expect MSCALE to hold in a very diverse range of problems, and if one could craft models violating MSCALE, these would have to be rather pathological (see Appendix \ref{s:mscaleviolation}). For example, in models where the tunneling rate can be computed perturbatively in the transverse field \cite{pietracaprina2016forward,baldwinlaumann2016,baldwinlaumann2017,scardicchio2017perturbation,baldwin2018quantum}, MSCALE holds by inspection. Lastly, we will also need for MSCALE to hold in a certain statistical sense even when applied to long operator strings, with $k/N$ finite.  This assumption may or may not hold in specific models but does in all the models considered in this work and a few others that we have checked.
\subsection{Exponential proliferation of weak resonances}
Assuming that MSCALE is true, we now consider the effect of oscillating fields on quantum annealing. Consider first a single oscillating term acting on spin $j$ at frequency $f_j$. We treat this term as a perturbation to $H \of{t}$, which drives a transition between two competing ground states $\ket{0}$ and $\ket{1}$, with Rabi frequency $ \Omega_j=\gamma_j \Delta_{min}$. Here, $\gamma_j$ is some small $O \of{1}$ constant which depends on the amplitude of the drive field and the matrix element of the corresponding spin operator. Note that the strength of the drive field term itself is not necessarily small, and the exponential suppression of the resulting matrix element is a consequence of the small gap near first order transitions. We further assume that the \textit{local} gap $\Delta_{local} \gg f_j$ so that the system remains in the subspace of $\ket{G_0}$ and $\ket{G_1}$ with negligible probability of being excited outside of it ($\Delta_{local}$ is defined as the minimum energy of excited states reached by single spin flip operations with $O \of{1}$ or $O \of{1/N}$ matrix elements).
This constraint on the applied frequencies is discussed more extensively in section~\ref{heatsec}. In sum, we assume the hierarchy of scales:
\begin{eqnarray}\label{scales}
\Omega_j \ll \abs{E_0 - E_1 }, f_j \ll \Delta_{local}
\end{eqnarray}
If energy difference between the two states sweeps across an energetic range $W$ over a time $t$, and the range includes $E_1 - E_0 = \pm f_j$ then a simple Fermi's golden rule calculation predicts that the two states will be mixed with a probability $P_{0 \to 1} \simeq 0.5 \of{1 - e^{-4 \pi \Gamma_{T} t}}$, where $\Gamma_T = \abs{\Omega_j^2}/W$. Note that if the primary phase transition at $E_1 - E_0 \simeq 0$ is crossed this will also contribute to $\Gamma_T$. Because the system may mix the states by either absorbing or emitting a photon, this result is equally valid for both positive and negative $\of{E_1 - E_0}$. Finally, since the system is being continuously driven, its asymptotic occupation of both states is equal to 1/2 as $t \to \infty$.

Now consider the addition of a second oscillating term with frequency $f_k$ at spin $k$. If the energetic range $W$ includes both $f_j$ and $f_k$, one would naively calculate a new mixing rate $\Gamma_T = \of{ \abs{\Omega_j^2} + \abs{\Omega_k^2} }/W$ since both processes contribute to the mixing of the two states. However, as shown in Fig.~\ref{RFQAschematic}, this is an underestimate, since are now higher order processes at work as well. Let the range of the sweep include $E_1 - E_0 \simeq f_j + f_k$ and assume the state begins in $\ket{G_0}$; at this point the system is near resonant with a process where the system absorbs a \textit{virtual} photon at site $j$ ($k$), virtually creating an excitation by flipping site $j$ ($k$), and then absorbs a second photon at site $k$ ($j$) which mixes that virtual excited state with the real ground state $\ket{G_1}$. The Rabi frequency for this process is $\gamma_{jk} \Delta_{min}$, where $\gamma_{jk}$ is a small number which scales as the product of $1/\Delta_{local}$ and the strength of the drive field but, per MSCALE, is \textit{not} on average expected to scale appreciably\footnote{In cases where $\Delta_{local}$ has inverse polynomial scaling in $N$, we expect that the bare drive amplitudes and/or frequencies $f_i$ would be similarly reduced if we were to remain in the vicinity of the phase transition, ensuring constant scaling; see our derivation of the quantum speedup in the Grover problem for an example of this effect.} with $N$. There are a total of four such resonances, all of which contribute in simple summation to $\Gamma_T$ (see section~\ref{bangbangsec}) provided that the spacing of the resonances is large compared to the (exponentially small) $\Omega_j, \Omega_k$ and $\Omega_{jk}$. The same argument can be made at third order if a third term is added at spin $l$, with eight third order resonant contributions.

The possibility of a true quantum speedup arises when $N$ oscillating sources are included. Let us assume that $K \leq N$ spins change configuration in the phase transition; oscillating terms applied to spins whose configuration does not change between $\ket{G_0}$ and $\ket{G_1}$ will probably not appreciably contribute to the mixing rate. At first order, there are $2K$ resonances, and second order $4 \binom{K}{2}$, at third order $8 \binom{K}{3}$, and so forth, and though the average strength of each resonance is expected to decrease exponentially in its order, the combinatorical explosion of terms more than balances this out and can increase $\Gamma_T$ by a factor which is exponential in $K \sim O \of{N}$.

To be more concrete, let us assume that the $m$th order term $\Omega^{(m)} \simeq \Lambda^m \Omega_0$, where $\Lambda < 1$ (here we assume $\Lambda$ incorporates the amplitude of the oscillating fields, matrix elements from the operators being driven, the local gap, and other details) and $\Omega_0 \equiv \Delta_{min}/2$ is the mixing rate from the primary phase transition itself. This form is inspired by MSCALE and the basic scaling structure of $m$-th order perturbation theory, and ignores subleading corrections. It is quantitatively true for the Grover/QREM problem classes (see section~\ref{grovcalcsec}) and likely correct in more general cases as well. As before, we assume the energy difference between the two ground states is averaged over a range $W$ (which could be the range annealed over in an adiabatic sweep, the range from which a pause point is guessed in annealing with a pause, or the width of the band of solutions in a population transfer algorithm). Fermi's golden rule predicts a total solution rate
\begin{eqnarray}\label{RFQAbaseform}
\Gamma_T = \frac{\Omega_0^2}{W} \sum_{m=1}^K \of{2 \Lambda^{2}}^m \binom{K}{m} \stackrel{K\to N}{=} \frac{\Omega_0^2}{W} \of{1+ 2 \Lambda^2 }^N.
\end{eqnarray}
A more detailed derivation of this result can be found in Appendix~\ref{bangbangsec} and below, for the case of Grover search. It assumes that the oscillating frequencies are small compared to the local gap $\Delta_{i,local}$, but \textit{large} compared to $\Delta_{min}$, the hierarchy of scales in Eq.~\ref{scales} which is generally easy to satisfy in hard quantum spin glass problems. For the problems considered in this work, this results in a polynomial quantum speedup (i.e. reduced difficulty exponent $\Upsilon>0$)
over both constant rate annealing (without oscillating fields) and classical search algorithms; whether or not there are useful problems where RFQA is capable of an exponential quantum speedup ($\Upsilon=0$, e.g. reducing an average exponential time to solution to a polynomial) is an issue %which will be 
left to future studies.

% \footnote{The result (\ref{RFQAbaseform}) also assumes that the adiabatic sweep is not perfectly uniform, lest a precision interference effect cause state rotations from positive frequency terms on one side of the transition be undone by negative frequency terms on the other side. This effect, which arises from reflection symmetry about the transition point in the limit that the frequencies $f_i$ are large compared to the effective (exponentially small) drive amplitudes $\Omega_i$, is frustrated by asymmetries in the trajectory or matrix elements across the transition, and by random noise. Thus, this concern is unlikely to matter in real systems, but can arise in theoretical simulations where such complexities are absent, and we employ non-uniform annealing schedules in all subsequent calculations in this paper.} 
\subsection{Choice of the driver Hamiltonian}
So far we have avoided specifying a physical model for our oscillating sources, and there are many possible formulations of RFQA. All of these modify the driver Hamiltonian and leave the problem Hamiltonian unchanged, and are compatible with variations in the annealing schedule, such as annealing with a pause or reverse annealing \cite{kadowaki2019experimental,marshall2019power,king2019quantum}. The simplest method, which we call RFQA-D (with D referring to the operator Direction or electric Displacement), is the focus of this work, and oscillates the \textit{directions} of the transverse fields in the driver Hamiltonian:
\begin{eqnarray}
\label{e:isospectral}
H_D = - \kappa \sum_{i=1}^N & & \left [ \cos \of{\barA \sin \of{ 2 \pi f_i t+ \varphi_i}} \sigma_i^x \right. \\
& & \left. +  \sin \of{\barA \sin \of{ 2 \pi f_i t+ \varphi_i}} \sigma_i^y \right ]. \nonumber
\end{eqnarray}
As we describe in Appendix~\ref{fluxqubitsec}, RFQA-D can be implemented in current flux qubit hardware using oscillating electric fields, and it has the elegant property that when combined with a problem Hamiltonian in the $z$ basis, it preserves the \textit{instantaneous} energy spectrum in evolution, since it is equivalent to the standard transverse field Hamiltonian through a simple basis rotation. This allows us to make a number of analytical predictions, and perform specialized numerical calculations, that would be difficult or impossible to formulate in other contexts; the predictions in this work are all based on RFQA-D \footnote{One could also oscillate the magnitudes of the transverse fields (RFQA-M), or the magnitudes and/or directions of transverse coupling elements (RFQA-C); more exotic variants of RFQA could be implemented using novel qubit designs \cite{kerman2019superconducting} or through simulation of RFQA evolution in a digital quantum computer.}.

% \subsection{Variants}

% A second method, RFQA-M, oscillates the \textit{magnitudes} of the transverse fields (assume the amplitudes $\barA_i < 1$):
%\begin{eqnarray}
%H_M = - \kappa \sum_{i=1}^N \of{1 + \barA_i \sin \of{ 2 \pi f_i t+ \varphi_i} } \sigma_i^x.
%\end{eqnarray}
% Unlike RFQA-D, this method does not preserve the instantaneous energy spectrum of the evolving system as the fields oscillate, making analytical predictions more difficult. Finally, one could also include tunable transverse coupler elements between qubits, and oscillate their magnitudes and/or directions; we call this very general class of methods RFQA-C. We will focus on RFQA-D in this work as it is the most natural for superconducting flux qubits; the application of other RFQA variants to interesting trial problems will be the subject of future studies. However, the basic mechanism, an exponential proliferation of weak resonances leading to accelerated tunneling between minima in hard optimization problems, is generic to all forms of RFQA, and we expect many of the phenomenological effects demonstrated here to occur in other problem classes as well. More exotic variants of RFQA could be implemented using novel qubit designs \cite{kerman2019superconducting} or through simulation of RFQA evolution in a digital quantum computer.

\section{RFQA and Oracle Problems}\label{oraclesec}

\subsection{Problem Hamiltonians}

To demonstrate the promise of the tunneling acceleration predicted in (\ref{RFQAbaseform}), we now consider the application of RFQA-D to two related oracle problem classes. The first class is an analog variant of the Grover problem \cite{grover1997,zalka1999,rolandcerf2002,yoderguang2014,dalzellyoder2017,jiang2017near}. Specifically, we consider a variant of the Grover problem with $M$ target states $\cuof{\ket{G_i}}$, but with small variations in the weight of each state:
\begin{eqnarray}\label{HGM}
H_{GM} = - \sum_{i=1}^{M} \of{\frac{1}{2}+\delta_i} \ket{G_i} \bra{G_i}.
\end{eqnarray}
Here, the $\ket{G_i}$ are $M$ random bitstrings, and $\delta_i \propto O \of{1/N}$ are random offsets; in our simulations we choose $\delta_i$ to be uniform random real numbers chosen from the range $\cuof{- 0.27/N, 0.27/N}$. There is no special significance to the precise size of the energetic range other than that it creates well-defined bands at low energy. 

The second class, which we refer to as a banded quantum random energy model (BQREM), is generated by the following sequence of steps. First, in step (i) we take the diagonal entries of the total spin operator $\frac{1}{2N} \sum_{i=1}^N \of{\sigma_i^z}$. In step (ii), we then randomly choose $M-1$ random bit strings which are not $\ket{0000 ... }$ (out of the $2^N$ total entries) and set their entries equal to $-1/2$; this ensures a ground state band of $M$ total states. For step (iii), we add to each of the $2^N$ entries an independently chosen random offset $\delta_i$ as in $H_{GM}$. Finally, in step (iv) we randomly shuffle the positions of all the $2^N$ entries. The resulting problem Hamiltonian has $N$ bands of width $\propto 1/N$ each, with a density of states in each level given by the binomial distribution. We choose this prescription instead of Gaussian random values to ensure a well-defined ground state band of precisely $M$ states in each problem instance. Note that, due to the randomization of excited state energies, when a transverse field is applied the perturbative corrections to the energies in the BQREM are state dependent and broaden each band to an inverse polynomial width even if all $\delta_i = 0$ \cite{faoro2018non}.

As with the standard Grover problem, information theoretic bounds ensure that no classical algorithm can find a solution to either of these problems in less than $O \of{2^N / M}$ time, and no quantum algorithm is capable of more than a square root speedup over this \cite{zalka1999}. However, because of the inverse polynomial energy uncertainty, we are unaware of any existing quantum algorithms which would maintain this speedup in the large $N$ limit. 

Though they have no realistic analog implementation, we choose to benchmark RFQA using the Grover problem and BQREM for a variety of reasons. First, they are among the few problem classes where rigorous speed limits can be defined for both classical and quantum approaches; given $M$ target ground states and a Hamiltonian energy scale which is $O \of{1}$, no classical algorithm can find one in less than $O \of{2^{N}/M}$ time, and no quantum algorithm can find one in less than $O \of{\sqrt{2^{N}/M}}$ time. Any average time to solution scaling in between these bounds thus represents a true quantum speedup. This speed limit comes from the lack of any ``guidance" in Hilbert space toward the solution states; adding basins of attraction to the problem Hamiltonian can reduce the difficulty exponent, though the time to solution is still exponentially long unless these basins are extensively wide \cite{atia2019high}. Second, their simple structure allows for the matrix elements and difficulty scaling to be predicted analytically, something which is typically impossible in more general cases. Further, the random energy model can be viewed as the limit of a spin glass with exponentially many multi-body interactions, and many phenomenological features of the phase diagram of these models are shared with more realistic spin glass Hamiltonians \cite{baldwinlaumann2016,baldwinlaumann2017}. 

We will now calculate the magnitudes and quantities of the matrix elements $\Omega_k$ for the Grover problem, and show that RFQA is capable of producing a real quantum speedup (though not an optimal one) for unstructured search, in a manner which tolerates realistic local noise and does not require exponential precision to succeed. 

\subsection{Analytical formalism}\label{grovcalcsec}

To predict the matrix elements, and, ultimately, quantum speedup, we fully diagonalize the system for a problem Hamiltonian strength $s$ near the transition point $s_c$. To do so, we construct a perturbation theory in $s H_{G1}$ using $H_D$ 
(with all $f_i=\phi_i=0$) as our unperturbed basis, and find the corrections to $\ket{G}$ by requiring that it is orthogonal to all other states to leading order in $2^{-N/2}$. At this order the full $2^N$ problem is reduced to mixing $N+1$ states only, as we demonstrate explicitly below.
%Formally, this is an overcomplete basis of $2^N + 1$ states, but since the spectral weight of the single extra state is drawn from the middle of the spectrum the overcompleteness has negligible effect on the low-energy physics. 
Our instantaneous Hamiltonian is:
\begin{eqnarray}
\label{eq:Hinst}
H = -\frac{1}{2N} \sum_{i=1}^N \sigma_i^x - s  \ket{G} \bra{G}
\end{eqnarray}
Let $c_i \equiv \bra{G} \sigma_i^z \ket{G}$, and  let $\ket{1} \equiv \frac{1}{\sqrt{N}} \sum_{i=1}^N c_i \sigma_i^z \ket{0}$, $\ket{2} \equiv \frac{1}{N} \sum_{i,j} c_i c_j \sigma_i^z \sigma_j^z \ket{0}$, and so on. We can then compute the transformed eigenstates as:
\begin{eqnarray}\label{Gstates}
\ket{0'} &=& \ket{0} + 2^{-\frac{N}{2}} \frac{s }{ \of{E_0 - E_G}} \ket{G} + O \of{2^{-N}}, \\
\ket{1'} &=& \ket{1} + 2^{-\frac{N}{2}} \frac{s N^{1/2} }{ \of{E_0 + 1/N - E_G}} \ket{G} + O \of{2^{-N}}, \nonumber \\
\ket{2'} &=& \ket{2} + 2^{-\frac{N}{2}} \frac{s N }{ \of{E_0 + 2 /N - E_G}} \ket{G} + O \of{2^{-N}} \nonumber \\
\ket{G'} &=& \ket{G} - 2^{-\frac{N}{2}} \of{ \frac{s  }{ \of{E_0 - E_G}} + 1 } \ket{0} \nonumber \\
& & - 2^{-\frac{N}{2}} \of{ \frac{s N^{1/2} }{ \of{E_0 + 1/N - E_G}} + 1 } \ket{1} \nonumber \\
& & - 2^{-\frac{N}{2}} \of{ \frac{s N }{ \of{E_0 + 2/N - E_G}} + 1 } \ket{2} + ... \nonumber
\end{eqnarray}
We will now restore oscillations %consider 
in one of the transferse fields (cf. Eqs. \ref{eq:Hinst} and \ref{e:isospectral}), expand to linear order in $\alpha$ and compute the resonant matrix elements for the mixing of $\ket{0'}$ and $\ket{G'}$.% through the oscillating $y$ fields.

Let us first consider the transition rate $\Omega_1$ for the mixing of $\ket{0'}$ and $\ket{G'}$ through an oscillating field Hamiltonian driving a single spin as $\alpha \frac{1}{2N} \sigma_i^y \sin \of{2 \pi f_i t}$ (equivalent to Eq. \ref{e:isospectral} at small $\alpha$). From (\ref{Gstates}), with $f_i = E_{G} - E_{0}$ we can immediately read off the matrix element $\Omega_1$ as
\begin{eqnarray}\label{Om1}
\Omega_1 &\simeq& \pm \frac{\alpha}{4 N} \bra{G'} \sigma_i^y \ket{0'} \simeq   \frac{\alpha s 2^{-\frac{N}{2}}}{ 8 N \of{E_0 + 1/N - E_G}} \\
& \simeq & \frac{\alpha}{4}  s_c  2^{-N/2}
\end{eqnarray}
as $\bra{1} \sigma_j^y \ket{0} = \pm i / \sqrt{N}$. Taking the limit of $\of{ E_G - E_0 } \ll 1/N$, reflecting the hierarchy of scales in Eq.~\ref{scales}, and $s \to s_c$ yields the rate of the second line, $\Omega_1 \simeq \alpha \Omega_{0} / 4$. We take this limit as we expect the drive frequencies inducing these transitions to decrease polynomially with $N$ (remaining large compared to $\Delta_{min}$, which decays exponentially), for reasons explained below in section~\ref{heatsec}. The denominator thus reduces to powers of $\of{1-s}$, which is constant as $N$ increases. 

Now imagine we drive two spins at amplitudes $\alpha$ and frequencies $f_1$ and $f_2$. If $\abs{f_1 \pm f_2} \simeq E_{G} - E_{0}$ the system will be resonantly driven between $\ket{0'}$ and $\ket{G'}$ through a two-spin process, where one spin absorbs an off-resonant photon, virtually exciting it into the $\ket{i}$ manifold, and the second spin then absorbs a second photon, promoting it to $\ket{G'}$ through the component of $\ket{G'}$ along $\ket{2'}$. Noting that combinatorics will provide a factor of two increase (from the order in which photons are absorbed),
\begin{eqnarray}\label{Om2}
\Omega_2 &\simeq&   \frac{ \alpha^2 s  2^{-N/2}}{8 N^2 \of{1/N - f_{1/2}} \of{E_0 + 2/N - E_G}} \\
&\simeq& \of{\frac{\alpha}{4}}^2 s_c  2^{-N/2}.  \nonumber
\end{eqnarray}
Again, the rate in the second line is in the limit $\of{ E_G - E_0 } \ll 1/N$ and $s \to s_c$.

We can further extend these results to three spins, driven at frequencies $f_{1/2/3}$. The same arguments yield
\begin{eqnarray}\label{Om3}
\Omega_3 &\simeq& \of{\frac{\alpha}{4}}^3 s_c  2^{-N/2},  \nonumber
\end{eqnarray}
as the factor of 6 from combinatorics balances the denominator of $6/N^3$. Extending this result to $m$ spins, we finally conclude:
\begin{eqnarray}\label{OmmG}
\Omega_m \simeq \frac{\alpha^{m}}{4^{m}} s_c   2^{-N/2} = \of{\frac{\alpha}{4}}^m \Omega_0 .
\end{eqnarray}
Analytically calculating these amplitudes is significantly more difficult in the BQREM, but given that both models obey MSCALE, and that the paramagnet structure survives to some degree near the paramagnet to spin glass phase transition \cite{farhigoldstone2008,baldwinlaumann2016,baldwinlaumann2017,baldwin2018quantum,faoro2018non,smelyanskiy2019intermittency}, we expect nearly identical scaling there. This is confirmed by the extensive numerical simulations we present in sections ~\ref{numericsec}-\ref{coolingsec}, where the performance boost from RFQA-D is nearly identical in both models, with BQREM generally tending to display a slightly larger speedup. 

Given these results we can now sum the effect of tone combinations at all orders and predict the time to solution for the Grover problem boosted by RFQA. At zeroth order, we have the minimum gap itself, with $\Omega_0 \simeq s_c  2^{-N/2}  = \Delta_{min}/2$. At first order we have a total of $2N$ contributions, $N$ from positive frequencies ahead of the transition and $N$ from negative frequencies after the minimum gap has been crossed. At second order we have $\binom{2N}{2}$ independent terms, as each contribution of two frequencies (positive or negative) drives an independent transition between the two states. At third order we have $\binom{2N}{3}$, and so on; summing all contributions as in Eq.~(\ref{RFQAbaseform}) results in a total solution rate of
\begin{eqnarray}\label{maxGrate}
\Gamma_T \of{\alpha} \simeq \frac{s_c^2 }{2^N W} \sum_{n=0}^{N} \of{\frac{\alpha^2}{8}}^{n} \binom{N}{n} = \frac{s_c^2}{2^N W} \left(1+\frac{\alpha^2}{8}\right)^N,
%\propto e^{ - \of{\log 2 - \frac{\alpha^2}{8 } } N }.
\end{eqnarray}
which demonstrates a reduced difficulty exponent 
\begin{equation}
\Upsilon=\log 2-\log(1+\alpha^2/8).
\label{eq:upsilon}
\end{equation}
As we shall see next, this expression, derived perturbatively, appears to be rather accurate at least up to $\alpha\sim 1$, in part due to the isospectral nature of the driver Hamiltonian we chose.
\subsection{Optimal drive amplitudes}\label{optalphasec}
The result in Eq.~\ref{eq:upsilon} is a clear demonstration that the difficulty exponent $\Upsilon$ is dynamically reduced. We will now return to the driver Hamiltonian in Eq. \ref{e:isospectral} to estimate the optimal value of $\alpha$, and thus, the ultimate quantum speedup for this problem.
%We can predict the maximum performance of RFQA-D in this problem 
from simple spectral analysis. We first note that since the oscillating terms $\barA_i \sin 2 \pi f_i t$ are enclosed in trigonometric functions in $H_D$ there is a nonlinear relationship between the raw amplitude $\barA_i$ and the physical driving amplitude $\alpha_i$ responsible for driving transitions. To predict the maximum performance of RFQA-D we must find the optimal value of $\barA$ (which we call $\barA_m$). To find $\barA_m$, we fourier transform $\sin \of{\barA \sin \of{ 2 \pi f_i t} }$:
\begin{eqnarray}
\sin \of{\barA \sin \of{ 2 \pi f_i t} } = c_1 e^{2 \pi i f_i t} + c_3 e^{6 \pi i f_i t} + ... + {\rm H. c.}
\end{eqnarray}
All of these terms contribute to the driven many-photon transitions, though in practice terms at fifth order and higher are negligible. We can therefore find $\barA_m$ by maximizing the sum $\sum_j c_i \of{\barA}^2$, since all these terms enter quadratically into the sum of many-photon transitions. This in turn amounts to maximizing the sum of squared Bessel functions $ \sum_{k=0}^{\infty} J_{2k+1}^2 \of{\barA}$; from that we arrive at an effective optimal %maximum 
drive amplitude
\begin{eqnarray}\label{maxA}
\alpha_{opt}^2 \simeq 1.4,
\end{eqnarray}
at a raw drive strength $\barA \simeq 0.59 \pi$. Values of $\barA$ larger than this are counterproductive, as they produce weaker coefficients at low orders and increase the possibility of generating off-resonant excitations; for the problems considered in this work numerics showed no benefit increasing $\barA$ beyond this value. Further, very large values of $\barA$ may violate some of the perturbative assumptions used to derive our results. 
Eqns. \ref{eq:upsilon} and \ref{maxA} together %Plugging $\alpha = \alpha_{opt}$ into the sum of transition rates predicted above in (\ref{maxGrate}) thus allows us to 
provide an estimate of the optimal performance of RFQA-D for a given problem\footnote{Of course, for problems with more structure, the optimal $\barA$ may depend on the problem class or even on the details of a given instance. However, our argument is generic enough to provide a good starting point for a broad range of cases.}. For the Grover problem, we arrive at an improved difficulty exponent (recall $\log 2\approx 0.69$) and 
an average solution rate:
\begin{eqnarray}\label{Grovrate}
\Upsilon\approx 0.532, \Gamma_T \of{\alpha_m} \simeq 2^{-0.769N}
%{-0.747N}.
\end{eqnarray}
Inverting this to get a final time to solution, we see that it is obviously worse than the optimal runtime of $2^{N/2}$ achievable by variable rate annealing \textit{with exponential precision}. Nonetheless, it still represents a quantum speedup, and one which requires no detailed knowledge of the instantaneous gap. We will take up the question of noise tolerance of RFQA in the next Section. 
\subsection{Parametric suppression of heating}\label{heatsec}

Our analysis thus far has assumed that the system remains in the manifold of competing ground states, and if the system is excited out of this manifold through a proliferation of few-body excitations it may heat to infinite temperature given long times \cite{d2014long,PhysRevB.97.245122}, thus ruining any chance of solving the problem. % absent some cooling mechanism that returns the system to a ground state. 
Fortunately, there is a straightforward way to prevent this from occurring. We will now show that, for an extremely broad class of problems, if we polynomially decrease the applied frequencies $\cuof{f_i}$ with increasing $N$, we can exponentially suppress the heating rate from RFQA itself, ensuring that local heating does not spoil our results.

Following Eq.~\ref{scales}, we observe that, near quantum first order transitions (and in many-body localized systems \cite{nandkishore2015many}), there is typically a divergence between the \textit{global} gap $\Delta_{min}$ (the exponentially small gap between the two competing ground states of the system, the two lowest eigenvalues of $H$) and the \textit{local} gap $\Delta_{local}$. As discussed earlier, we define $\Delta_{i,local}$ as the minimum energy difference between the competing ground states and the lowest lying states which can be reached through local operations on spin $i$ with matrix elements that are at least $O \of{1/N}$. This minor update to the definition reflects that the local gap will vary from site to site in real problems. If we consider a single tone with frequency $f_i$, the quantum adiabatic theorem promises that if $f_i \ll \Delta_{i,local}$, the rate of producing off-resonant quasilocal excitations will be exponentially suppressed (see \cite{PhysRevLett.119.060201} for an extensive discussion). Since the oscillating sources are independent and large combinations of them contribute to transitions between competing ground states but not local (few-body) excitations, we can directly estimate the performance degradation from off-resonant heating. 

Let the generalized driving amplitude be $\tilde{\alpha}$, which includes the transverse field strength, $O \of{1}$ matrix elements for the local excitations, and all other details. Now consider a single cycle of evolution with $T = 2 \pi / f_i$ for a single oscillating field. The adiabatic theorem states that the probability of producing a local excitation can be stated approximately as $P_{i} \simeq e^{- 2 \pi \Delta{i,local}^2 / \tilde{\alpha} f_{i}}$. The error rate is thus given by $\Gamma_{i} = f_i P_i$. Since there are $N$ spins, for a total evolution time $T_{f} \of{N}$ we have
\begin{eqnarray}\label{heaterr}
P_{tot} \simeq N   \avg{f_i e^{- 2 \pi \Delta_{i,local}^2 / \tilde{\alpha} f_{i}} } T_{f} \of{N}.
\end{eqnarray}
Here the average is over all spins $i$. This is a simplified expression which leaves the $N$ dependence of $\Delta_{i,local}$ implicit. For the simulations detailed below, $\Delta_{i,local} \simeq 1/N$, but $\tilde{\alpha} \propto 1/N$ as well since it includes the transverse field strength, and consequently our frequency scaling choice $f_i \propto 1/N^2$ results in a local heating rate which is $O \of{ \exp \of{-N}}$. Reflecting this, in our numerics we observed negligible heating at large $N$, even for exponentially increasing runtimes. Further, real quantum annealing systems using flux qubits are coupled to cold baths by default\footnote{This is not the case for proposals to engineer quantum annealing or similar continuous-time analog quantum optimization protocols in more general, driven quantum information systems, such as trapped ions or transmon qubits, where the states are generated and manipulated through oscillating fields. In general the ``bath'' in those systems can be well approximated by infinite temperature\cite{kapit2017review}  and would pose severe challenges at large $N$ unless the system-bath coupling rates are extremely small.}, and the resulting local cooling may rapidly correct these errors, rendering this entire issue moot.

\subsection{Closed system numerical results: accelerated paramagnet to Grover state transitions}\label{numericsec}

\begin{figure}
\includegraphics[width=3.25in]{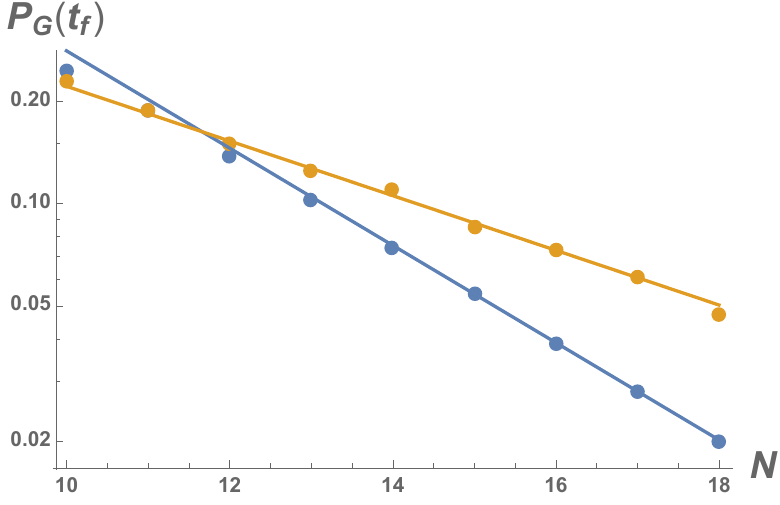}
\caption{Final average success probability of finding a single Grover state using the bang-and-wait protocol, with $s$ randomly chosen in the range $\cuof{0.35,0.55}$ and a wait time $t_f \of{N} = 0.9375 \times 2^{N/2}$, for uniform transverse field (blue points), and RFQA-D (gold points; additional parameters are given in the text). The straight lines correspond to numerical fits of $P_G \of{t_f} \simeq 7.55 \times 2^{-0.475 N}$ for uniform transverse field and $1.405 \times 2^{-0.269 N}$ for RFQA-D, confirming our predictions that any quantum speedup vanishes for uniform transverse field but persists when the transverse fields are oscillated.}\label{groverGJ}
\end{figure}

To verify the claims in Eqs. (\ref{Gstates}-\ref{maxGrate}), we numerically simulate the bang-and-wait algorithm described in appendix~\ref{bangbangsec}, for a single Grover state with $N$ running from 10 to 18. The results of our simulations are shown in Fig.~\ref{groverGJ}; in these studies we initialize the system in the paramagnetic ground state, jump to a randomly chosen $s$ between 0.35 and 0.55 (the phase transition point sits approximately at $s_c \simeq 0.5 - 1/N$; each datapoint is averaged over 960 random choices of $s$ in this range), and then wait a time $t_f \of{N} = 0.9375 \times 2^{N/2}$ before measuring the state. For RFQA we choose the optimal $\barA = 1.84$ and choose each oscillating frequency with a random magnitude in the range $\cuof{1/N^2, 2/N^2}$ and random phase. Were we to choose $s_c$ exactly for each $N$ we would recover the Grover state with $O \of{1}$ probability without using RFQA, but if we assume that we do not know the location of this point (as would be the case for most real problems), averaging over random guesses reduces the success probability to $P_G \of{t_f} \propto O \of{2^{-N/2}}$ (best fit to our data is $7.55 \times 2^{-0.475 N}$, likely due to higher order effects), which when combined with the $2^{N/2}$ runtime erases the quantum speedup. In contrast, with RFQA-D the proliferation of multi-photon resonances increases the transition rate, which in turn causes $P_G \of{t_f}$ to decay much more slowly with $N$, with a numerically extracted fit of $P_G \of{t_f} \simeq 1.405 \times 2^{-0.269 N}$, very close to the $2^{N/4}$ advantage predicted analytically. These results confirm the validity of our analytical calculations in this problem, and further bolster our expectations for a quantum speedup with oscillatory transverse fields.%by adding oscillating fields to transverse field quantum annealing.

\subsection{Closed system numerical results: quantum accelerated spin glass thermalization}\label{poptransfersec}

Recently, multiple authors \cite{baldwin2018quantum,faoro2018non,kechedzhi2018efficient,smelyanskiy2018non,smelyanskiy2019intermittency} have considered the problem of tunneling between minima in the Grover problem and quantum random energy model, and demonstrated quantum advantages to tunneling via a transverse field, with the average time to find one of $M$ total minima scaling faster than the $2^{N}/M$ limit of classical approaches. In each of these approaches the system is initialized in a known classical minimum state, a transverse field is applied for some long time $t$, and the system is then measured to see if another minimum has been found. In all these cases a quantum speedup is found, and if the transverse field strength is allowed to increase as $\sqrt{N}$ it can become asymptotically optimal.

However, in all of these algorithms the system is only capable of finding minima which are exponentially close in energy to the initial state, and local potential noise (which we expect to scale as $\sqrt{N}$ in real analog implementations) would entirely erase the quantum advantage of these protocols, just as it does in the Grover problem. Further, the increasing transverse field strength poses a significant challenge if we consider the system as a proxy for more realistic problems. Even if we were able to engineer the Grover/QREM Hamiltonian, having $H_D$ increase as $\sqrt{N}$ means that the ground states of $H_P$ are far from the true, paramagnetic ground state of the combined Hamiltonian. Consequently, the system's coupling to a cold bath would have to be extremely weak in order to ensure that the system does not relax to a paramagnetic ground state (and away from the band of Grover solutions). Thus, even if we take the QREM as a proxy for more realistic spin glass problems, the question of whether or not quantum tunneling between minima is capable of providing a provable quantum speedup over classical methods, \textit{when realistic noise sources are taken into account}, remained an open one.
%Until now. 

In this Section, we will consider the application of RFQA-D to population transfer in the Grover problem (as specified in Eq.~\ref{HGM}) and the BQREM, and show that when the transverse fields are oscillated, the system is capable of finding one of the $M-1$ other ground states in an $M$-state band of width $W \propto 1/N$ in a time which emprically scales approximately as $c_M 2^{0.76 N}$ for the Grover problem and $d_M 2^{0.75 N}/N$ for the BQREM, both very close to the $2^{0.77N}$ found for mixing between the paramagnet and Grover ground states in the previous section. These results are found with a transverse field strength chosen so that the paramagnet ground state has nearly equal energy to the problem solutions. They are thus robust against (and as we shall show in Sec.~\ref{coolingsec}, likely enhanced by) coupling to a low temperature bath.
%\begin{eqnarray}\label{HGM}
%H_{GM} = - \sum_{i=1}^{M} \of{\frac{1}{2} + \delta_i } \ket{G_i}\bra{G_i}.
% \end{eqnarray}

To attack these problems, we initialize the system at $t = 0$ in one of the $M$ ground states (in a band of width $0.54/N$) of $H_P$ with the driver Hamiltonian $H_D$ turned off, linearly ramp the transverse field strength from zero to a fixed value $\kappa_c$ in $O \of{N}$ time, wait a time $t_W \propto 2^{0.75 N}$ with both Hamiltonians turned on (this runtime scaling choice is optimal for this protocol; see appendix~\ref{optruntimesec}), and then measure the state. At all times when the transverse field strength is nonzero the field directions are oscillating with random signs and the optimal amplitude $\barA = 1.84$ predicted earlier in Section~\ref{optalphasec}. As in previous cases the $N$ frequencies are randomly chosen with $f_i \in \pm \cuof{1/N^2, 2/N^2}$. We choose $\kappa_c = \frac{1}{2} + O \of{1/N}$ for each $N$ such that the paramagnetic ground state energy is equal to the \textit{average} location of the center of the problem Hamiltonian's ground state band. Here the average is taken over random instances; the choice of $\kappa_c$ in each simulation does not depend on any knowledge of a particular $H_P$. The individual state energies, and thus bandwidth and center location, vary randomly from instance to instance. We find this prescription to be the most effective, and as argued above it would not be disrupted by couplings to a cold bath. As in \cite{baldwin2018quantum,faoro2018non,kechedzhi2018efficient,smelyanskiy2018non,smelyanskiy2019intermittency} our method is robust to small variations in transverse field strength. Our results are averaged over 800-1600 random choices of $H_P$ and the frequencies $f_i$ for each data point.

\begin{figure}
\includegraphics[width=3.25in]{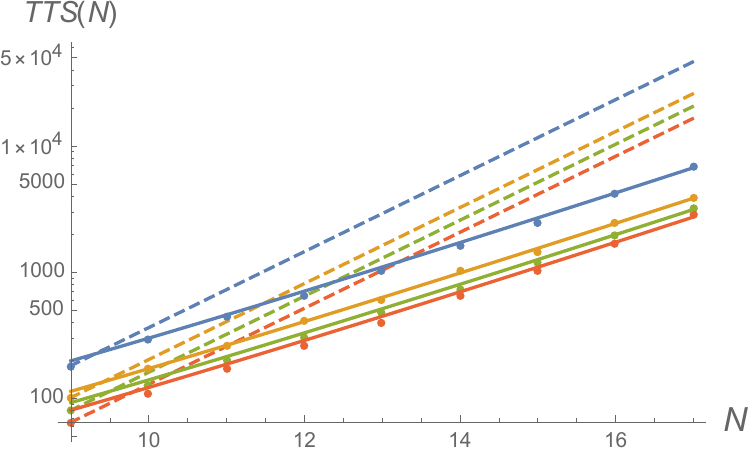}
\includegraphics[width=3.25in]{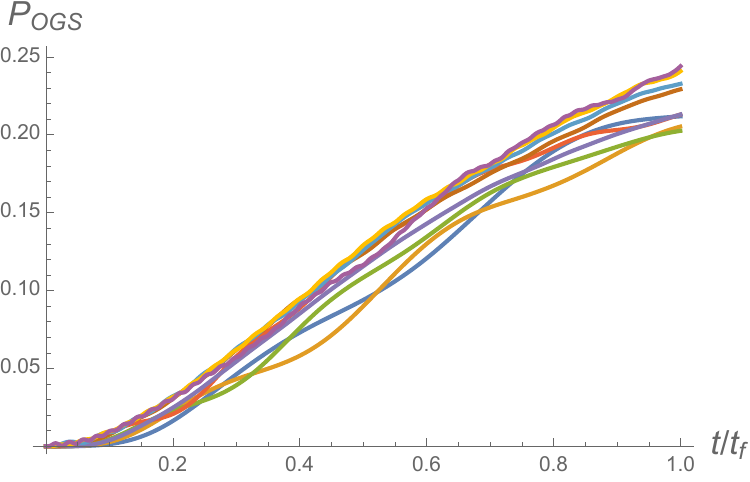}
\caption{Top: Average time to find another ground state in the banded QREM using RFQA, for $M=\of{2,3,4,5}$ (blue, yellow, green, red) computed from the final success probability for a runtime $t_f = 0.56 N + 0.0092 \times 2^{0.75 N}$ (the linear part corresponds to ramping the transverse fields from zero to $\kappa_c$; they are held at constant magnitude for the exponentially long time), with $N$ running from 9 to 17 and other parameters stated in the main text. Dots represent data, and the solid curves are fits to $c_M \times 2^{0.75 N}/N$, with $c_M$ the only free parameter. Fits which allowed the exponent to vary or removed the polynomial denominator consistently returned a scaling of $2^{\of{0.75 \pm \epsilon} N}$, with $\epsilon \leq 0.06$. Dashed lines plot $c_M' 2^{N}$, for comparison purposes. Since no classical algorithm can find another ground state in less than $O \of{2^N / M}$ time, these results demonstrate an unambiguous quantum speedup. Bottom: An example set of traces of success probability vs. time for $M=4$ total ground states with $N$ running from 9 to 17 (in ascending order in $N$, these curves are colored dark blue, gold, green, red, light purple, brown, light blue, yellow and dark purple). Runtimes are rescaled at each $N$, so curves lying on top of each other represent a quantum speedup. }\label{QREMnonoise}
\end{figure}

\begin{figure}
\includegraphics[width=3.25in]{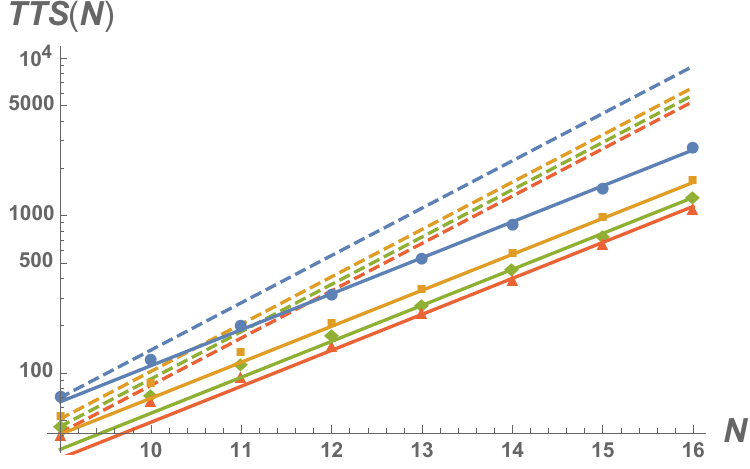}
\caption{Average time to find another state in the Grover problem using RFQA, for $M=\of{2,3,4,5}$ (blue, yellow, green, red) computed from the final success probability for a runtime $t_f = 0.56 N + 0.0092 \times 2^{0.75 N}$ (the linear part corresponds to ramping the transverse fields from zero to $\kappa_c$; they are held at constant magnitude for the exponentially long time), with $N$ running from 9 to 16 and other parameters stated in the main text. In contrast to the BQREM in FIG.~\ref{QREMnonoise}, data is best fit by $TTS \of{N} \simeq c_M 2^{0.76N}$ (without a $1/N$ prefactor). Dashed lines again plot $c_M' 2^{N}$, for comparison purposes.}\label{grovtherm}
\end{figure}

In Figs.~\ref{QREMnonoise} and \ref{grovtherm}, we present the results of extensive numerical simulations for the BQREM and Grover problems, verifying the analytical predictions made earlier. In both cases the success probability is approximately constant at $t_f \propto 2^{0.75N}$, leading to a quantum reduction of the difficulty exponent of the time to solution (TTS). Possible polynomial prefactors lead to some uncertainty in the exponent when fitting $TTS \of{N}$; averaging over $M$ ranging from 2 to 5, for the Grover problem we found a best fit of $TTS \of{N} \simeq c_M 2^{0.76N}$, whereas for the BQREM the best fit includes a $1/N$ prefactor and $TTS \of{N} \simeq c_M 2^{0.75N} / N$. Fits where the exponent was allowed to vary, with or without a $1/N$ prefactor, consistently returned a scaling of $2^{\of{0.75 \pm \epsilon} N}$, with $\epsilon \leq 0.06$.

We conclude from these results and the analytical arguments of the previous section that RFQA is capable of accelerating ``thermalization" in closed-system quantum spin glasses, where the transverse field term induces multi-qubit tunneling processes which allow the system to find other minima which are near in energy to the initial state but an extensive $O \of{N}$ number of spin flips apart. Our RFQA method significantly improves on previous results (at least in analog settings), in that while the quantum speedup found is less than the provably optimal square root scaling, we recover it even when the bands of solutions have inverse polynomial, rather than exponentially small, width. These are important results, but the ultimate focus of this paper is on open system effects, and we have not yet addressed how our derived quantum speedup will respond to a realistic noise model. We begin addressing these issues now.

\section{RFQA in open quantum systems}
\label{noisysec}
This Section considers two important modes by which the environment couples to the quantum annealer -- classical noise and cold bath. Both of these are present, and arguably dominant, in experimental realizations. Fortunately, neither lead to strong entanglement between the annealer and the environment, which allows for considerable understanding of the negative and positive effects. 
Nothing we have presented so far is tied directly to any particular hardware realization, and could even be implemented in a digital quantum computer through trotterization or similar schemes. To consider an open system, we need to specify a physical hardware platform and noise model. Reflecting their popularity for quantum optimization, we choose superconducting flux qubits. The single qubit noise model for flux qubits consists of unitary flux noise along $z$ (with an approximately $1/f$ power spectrum) and lossy entangling coupling to degrees of freedom in a cold bath. We will consider both of these effects independently in detail, and then collect our results as a complete picture to argue for the plausibility of fault welcoming quantum computing. As we shall see, phase noise is uniformly harmful but unlikely to erase the quantum speedup. 
%(though it may reduce it relative to the case where it is absent). 
The effect of cold bath coupling is much more subtle, in that the RFQA oscillations induce a quantum acceleration of bath-assisted multi-qubit tunneling% (and thus, phase transitions)
, which may dominate the harmful open system effects and lead to fault welcoming behavior; see Sec. \ref{sec:fwqc} for an extensive discussion.
\subsection{Peformance degradation in a quantum random energy model with single qubit $1/f$ noise}\label{1overfsec}
%Having studied the closed system problem in detail, we now consider RFQA in open systems. So far, vior.
We begin by considering phase noise, which in quantum annealing applied to spin glasses has two primary effects. The first, a slow, continuous drift in the energies of competing local minima, is roughly expected to reduce the solution probability by a factor 
\begin{eqnarray}\label{1overfloss}
P_{OGS} \of{t} \to  \frac{P_{OGS} \of{t} \times \of{M+1}}{M+ 1+ g_{int} \of{ c \of{t} \sqrt{N} \avg{\delta h_{i} }_{RMS} } },
\end{eqnarray}
where $N$ is the number of spins, $M$ is the number of classical solution states in the ground state band, $g_{int} \of{E}$ is the total number of states with energy no greater than $E$ above the ground state, $c \of{t}$ is a small dimensionless prefactor which increases at most logarithmically in $t$ and $\avg{\delta h_{i} }_{RMS}$ is the RMS average deviation of the single qubit energies by $1/f$ noise. This simply reflects the fact that $1/f$ noise can bring many local minima (and their low-lying excitations) into energetic competition with the true ground state in a large system, and the system will attempt to explore all of them as they pass in and out of resonance. Generically, we expect that $g_{int} \propto e^{d \sqrt{N}}$, for some small $d$.

\begin{figure}
\includegraphics[width=3.25in]{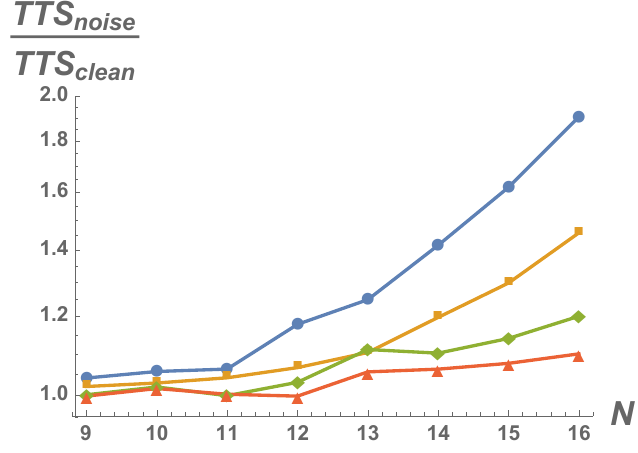}
\includegraphics[width=3.25in]{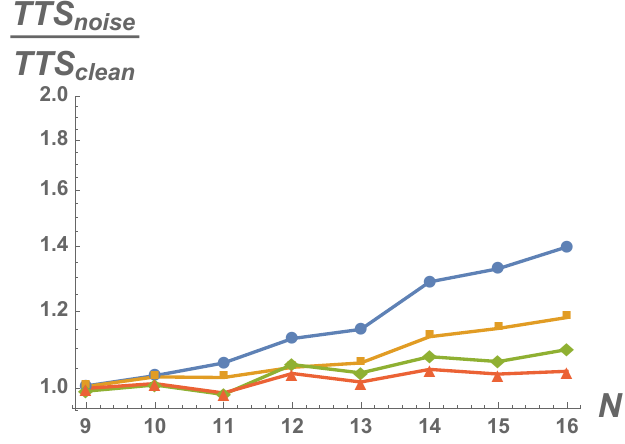}
\caption{Relative performance degradation from $1/f$ noise for population transfer in the BQREM, measured by average time to solution, with $M=4$ and all non-noise parameters identical to those used in Fig.~\ref{QREMnonoise}. The curves in the top panel correspond to $T_{2R} = 25 N/10$ (blue circles; all connecting lines are included to guide the eye), $T_{2R} = 25 N/10$ with the BQREM's first excited band counted as ``solutions" for the purposes of computing TTS (gold squares), $T_{2R} = 50 N/10$ (green diamonds), and $T_{2R}= 50 N/10$ including the first excited band (red triangles). The curves in the bottom panel plot the same quantities with an upper frequency cutoff included in the $1/f$ noise to mitigate local diabatic heating (see text for details). As seen in the plots, for the strongest noise the quantum speedup essentially disappears, but for weaker noise or an upper cutoff the speedup is slightly reduced by not eliminated (the TTS ratio must increase as $2^{N/4}$ for the quantum speedup to be completely cancelled out). Further, for strong noise without an upper cutoff much of the performance degradation comes from diabatic heating to high energy states, a mechanism which is unlikely to occur in more realistic problem Hamiltonians and would be largely ``corrected" by local cooling from interaction with a cold bath.}\label{QREM1overf}
\end{figure}

We do not expect this effect to erase the quantum advantage of RFQA, at least for problems with exponential difficulty scaling. In real quantum annealers, one would likely never run a single anneal for exponentially increasing time, and would instead sample the output of an exponentially growing number of trials, each of which runs for polynomial time. In this limit the success probability for a given trial is expected to increase linearly with time (with slope $\Gamma_T$) and the speedup in Eq.~\ref{RFQAbaseform} suggests a total time to solution which scales as $TTS \of{N} \propto \exp \of{\Upsilon N + d \sqrt{N}}$. If RFQA reduces the difficulty exponent $\Upsilon$ by an $N$-independent amount relative to constant schedule annealing and/or classical algorithms, as it does here, the correction from Eq.~\ref{1overfloss} is subleading and thus does not erase the quantum speedup. Note that in the numerical simulations discussed below, to make more direct comparison to the closed system results and for reasons discussed in appendix F, we \textit{did} run each instance for exponentially increasing time.

The second effect of $1/f$ noise, local heating, is more serious for our simulations. Though suppressed by the strength of the transverse field in the paramagnetic phase, and suppressed by the fact that the eigenstates are superpositions of small numbers of $z$ basis states in the spin glass phase, a randomly fluctuating $\sigma^z$ field can diabatically create spontaneous local excitations at finite energy. When these excitations occur they kick the system out of the low-energy manifold near the phase transition and thus, since the oscillating fields only induce transitions in a fairly narrow energy range, drop the success probability to zero. Even if the heating rate is very small, given exponentially increasing runtimes these excitations will proliferate given large enough $N$ and $t_f$ -- this effect is clearly observed in the top panel of Fig. \ref{QREM1overf}.

However, in real quantum annealers, local relaxation via a cold bath occurs at much faster rates than this, correcting this error source and in the end renormalizing the difficulty exponent somewhat without fully eliminating the quantum advantage.The combined effect of diabatic heating from $1/f$ noise followed by local cooling may be estimated as follows. The dominant process in the parameter ranges we consider appears to be mixing with residual paramagnet states; by the estimates in \cite{faoro2018non} the squared matrix element between QREM local minima $\abs{\bra{G_i } \sigma_1^z \ket{G_j}}^2$ decays more quickly than $2^{-N}$ and is thus ignorable. So let us consider mixing with low-energy paramagnetic states-- for an excitation of energy cost $E$ the transition rate from $1/f$ noise scales as $1/E^2$, whereas the relaxation rate from a cold bath is roughly energy independent. So we expect these processes to be highly suppressed, and even if we consider the worst case assumption of $E \sim O \of{1}$, and ignore that there are very few states in this range, then we expect a reduction in the probability of hitting the solution by a factor which scales approximately as $\of{ 1 - \Gamma_{ex,1/f}/\Gamma_{rel,bath} }^N$, where $\Gamma_{ex,1/f}$ is the rate of creating excitations through $1/f$ noise at a given spin, and $\Gamma_{rel,bath}$ is the rate of relaxation for the same transition via interaction with the cold bath. In general, we expect $\Gamma_{ex,1/f}/\Gamma_{rel,bath} \ll 1$, so this does produce an exponential reduction in the solution probability but only a small, $\mathcal{O}(\Gamma_{ex,1/f}/\Gamma_{rel,bath})$ increase in the difficulty exponent, $\Upsilon$.
%, though potentially with a scaling exponent which is much smaller than $\of{ \ln 2}/4$ so the an overall quantum speedup can survive. 
Further, if $\Gamma_{ex,1/f}$ scales as $1/N$ or $1/N^2$ this contribution will be an irrelevant constant at large $N$.

To test the effect of $1/f$ noise on our system, we repeat the numerical experiments of the previous section (cf. Fig. \ref{QREMnonoise}%population transfer in the BQREM
) with $1/f$ noise added, using the prescriptions detailed in Appendix \ref{sec:1overfsimulation}.  
We characterize $1/f$ noise by its Ramsey time chosen to scale with $\sim N$ to match with $1/N$ scaling chosen for problem Hamiltonian scales (this is for pure convenience)\footnote{As a side note, it is worth comparing these estimates to the real experimental parameters observed in the d-Wave quantum annealers \cite{dwave2019whitepaper}, the only commercially available quantum hardware. The maximum transverse field strength $\Delta / h$ is around 5 GHz; we shall use 2.5 GHz in our comparison since we are considering dynamics near the phase transition point. The Ramsey $T_2$ of these qubits varies somewhat from one report to the next, with a typical value of around 15 ns. In our simulations, at $N=10$ the transverse field strength is set to 0.05, so equating that with 2.5 GHz results in a noise $T_{2R}$ of 750. In other words, our strong noise simulations ($T_{2R} \simeq 25$) correspond to an isolated coherence which is around a factor of thirty times \textit{lower} than the d-Wave flux qubits.}.
%The results of these simulations are shown in Fig.~\ref{QREM1overf} (top panel). 
We observe in Fig.~\ref{QREM1overf} (top panel) that that for the strongest
noise considered the quantum speedup begins to vanish by $N=13$ or so, whereas for weaker noise the time to solution is only modestly affected. Furthermore, we may assess the relative significance of energy drift vs. adiabatic heating degradation channels by including the first excited state in the analysis -- for stronger noise the system appears to have run away from the ground state and low lying levels, unlike weaker noise where considerable probability resides in the first excited state. 
To further explore the significance of diabatic heating and therefore assess the promise of its mitigation with the cold bath we repeat the simulation with a truncated $1/f$ noise spectrum, with upper cutoff placed below the gap to the first excited band (and appropriately adjusted noise amplitude to keep $T_{2R}$ unchanged).  The results are consistent with a our expectation of suppressed heating. A proper simulation of this effect is prohibitively expensive\cite{jaschke2019thermalization}, unfortunately.

\subsection{Performance boost via %quantum acceleration of 
bath-assisted multi-qubit tunneling}\label{coolingsec}

Rather remarkably, beyond accelerating mixing in the vicinity of phase transitions, RFQA can exponentially increase the rate of \textit{bath-assisted} multi-qubit tunneling as well. If the bath is sufficiently cold \cite{albash2017temperature}, the arguments from earlier in this work predict that bath-assisted relaxation can find the true ground state more quickly than mixing near the phase transition itself by a factor proportional to the number of such transitions, which can be as large as that number of q-bits $\propto N$ . While only a polynomial (prefactor) speedup, this may be a significant boost to the performance. To understand why this occurs, we model the bath \cite{gardinerzoller} as a large collection of oscillators or free spins, each weakly coupled to a primary qubit. As before, will illustrate the speedup mechanism using the BQREM as our problem Hamiltonian, though the arguments we present can easily be generalized to more realistic problems. 
\subsubsection{Spin-bath model and theoretical analysis}
\begin{figure}
\includegraphics[width=3.25in]{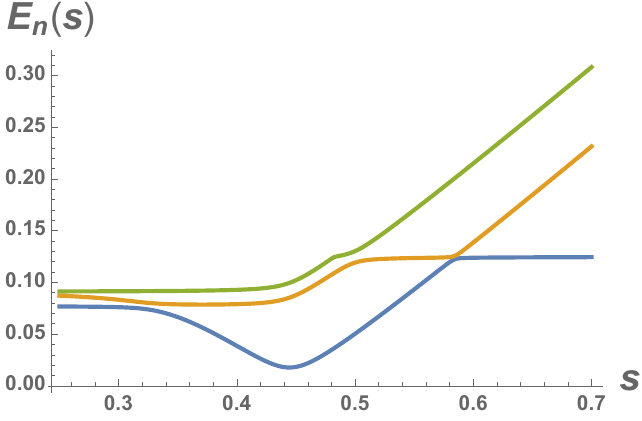}
\caption{First three excitation energies of the Grover problem with an additional bath spin as a function of annealing parameter $s$, calculated numerically for $N=12$. The primary avoided crossing at $s \simeq 0.45$ is present as normal, but if this avoided crossing is diabatically missed, the system experiences a second avoided crossing when $E_0 - E_G = \omega_B$, the energy of the bath spin (this occurs at $s \simeq 0.58$ in this example). Due to the weak coupling between the bath spin and the primary system, the gap at this avoided crossing is substantially smaller than it is at the primary phase transition, but, per MSCALE, it is smaller by a factor which is constant as $N$ increases. Further, all the arguments which yield the RFQA speedup around the primary ground state phase transition apply to the bath-assisted one as well, giving a quantum acceleration of bath-assisted multi-qubit tunneling; as shown in Fig.~\ref{QREMbath}, this gives a nearly identical quantum speedup to our closed system results from earlier in this paper.}\label{grovbathevs}
\end{figure}

Consider first a single additional spin with excitation energy $\omega_B > 0$, coupled weakly to one of the primary qubits through a local spin coupling (which could be any of $x$, $y$ or $z$) of strength $g_{QB}$:
\begin{eqnarray}\label{bathH}
H_{tot} = H_D + s H_G + g_{QB} \sigma_1^{a} \sigma_{B}^x + \frac{\omega_B}{2} \sigma_{B}^z,
\end{eqnarray}
where $a$ denotes one of the three components of the q-bit spinor. Provided that $g_{QB}$ is weak compared to the base value of the transverse field ($1/2N$ in our simulations), before the phase transition the bath spin has no effect on the physics and the system evolves as normal. However, as shown in Fig. \ref{grovbathevs}, after the groundstate transition, the bath spin induces a second, weaker transition in the first excited state as $s$ increases, located near the point where $E_0 - E_G = \omega_B$. Let the primary system be $P$ and the bath spin be $B$. This transition mixes the states $\ket{0'_P,0_B}$ and $\ket{G'_P,1_B}$, with a mixing rate $\Omega_{0B}$ and gap width $\Delta_{min,B} = 2 \Omega_{0B}$, which can be calculated using the perturbation theory of the previous sections. As predicted by MSCALE (and inferred for this problem from the results in section~\ref{grovcalcsec}), $\Omega_{0B}$ has the same large-$N$ scaling as the primary transition matrix element $\Omega_0$, though it is smaller by an $O \of{g_{QB} / \kappa}$ prefactor, where $\kappa$ is the strength of the transverse field. Consequently, if the system diabatically crosses the phase transition point and remains in the paramagnetic state $\ket{0_P}$, the bath spin can relax the system to the Grover state, providing an additional mechanism for solving the problem. 

The bath-assisted relaxation rate scales as $\abs{\Omega_{0B}^2} \propto 2^{-N}$, so it does not present a polynomial speedup over classical random guessing, though since each spin could couple to a bath independently, a (logarithmic) factor of $N$ speed increase is possible. Now imagine we include the oscillating fields of RFQA-D. At first order, there are $2N$ points where one of the applied fields is resonant with the new transition, and each of these matrix elements $\Omega_{1B,i}$ is dominated by the virtual mixing of $\ket{\tilde{1}'_P,0_B}$ with $\ket{G'_P,1_B}$ (where $\ket{\tilde{1}'_P}$ is the appropriate single excitation manifold in the paramagnetic phase). The perturbative analysis in Eqns.~(\ref{Om1}-\ref{maxGrate}) predicts that if $g_{QB}$ and $\omega_B$ are both polynomially small, $\Omega_{1B,i}$ will be smaller in magnitude than $\Omega_{0B}$ by a factor $\propto \alpha/4$ in the large $N$ limit. At second order, we have $2 \binom{N}{2}$ terms, each weaker by $\of{\alpha/4}^2$, and so on; when these contributions are all summed up either through slowly varying $s$ or averaging over random pause points, the resulting energy-averaged relaxation rate should be larger than $\abs{\Omega_{0B}^2}/W$ by a factor $\propto 2^{0.25N}$, the same quantum speedup recovered in the primary (closed-system) transition.

\begin{figure}[h!]
\includegraphics[width=3.25in]{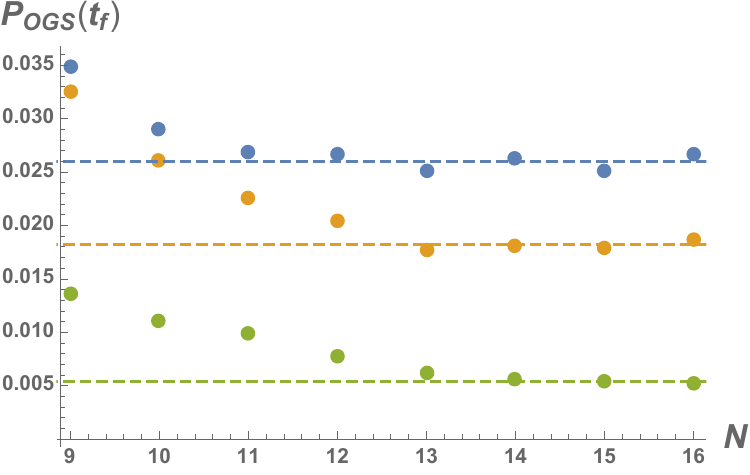}
\includegraphics[width=3.25in]{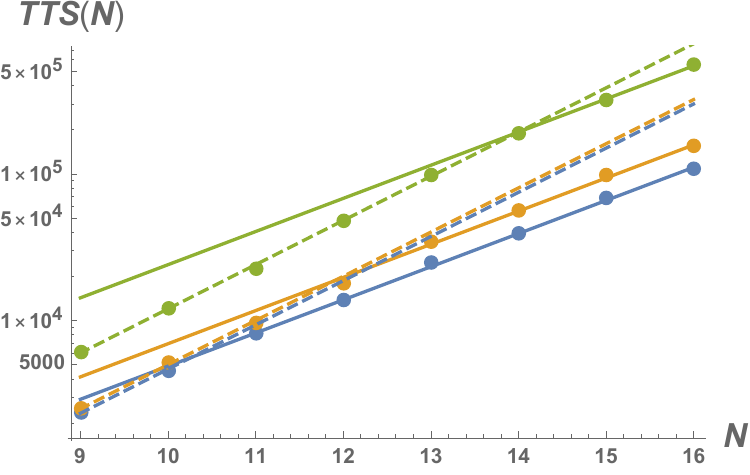}
\caption{Quantum acceleration of bath-assisted phase transitions the BQREM, using RFQA-D with the simulation parameters described in the text, for $N$ running from 9 to 16 with a single auxiliary bath spin. In all three plots blue, gold and green points and curves correspond to a bath spin which couples to the primary system with a local $\sigma^x$, $\sigma^y$ or $\sigma^z$ operator, respectively. Top: probability of finding the primary system's true ground state at $t = t_f \of{N} = 0.56 \of{N+1} + 0.0093 \times 2^{0.75 N}$; dashed lines are the average of the final three points. A constant final success probability corresponds to a polynomial quantum speedup relative to classical search algorithms. Bottom: average time to solution (TTS) as a function of $N$, taken from the success probability at $t_f \of{N}$. Solid lines represent fits of the last three data points to $c_i \times 2^{0.75 N}$, and dashed lines extrapolate the classical $2^N$ scaling from $N = 9$. In all three cases the asymptotic TTS, up to a prefactor, is well approximated by the quantum speedup computed in Eq.~\ref{maxGrate}, confirming both our analytical theory and the MSCALE conjecture (which can be proven to be true in this system).}\label{QREMbath}
\end{figure}

When multiple bath degrees of freedom are included, each one induces a new %phase 
transition in a low-lying excited state, which in turn increases the chance of ending up in the true ground state even if the primary phase transition is diabatically crossed. Since the bath degrees of freedom are independent and coupled weakly to the primary system, these transitions cannot interfere with each other as they correspond to exciting separate, uncoupled degrees of freedom; however, for a given transition, self-interference of close frequencies remains possible (see the discussion in appendix~\ref{sisec}). Fortunately, since the base matrix element $\Omega_{0B}$ is substantially smaller than the primary transition matrix element $\Omega_0$, while the local gap $\Delta_{local} = 1/N$ is unchanged, it is substantially easier to choose frequencies large compared $\Delta_{min}$ but small compared to $\Delta_{local}$ for small $N$ in the range of straightforward classical simulation. 

% \subsection{A simple example}

\subsubsection{Numerical results for the BQREM}

To demonstrate a quantum speedup for bath-assisted relaxation, we prepare the BQREM Hamiltonian with a single ground state at an energy  $E_i = -1/2 + \delta_i$ and then randomly select an additional state to be the new, ``true" ground state with an energy $E_j = -1/2 - 2/N + \delta_j$, where the shifts $\delta_{i/j}$ are randomly chosen from the range $\cuof{-0.27/N,0.27/N}$ as before. We then include a single bath ``spin" with excitation energy $\omega_B = E_i - E_j + 2 \delta_k$. These energies are chosen to isolate the effect of the bath spin; the broad separation in energy of the two states ensures negligible direct mixing between them even with oscillating fields turned on, and the bath spin's inverse polynomial detuning matches the inverse polynomial precision considered everywhere else in the paper. The bath spin is coupled to a single primary qubit with a $\sigma^x$, $\sigma^y$ or $\sigma^z$ coupling $g_{QB} \simeq 0.3/N$.

As before, we simulate a population transfer algorithm, where at $t=0$ the system is initialized in the ``false" ground state, with the bath spin in its ground state and with all transverse fields, and the system-bath coupling, turned off. The transverse fields and the system-bath coupling (which is of the form $g_{QB} \sigma_{1}^{x/y/z} \sigma_{B}^x$ are then ramped from zero to their fixed values over a time $t_r \simeq 0.56 \of{N+1}$ and then the system waits a time $t_f \of{N} = 0.0093 \times 2^{0.75 N}$ with the oscillating fields turned on. The results of these simulations can be be seen in FIG.~\ref{QREMbath}. For all types of coupling the success probability at $t=t_f \of{N}$ asymptotically approaches a constant value as $N$ grows, yielding the same quantum speedup seen in the closed system.

\section{Summary of results and an outlook for ``Fault Welcoming'' quantum computing}
\label{fwsec}
We now conclude with a summary and an extended discussion of broader context of our results, organizing it into five themes: (i) the notion of error correction for annealers, (ii) quantum equilibration and thermalization as a resource, (iii) a re-examination of new results on RFQA with and without the environment through the lens of (i) and (ii); (iv) a proposed definition of ``fault-welcoming quantum computing'', and, finally, (v) we argue that our results suggest that fault tolerant and possibly even fault welcoming adiabatic quantum computing is realistic.
\subsection{Error correction}
\label{sec:errorcorrection}
As mentioned in the introduction, the necessity of error correction is a basic feature of the gate model of quantum computation. The threshold theorem and many promising examples of topological error correction codes allow for the formulation rigorous bounds and estimates of overhead, and critically, these estimates are all independent of the underlying algorithm one would wish to run on the error corrected machine. One might naively expect that quantum annealing might simply borrow much of this well developed formalism, but unfortunately such a notion fails, in part because the intrinsic nonlocality of error correction requires high order interactions that are largely unrealistic in an analog context \cite{sarovaryoung2013,youngsarovar2013}. And indeed, for specific examples such as the Grover problem, it is relatively easy to convince oneself of AQC's \emph{fragility}, e.g. to finite precision in control parameters and random noise. There is no fundamental physical law, however, that forbids realizing a quantum computational advantage in systems interacting with the environment -- in fact, this may well be a generic situation provided the quantum advantage is not (exponentially) fine tuned and somewhat \emph{robust}, as is the case of RFQA developed in this work. Going further, no law even requires the totality of environmental effects to reduce quantum computational performance relative to an idealized closed system equivalent. It is thus one of key aims of this paper to argue that while traditional error correction of arbitrary code %in AQC 
is not \textit{possible} in AQC, it is also not \textit{necessary}, and a scalable quantum advantage can be realized in the face of realistic noise.

\subsection{Quantum thermalization and annealing}
\label{sec:resonantthermalization}

Our route to a scalable quantum advantage runs through the notion of quantum thermalization as a computational resource. While the past decade has seen significant advances in understanding quantum thermalization and how it may break down \cite{RNDAH2015} much of the progress has focused on typical behavior of highly excited many-body states \cite{Kaufman794}. One expectation particularly relevant to the present setting is that a many-body system heats up to infinite temperature when irradiated externally -- for a single monochromatic drive this is known as ``Floquet thermalization'' \cite{Rigol2014,RM2018,Rigol2019}.  Clearly, such runaway process must be avoided if a quantum annealer is to succeed and, indeed, adiabatic theorem limits heating of gapped groundstates to exponentially slow rates (see Sec. \ref{heatsec}). There are other processes, however, whereby perturbations induce (or enhance) matrix elements among nearly degenerate states and thereby induce dynamics that does not appreciably increase mean energy.  This restricted notion of ``resonant thermalization" is clearly highly beneficial to quantum annealing and is the essence of both the population transfer works discussed earlier (where it is discussed in the language of non-ergodic extended states), and RFQA (see Section \ref{grovcalcsec}).

And as a resource, quantum thermalization has significant computational power. To see this, let us consider two ways by which an environment might assist AQC: a zero temperature bath, and a low temperature bath which instantly thermalizes the system. That a zero temperature bath would assist AQC is relatively obvious; the action of the bath would monotonically reduce the system's energy at all times, and if the goal is to find the ground state of a problem Hamiltonian these processes can only help that effort. However, as discussed in Appendix.~\ref{sec:EAss}, per MSCALE this is expected to only provide an $O\of{N}$ enhancement in general problems, where $N$ is the number of qubits. More intriguingly, consider a hypothetical \textit{instantly thermalizing} quantum annealer \cite{albash2017temperature}, which for any problem Hamiltonian is always in perfect thermal equilibrium with its environment at all points in time. Such a system is impossible to construct. If the environmental temperature could be held low enough, then the system would find the ground states of NP problem Hamiltonians in constant time with constant or at least inverse polynomial probability, and somewhat higher temperatures could still find approximate solutions to NP problems in polynomial time, collapsing the polynomial hierarchy in either case \cite{sahni1976p}. Since $\rm{ BQP \neq NP}$ has not been formally proven this may be possible, but the instantly thermalizing machine would find the ground state of a Grover or QREM oracle Hamiltonian in less than $O \of{2^{N/2}}$ time, which rigorous proofs \cite{zalka1999,farhigoldstone2008,farhi2010unstructured} forbid \footnote{We expect that the speed limit in this system arises from MSCALE. To exceed the $2^{N/2}$ bound the system-bath coupling likely must increase exponentially in system size, which would eventually break down the perturbative assumptions required to separate the system from its bath, and in turn freeze evolution through continuous measurement and/or exponentially dilute the probability of finding the solution state as an eigenstate of the combined system, depending on the type of coupling.}. But if an instantly thermalizing quantum device is impossible, and systems which thermalize in finite (if potentially very long) timescales are ubiquitous, then it is worth considering if a device with \textit{accelerated} thermalization could bridge the gap, yielding a quantum speedup without violating any fundamental bounds. To do so, we now review the key results of this paper, posit a definition for ``Fault-welcoming" quantum computing, and then consider the sum of all realistic noise effects to explore whether RFQA might be a realistic route to such behavior.

\subsection{Summary of the new results of this paper}
\label{sec:summary}
\begin{itemize}
\item
The central result of the paper is an explicit demonstration of RFQA -- a resonant enhancement of transition rates among competing groundstates in a structureless Grover problem.  This advance consists of a detailed perturbative construction, supplemented with analytic optimization to maximize the effect, analytic argument for suppressed heating and numerical confirmation of all these effects in both Grover and another similarly structureless banded quantum random energy model (BQREM), akin to quantum spin glasses.  This construction was motivated by an observation that the celebrated quantum speedup in the Grover problem is exponentially fragile. By contrast, RFQA operates on polynomial fine-tuning and its success unsurprisingly requires a careful organization of coupling scales and driving frequencies, all polynomially small. RFQA is an example of resonant thermalization defined above.
\item
To facilitate the construction of RFQA we conjectured that generic few particle matrix elements between target states have identical scaling to that of the minimal gap when such states cross -- this is based on several explicit cases and seemingly general rule of thumb among workers in the field. This MSCALE conjecture also allows us to identify an important channel by which a cold bath is able to generate a large boost to RFQA, via a prefactor equal to the number of qubits.  This result is demonstrated at the perturbative level analytically and verified numerically.
\item
Lastly, and directly addressing points raised in Sec. \ref{sec:errorcorrection}, we verified numerically and argued analytically that the quantum advantage generated by RFQA can survive despite external $1/f$-type classical noise.% We have observed that while weak noise does not affect performance (in line with analytic arguments), sufficiently strong noise degrades the advantage but does not eliminate it entirely.
\end{itemize}
\subsection{A heuristic definition for Fault Welcoming Quantum Computing}
\label{sec:fwqc}
Playing off the common notion that any interaction with the environment is a fault that corrupts pristine quantum evolution and therefore needs to be corrected, we would like to advance the notion of a fault-welcoming quantum computing.  A \textit{Fault Welcoming Quantum Computer} is an extensible quantum computing system which, for a given class of problems, displays a quantum speedup (defined by scaling with problem size $N$) over the best known classical algorithms, and when \textit{all} realistic uncontrollable noise sources are considered, performs better (in average time to solution, relative quality of solutions, etc.) than it would if all uncontrollable noise sources were absent.

It is worth pausing to unpack this definition piece by piece. First, ``for a given class of problems" is included to reflect that near-term (so-called NISQ-era \cite{preskill2011}) quantum devices are not necessarily expected to be universal quantum machines, but may nonetheless perform extremely well for specific classes of problems (quantum annealers, of course, being the canonical example here). Second, ``displays a quantum speedup," recognizes that while a physicist or quantum computer engineer might greatly appreciate seeing a problem solved with quantum hardware, the average user only cares about the cost, runtime and quality of solutions. Absent a quantum speedup the vastly lower costs of classical machines make quantum hardware a poor choice. A quantum system which reaps a net benefit from open system effects but still exhibits worse scaling than state of the art classical routines is likely of little practical use. And finally, ``when \textit{all} uncontrollable realistic noise sources are considered" is chosen to note that while one might make design choices to reduce the strength of some noise channels relative to others, it does not seem realistic to completely zero out some couplings between a system and its environment while preserving or even amplifying others. Further, we focus only on uncontrollable noise sources, and not open system effects induced intentionally for things like measuring and resetting qubits that are part of the normal operation of a quantum computer. 

Before we proceed to discuss RFQA as a step towards FWQC we must acknowledge a large body of prior work on environment assisted quantum computing, which is reviewed in Appendix \ref{sec:EAss}. These results, while often encouraging, all stop short of establishing what we would consider fault welcoming behavior. 

\subsection{Noise resilience and prospects for fault-welcoming quantum computing with RFQA}
The importance of avoiding fine tuning requirements in the adiabatic protocol has been emphasized throughout this work, and RFQA clearly demonstrates such robustness.  Unlike numerical proof-of-principle simulations, e.g. in Fig. \ref{QREMnonoise}, realistic implementations of RFQA (see, e.g. Sec. \ref{fluxqubitsec}) are likely to steer clear of very long running times of order $t_f\sim e^{\Upsilon N}$ and rather work in the regime $t/t_f \lll 1$, likely vanishing in $N$. Collecting data over short repeated runs should produce sufficient statistics to demonstrate the advantage, since we expect the probability of success to be a simple (linear in general cases, as seen from Appendix~\ref{bangbangsec}, and at worst quadratic) function at small $t/t_f$. We make this assumption moving forward as we consider noise.

While unstructured, random error such as the depolarizing noise commonly studied in the gate model leads to an incoherent random walk in Hilbert space and consequently, the complete failure of any protocol without error correction, this is fortunately not the empirical noise model of superconducting flux qubits. This consists of classical $1/f$ noise and a low temperature bath, both of which can be considered completely independent noise sources for each qubit \footnote{On physical grounds we expect that short runtimes $t/t_f\lll1$ should help further suppress the influence of other environmental effects.}. In the absence of noise, the success probability scales as $t / t_f$, or $P \of{t} \propto t e^{- \Upsilon N}$, where the total difficulty exponent $\Upsilon$ includes the RFQA speedup and is captured in Eq.~\ref{RFQAbaseform}. Let us now consider $1/f$ noise, which has two primary channels in RFQA. The first is that, at relatively short times the ground state manifold will be smeared by random fields over a window of energy on the order $\sqrt{N}$, and since the system will explore all of these states at approximately equal rates this reduces the success probability by a subleading factor, $P \of{t} \propto t e^{- \Upsilon N - c \sqrt{N}}$, for some small $c$. The second, diabatic heating by the high frequency tail of the noise spectrum, can lead to a runaway from the ground state manifold at long times and is thus more serious. However, since diabatic heating is a local process in realistic, structured problems, we expect that the coupling to a cold bath will remove these excitations much more quickly than they are created. If we combine the diabatic heating rate and the rate of creating local excitations by absorbing energy from the cold (but not necessarily zero temperature) bath into a single (small) rate $\Gamma_E$, and define the corresponding cooling rate from the bath as $\Gamma_C$, then a reasonable \textit{worst case} estimate that ignores the likely slow rate of this heating will be
\begin{eqnarray}\label{worstcaseP}
P_{{\rm (worst \; case)}} \of{t} \propto t e^{- \Upsilon N - c \sqrt{N}} \of{ 1 - \frac{\Gamma_E}{\Gamma_C}}^N .
\end{eqnarray}  
As we expect $\Gamma_E / \Gamma_C \ll 1$, if the RFQA-induced reduction of the difficulty exponent $\Upsilon$ is fairly substantial (as it is for the Grover problem), these noise effects cannot erase the quantum speedup, and we have thus derived a quantum speedup capable of tolerating experimental noise.

Pushing further, we note that this estimate ignored a potentially critical channel, namely the RFQA dressing of bath-induced phase transitions derived in Sec.~\ref{coolingsec}. In that section we took the particularly simple case of a single ``bath'' spin coupled locally to a qubit in the Grover or BQREM problem, Eq. \label{bathH}. The net effect in the model we considered, with $N$ bath spins (one per qubit) is that RFQA acquires an order $N$ prefactor improvement (without changing the difficulty exponent $\Upsilon$). However, this is not a realistic description of the bath in solid state systems, and we must instead consider a very large distribution of bath spins with correspondingly weak couplings for each qubit, producing a continuous density of states $\rho \of{ \omega}$. The interaction of the system with each one of this vast bath of spins should be similarly dressed (and thus amplified) by RFQA, but this is a much more complicated case and a detailed consideration of it is beyond the scope of this paper. Instead, we merely state that it is quite plausible that a set of circumstances exists where an RFQA-amplified interaction with this continuous distribution \textit{could} further reduce the difficulty exponent $\Upsilon$. If this effect dominates local heating the system would be faster for being open than an idealized, noise-free copy, and thus, fault-welcoming. Such a case was envisioned by Amin \textit{et al} over a decade ago \cite{aminlove2008}, but was subsequently argued against by Wild \textit{et al} \cite{wild2016adiabatic}, who showed that longitudinal couplings to any finite temperature bath could erase the quantum advantage in unstructured search. The addition of RFQA would complicate this picture significantly, but elucidating its effects will be left for future work.

In summary, we proposed a deceptively simple modification to quantum annealing, where each transverse field experiences low-frequency coherent oscillations in its direction, and showed that it is capable of providing a quantum speedup by accelerating multiqubit phase transitions. We argued that this speedup is robust to at least moderate amounts of local noise, and further showed analytically and numerically that the RFQA speedup mechanism also accelerates bath-assisted transitions, and consequently amplifies the influence of the helpful noise channel in solving hard optimization problems. These results provide tantalizing hints of fault welcoming behavior but do not firmly establish it. This would require careful consideration of more realistic models, architecture-dependent issues such as the overhead of embedding logical problems in qubit grids with short-ranged connectivity, and so on. Experimental implementations of RFQA (along the lines outlined in the paper or otherwise) are especially interesting as they are likely to illuminate and guide further progress on developing the ideas outlined in this work.
%Further, our method has a fairly straightforward experimental implementation. While it ultimately remains to be seen whether or not quantum algorithms can provide scalable quantum speedups for realistic, useful problems in absence of error correction, we hope that we have convinced the reader by the end of this relatively long paper that RFQA in flux qubits applied to hard spin glass problems is an excellent candidate for observing such a quantum advantage. 

\section{Acknowledgements}

EK's research was supported in early stages by the Louisiana Board of Regents grant LEQSF(2016-19)-RD-A-19 and by the National Science Foundation Grant No. PHY-1653820. V. O. acknowledges support from NSF Grant No. DMR-1508538. EK would also like to acknowledge the hospitality of the International Center for Theoretical Physics in Trieste, Italy, as well as the Kavli Institute of Theoretical Physics in Santa Barbara, CA. We would like to thank Alex Burin, Yu Chen, Steven Girvin, Sarang Gopalakrishnan, Nicole Yunger Halpern, Sergey Knysh, Antonello Scardicchio and Zhijie Tang for useful discussions.

\appendix

\section{Derivation of the RFQA speedup and ``Bang-and-wait" search algorithms}\label{bangbangsec}

In the main body of the text, we proposed that if the $m$th order resonances have an average strength $\Omega_m \propto \Lambda^m \Omega_0$, the mixing rate at a phase transition from RFQA takes the form:
\begin{eqnarray}\label{RFQAbaseform2}
\Gamma_T = \frac{\Omega_0^2}{W} \sum_{m=1}^N \of{2 \Lambda^{2}}^m \binom{N}{m} \simeq \frac{\Omega_0^2}{W} \exp \of{2 \Lambda^2 N}.
\end{eqnarray}
To derive this, we will consider a single spin as a proxy for the two competing ground states, and show that the average rate of mixing in a Landau-Zener-like sweep across $N$ oscillating fields scales as the sum of the squared Rabi frequencies of all $N$ tones. To do so, we reconsider the LZ transition from the perspective of Fermi's Golden rule, and consider the minimum gap $\Delta_{min} = 2 \Omega$ as a perturbation which causes decay from $\ket{0}$ to $\ket{1}$, with energy transferred into an environment with a Lorentzian density of states peaked about $\epsilon = 0$ with narrow, ficticious width $\Gamma'$ (which we can take to zero later). We assume that we sweep from $z$ bias $\epsilon = -W/2$ to $+W/2$ in a time $t_f$. Assuming the sweep is quick enough that we can linearize the transition probability, and that the linewidth is narrow compared to the range of the energy sweep, $\Gamma' \ll W$ we obtain
\begin{eqnarray}\label{P1f}
P_1 \of{t_f } &\simeq&  \int_{0}^{t_f} dt \abs{\Omega}^2  \frac{ \Gamma' }{ \frac{ \Gamma'^{2} }{4}  + W^{2} \of{ \frac{t}{t_{f}} - \frac{1}{2} }^2 } \\
&=& \frac{ 4 \abs{\Omega}^2 \arctan \of{W/\Gamma'}}{W} t_f = 2 \pi \frac{\abs{\Omega}^2}{W} t_f \nonumber
\end{eqnarray}
If we time average this result we obtain the mean transition rate $\Gamma_{01} = 2\pi \abs{\Omega}^2 / W$; exponentiating this recovers the Landau-Zener result.

Now imagine that instead of a simple $H \of{t} = \epsilon \of{t} \sigma^z / 2 + \Omega \sigma^x$, we instead have a more complex oscillatory driving element:
\begin{eqnarray}\label{LZosc}
H \of{t} = \epsilon \of{t} \frac{\sigma^z}{2} + \of{ 2 \sum_{i=1}^{N} \Omega_i \cos \of{2 \pi \omega_i t + \phi_i} } \sigma^x
\end{eqnarray}
For a single tone ($N = 1$), one can recover the adiabatic result (\ref{P1f}) by a simple rotating frame transformation. But for larger $N$, the same Fermi's Golden rule argument applies, provided that the frequencies $\omega_i$ are well-separated compared to the amplitudes $\Omega_i$ and that all the $\omega_i$ are contained within the energetic range $W$ swept through in a time $t_f$. Taking into account that the rate of transition from $\ket{0}$ to $\ket{1}$ is the same as the rate to be driven back from $\ket{1}$ to $\ket{0}$, assuming that the system begins in state $\ket{0}$ at $t=0$, we arrive at a final excitation probability $P_1 \of{t_f}$ given by:
\begin{eqnarray}\label{P1}
P_1 \of{t_f} &=& 0.5 \of{1 - \exp \of{- 4 \pi \Gamma_T t_f} }, \\
\Gamma_T &\equiv & \frac{  \sum_{i=1}^{N} \abs{ \Omega_{i}  }^{2} }{W}. \nonumber
\end{eqnarray}
This matches (\ref{P1f}) for short ramp times, and is also valid in the large $t_f$ limit, though unlike the adiabatic LZ problem the long time asymptotic state is an incoherent mixture of $\ket{0}$ and $\ket{1}$ with equal probability. Tones $\omega_i$ which do not lie in the energetic range $W$ do not contribute to the transition probability. We thus conclude that the mixing rate between states for a single spin in a slowly varying $z$ field, subject to $N$ weak transverse oscillating fields, scales as the sum of the squared Rabi frequencies of all tones. In more complex, multi-qubit problems, the $m$th order resonances are all included as additional oscillating sources in Eq.~\ref{P1}, yielding Eq.~\ref{RFQAbaseform2} as claimed earlier.

One caveat to the above analysis is that it can break down when combined with a perfectly uniform adiabatic sweep, due to interference effects. Specifically, as seen in studies of single oscillating sources in two-state Landau-Zener transitions, if the matrix elements $\Omega_j$ are symmetric on either side of the phase transition, the annealing schedule is uniform (so that the rate of variation of $H \of{t}$ does not change across the phase transition), and the system is undisturbed by random noise, then due to a reflection symmetry rotations between the states from positive frequency oscillating terms crossed before the transition can be identically canceled by negative frequency contributions on the other side of it, erasing the acceleration promised by the proliferation of oscillating sources. We noticed this issue in our own simulations of the Grover problem; when we simulated constant schedule annealing we were unable to recovery any quantum speedup, with or without RFQA. 

Fortunately, this interference effect is a precise cancellation that can be easily eliminated in many ways. The simplest is to simply consider non-uniform annealing schedules, and shortly we will formally derive the speedup (\ref{RFQAbaseform}) in the case of annealing where global parameter adjustment pauses and the system is allowed to evolve under the influence of the oscillating fields and an otherwise constant Hamiltonian for potentially long periods of time. These schedules allow the full RFQA speedup to be realized in a broad range of problems. Further, in real analog systems, random noise will cause fluctuations in the annealing schedule and local Hamiltonian parameters which scramble any self-interference effects of this type. Likewise, in real problems the matrix elements may not be symmetric across the phase transition, further mitigating this concern. But while we do not necessarily expect constant schedule annealing to fail in real implementations, we felt it worth pointing out here, lest readers wonder why we do not consider the simplest and most widely studied form of quantum annealing in our simulations.

The scaling form (\ref{RFQAbaseform2}) is generic up to prefactors, and we expect it to arise in any algorithm where the phase transition is energetically averaged over, whether through a uniform sweep, random guessing, or longitudinal fluctuations from single qubit noise. To demonstrate a quantum speedup in the Grover problem, and in all other examples considered in this work, we simulate a ``bang-and-wait" algorithm \cite{yang2017optimizing} to induce transitions between a chosen initial state and the target state which solves the problem. If we choose the initial state to be the simple transverse field paramagnet this algorithm is comprised of the following steps:

(i) The system is initialized in the paramagnetic ground state of $H_D$, $\ket{0_D}$.

(ii) Choose a value $s$ from a range $\cuof{ s_{min}, s_{max}}$ which is presumed to include the phase transition point $s_c$ (for the Grover problem, $s_c = 0.5 + O \of{1/N}$). We choose a random $s$ as in more general problems we do not expect to know the location of the transition(s), and in any realistic analog implementation it would be obscured by $1/f$ noise. Choosing $s_c$ exactly in noise-free evolution recovers the full Grover speedup (with or without oscillating fields) and is thus optimal; however, it requires exponentially growing precision and such a resource is capable of reproducing the Grover speedup in fully \textit{classical} algorithms \cite{hen2019quantum}. Throughout this work, we assume only inverse polynomial precision in all quantities. Once $s$ is chosen, begin evolving the system under $H_D \of{t} + s H_G$, either by ``jumping" instantaneously to this point or linearly ramping from $s=0$ over a short ramp time scale.

(iii) Evolve for a time $t_f$ and measure the state. Let $\Omega_0 = \Delta_{min}/2$ be the transition matrix element between $\ket{0_D}$ and $\ket{G}$ when the two states are at resonance, and then measure the state. If we ignore the oscillating fields for the moment, the probability of finding the system in the solution state $\ket{G}$ depends on the energy difference $\epsilon$ (which is set by the choice of $s$) between $\ket{G}$ and $\ket{0_D}$, and is given by
 \begin{eqnarray}
P_{base} \of{\epsilon} \simeq \sin^2 \of{\sqrt{\epsilon^2 +  \Omega_{0}^2} t_f} \frac{\Omega_{0}^2}{\epsilon^2 +  \Omega_{0}^2} .
\end{eqnarray}
Of course, we must average this function over energies in a range $W$ (which is proportional to $\of{s_{max}-s_{min}}$), as we do not know the exact location of $s$. The average value of $P_{base}$ depends on the choice of $t_f$, and is characterized by two regimes: $t_f \ll 1/\Omega_0$ and $t_f \propto 1/\Omega_0$. In the first regime, $P_{base}$ is a relatively flat function since the otherwise sharp peak at $\epsilon \simeq 0$ is suppressed by $\sin^2 \Omega_0 t_f \ll 1$. In this regime, the average value of $P_{base}$ increases linearly with time, and the average probability of solution scales very roughly as $\Omega_0^2 t_f / W$. Similarly, in the second regime where $t_f \propto 1/\Omega_0$, $P_{base}$ is sharply peaked near $\epsilon = 0$, and is $O \of{1}$ if $\abs{\epsilon} \leq \Omega_0$, and $O \of{\Omega_0^2 / \epsilon^2}$ outside of it. Since the probability of guessing an $s$ in this range is $\Omega_0 / W$, we arrive at an \textit{average} value of $\Omega_0 / W$. However, the time to reach this value is $\propto 1/\Omega_0$, so the time to solution again scales as $W/\Omega_0^2$, and the scaling of both choices is thus identical, and identical to AQC with constant-rate annealing. Without assuming exponential precision this algorithm thus does not recover any quantum speedup in the Grover problem; however, we will see shortly that this is not the case for RFQA.

In RFQA, this algorithm is modified to include oscillations in the various driver Hamiltonian terms during evolution. Under this modification, the arguments of the previous sections predict that, so long as we can ignore off-resonant local heating and interference between drive fields at different frequencies (both these assumptions are discussed in depth below), the solution probability $P$ will be modified to the final form
\begin{eqnarray}\label{PRFQA}
P_{RFQA} \of{\epsilon} = \sum_k  \frac{\abs{M_k}^2 \sin^2 \of{\sqrt{\of{\epsilon - \tilde{f}_k}^2 + \abs{M_k}^2} t_f} }{ \of{\epsilon - \tilde{f}_k}^2 + \abs{M_k}^2} . 
\end{eqnarray}
Here, $M_k$ is the matrix elements of a given $n$-photon transition at total summed frequency $\tilde{f}_k$, and there are exponentially many terms in the sum. Repeating the analysis of the previous paragraph leads us to conclude that the solution probability, at short times, scales as 
\begin{eqnarray}\label{RFQArate}
P_{soln} \of{t_f} \propto \frac{t_f}{W} \sum_k \abs{M_k}^2.
\end{eqnarray}
This equation is generic, and as shown in the main text, it can model transition rates due to interaction with a bath as well. 

Formally, the bang-and-wait protocol we study here is equivalent to annealing with a pause, in the limit that the pause is much longer than the annealing time on either end of it. Likewise, when we study initializing the system in a classical minimum and searching for others, this protocol is equivalent to the population transfer schemes studied in \cite{baldwin2018quantum,faoro2018non,kechedzhi2018efficient,smelyanskiy2018non,smelyanskiy2019intermittency}, or the ``reverse annealing" employed in \cite{king2019quantum}. However, in all those cases the Hamiltonian is constant at the pause point; in our scheme the transverse fields are oscillated. 

\section{On violation of MSCALE}
\label{s:mscaleviolation}
While we imagine that one could craft models which violate MSCALE, we expect it to be true in an extremely broad class of cases. For example, consider crossings between minima deep in a quantum spin glass phase, which are believed to bottleneck the performance of AQC at large $N$ \cite{altshulerkrovi2010,knysh2016}. If the minimum gap in such a crossing can be computed approximately in high order perturbation theory (using $H_P$ as the base Hamiltonian and $H_D$ as the perturbation) \cite{pietracaprina2016forward,baldwinlaumann2016,baldwinlaumann2017,scardicchio2017perturbation,baldwin2018quantum}, then MSCALE can be shown to be true by inspection. This is because the matrix elements have identical perturbative structure to the tunneling matrix element that sets the minimum gap, albeit with up to $k \ll N$ fewer terms in the product. Similarly, we showed analytically in section~\ref{grovcalcsec} that MSCALE is true for the Grover problem, and numerically in section~\ref{poptransfersec} that it is also true in the quantum random energy model.

Further, if we restrict ourself to operators which are constituent parts of the many-body Hamiltonian, or simple rotations thereof, MSCALE being false would have curious implications for the value of the minimum gap itself. If MSCALE is true, then small perturbations to the Hamiltonian imply small (percentage-wise) variations in $\Delta_{min}$. However, if it is false, then small variations in local terms could produce huge changes in the tunneling matrix element, which, though not forbidden on thermodynamic grounds \cite{laumannmoessner2012}, strikes us as unphysical since the competing ground states in the phase transition do not have similarly chaotic responses to local perturbations\footnote{It is important to distinguish changes in a given pair of ground states' \textit{configuration} from their \textit{energetic heirarchy} in parameter space. Near a first order phase transition, where the gap is exponentially small, small perturbations can change the energetic hierarchy of competing states. However, if the system is in either state, its qualitative character will not change, and one would need to wait an exponentially long time to detect the mixing. The tunneling matrix element between the two states, which is set by their structure and the many-body Hamiltonian, should not be exponentially sensitive (percentage-wise, since it is typically exponentially small to begin with at first order transitions) to changes in these local energies.}. Likewise, one would expect that if the matrix elements of some local operators are allowed to decay with a slower exponent than $\Delta_{min}$ as $N$ increases, noise would almost always solve the problem more quickly than in a closed system, because noise can be modeled as collections of random local operations. We are unaware of any physical evidence for this fairly absurd possibility.

\section{Self-interference}\label{sisec}

The analysis of the previous section made two key assumptions: that the $m$-photon transitions induced by different collections of oscillating fields do not interfere with each other, and that system remains in the two-level manifold spanned by the competing ground states. The latter assumption (of no local heating) was discussed in detail in the main text; we consider the non-interference concern here.

For the problems considered in this work, we can appeal to the optimality of Grover's algorithm to infer when frequency crowding might cause the non-interference assumption to be violated. Specifically, consider two driven processes with effective Rabi frequencies $\Omega_1$ and $\Omega_2$, at total frequencies $f_1$ and $f_2$. If the separation between these frequencies is greater than the Rabi frequency of either process $\of{\abs{f_1 - f_2} > \abs{\Omega_{1,2}}}$, then interference between the two processes is not possible and they can be summed independently in Eq. (\ref{RFQArate}). If on the other hand the two frequencies are very close to each other they will interfere (in a way which could be constructive or destructive), but for just two processes averaging over random phases for the two processes produces an identical result to simply summing their squares independently. The danger comes when exponentially many processes must be considered. Specifically, the optimality of Grover's algorithm ensures that, for any range of frequencies with a width $2^{-N/2}$, the total Rabi frequency of all contributions in that frequency range cannot on average be larger than $O \of{ 2^{-N/2}}$, since that would imply a superoptimal solution to the Grover problem if we were to assume exponential precision in applying $H_G$ and jump to exactly that known spot. This in turn means that if we choose the frequency range $\cuof{f_{min},f_{max}}$ to be too small, many of the contributions to (\ref{RFQArate}) will be wasted in interference, and we will not recover the full quantum speedup of RFQA\footnote{The degree to which this self-interference effect will occur in more realistic problems, where no such optimality constraint exists, is not clear in general and likely depends on the details of the problem class.}. Clearly, $f_{min}$ and $f_{max}$ must be at least inverse polynomially large in $N$ for RFQA to succeed.

\section{Setting up simulations of $1/f$ noise, comparison to Dwave's flux qubits and peculiarities of QREM}
\label{sec:1overfsimulation}
Since its magnitude diverges at infinite time and/or zero frequency, care must be taken when simulating $1/f$ noise. To simulate its effects on RFQA, we used the following prescription. First, we chose a ``noise time" $t_N$ to be longer than any time window used in our simulations. We then generated individual $1/f$ traces by randomly generating $t_N/dt$ Fourier components, each with a random phase and amplitude which has an average value that scales as $S_0/f_n$, Fourier transformed the result to construct the coefficient $h_z \of{t}$, then subtracted the initial value $h_z \of{t} \to h_z \of{t} - h_z \of{0}$ to avoid complications from the diverging low frequency part. To reflect that qubits are not calibrated individually from one trial to the next in real experiments, we started the ``clock"  $h_z \of{t} \to h_z \of{t + t_0}$ near the middle of its time window when incorporating it in the simulation. Finally, to set the overall strength $S_0$, we computed the Ramsey $T_2$ for a single qubit experiencing $1/f$ noise by averaging over several thousand random traces generated by this method, and adjusted $S_0$ accordingly to arrive at the desired value. Our results were averaged over hundreds of individual sets of $\cuof{ h_{iz} \of{t} }$ to obtain good statistical accuracy.

The reason our strong noise simulations erase the quantum speedup as $N$ gets large is subtle, and has to do with a peculiarity of the QREM/Grover problem which is not found in more realistic spin glass problems. Specifically, as argued by Smelyanskiy \textit{et al} \cite{smelyanskiy2019intermittency}, residual bands of paramagnet-like states persist when the energy scales of the driver and problem Hamiltonian are nearly equal, and the squared matrix element between a low energy QREM state and any paramagnet state is $O \of{2^{-N}}$. The transition rate between a QREM minimum and one of these states is suppressed by the inverse energy difference, but this polynomial suppression is not enough to prevent heating to very high energy states, given that there are exponentially many such states which can be mixed diabatically through the high frequency tail of local $1/f$ noise. This error channel -- direct mixing with mid-spectrum states through local, low-frequency dominated phase noise -- is not known to occur in more realistic spin glass problems constructed from sums of few-body interaction terms, since the paramagnetic band structure generally does not survive in the spin glass phase in those models. Note also that paramagnetic states can relax rapidly back to the paramagnet ground state through simple local operations, so interaction with a cold bath would still effectively correct this error channel (as the paramagnetic ground state will mix directly with the QREM ground state band through RFQA), but again, such simulations would be prohibitively expensive for classical machines.

\section{Implementation in superconducting flux qubits}\label{fluxqubitsec}

In this section we show that our method has a straightforward implementation in superconducting flux qubits. The Hamiltonian of the superconducting flux qubit consists of the parallel combination of a SQUID, a capacitor, and an inductor. If we assume that the capacitance is $C$ and the capacitive coupling to an external applied voltage is $\gamma C$, the SQUID junctions have identical energy $E_J$, external magnetic fluxes $\Phi_S$ and $\Phi_L$ (expressed as multiples of the flux quantum $\Phi_0 = h /2 e$) are threaded through the SQUID and central loop, and the inductive energy is $E_L$, we can be express the flux qubit Hamiltonian in terms of the superconducting phase $\phi$ across the combination of circuit elements as
\begin{eqnarray}
H_{FQ} &=& \frac{\of{Q - \gamma C V_{ext} }^2 }{2 C \of{1+\gamma}} + \frac{E_L \phi^2}{2} \\
& & - E_J \sqof{\cos \of{\phi + \Phi_L} + \cos \of{\phi + \Phi_L + \Phi_S} }. \nonumber 
\end{eqnarray}
The flux qubit is quantized by using the canonical commutation relations to rewrite the charge operator in the phase basis ($Q \to - 2 e i \partial/\partial \phi$), and at the flux qubit operating point $\Phi_L = \pi$ with $E_J > E_L/2$, $H_{FQ}$ reduces to the Hamiltonian of a particle in a one dimensional double well with minima at $\pm \phi_m$, with the ground and first excited states defined by the symmetric and antisymmetric combinations of the two well states. If we define the left and right wells (corresponding to CW and CCW circulating persistent currents since $I \of{\phi} = I_c \sin \phi$) to be $\ket{0}$ and $\ket{1}$ in the $z$ basis, then quantum tunneling between the two wells produces an energetic splitting $2 \Delta$ between the two combinations.

Now, if $V_{ext}$ is a static quantity it can be removed from $H_{FQ}$ via a unitary gauge transformation, $\ket{\psi} \to \exp  \of{i \frac{\gamma C V_{ext}}{2e} \phi} \ket{\psi}$. Equivalently, if we do not perform such a transformation, the external voltage produces a complex phase difference of approximately $\theta \simeq 2 \phi_m \frac{\gamma C V_{ext}}{2e}$ between the two persistent current eigenstates. If we take these two states to be our logical basis, this transforms the tunneling term $\Delta \sigma^x$ in the flux qubit Hamiltonian to the linear combination $\cos \theta \sigma^x + \sin \theta \sigma^y$, where $\theta$ is the complex phase difference between the two well states. Again, for static fields this has no effect on the physics of the system since it can be gauged away, but for time-varying $V_{ext}$ the consequence of rotating the transverse field between $x$ and $y$ cannot be eliminated and may have dramatic impacts on the system's evolution. Thus, the transverse field oscillations of RFQA-D can be straightforwardly engineered by including capacitive couplings between the qubits and external voltage sources, something which is ubiquitous in radio-frequency driven platforms such as transmon circuits, but has not to our knowledge been included in quantum annealers so far.

\section{Optimal runtimes and choice of $\kappa_c$ for BQREM and Grover population transfer}\label{optruntimesec}

In the ``Analytical formalism" subsection, we derived a simple exponential decay law for the mixing between the paramagnetic ground state and the Grover state, using the analytically calculated matrix elements and final expression (\ref{maxGrate}). Reflecting MSCALE, the $n$-photon matrix elements elements are all proportional to the primary resonant tunneling rate between $\ket{G'}$ and $\ket{0'}$, which in turn scales as $2^{-N/2}$. Importantly, we did \textit{not} calculate the direct resonant mixing between two distinct Grover states, a problem that was exhaustively studied by Smelyanskiy et al \cite{smelyanskiy2018non}, who found that the direct matrix element $\Omega_{ij}$ between two Grover states at equal energy could only obtain the optimal $2^{-N/2}$ scaling when the transverse field strength was boosted relative to the magnitude of the Grover Hamiltonian by a factor $\propto \sqrt{N}$. If the transverse field and Grover Hamiltonians have near equal magnitude the scaling is substantially worse than $2^{-N/2}$ (though still better than the classical limit of $2^{-N}$), a trait observed in the QREM as well \cite{faoro2018non,smelyanskiy2019intermittency}.

Operating the system with a huge transverse field is disastrous for attempting to prove fault-welcoming behavior, since the solution states are far from the true ground state of the system and any interaction with a cold bath will cause the system to decay into the paramagnetic manifold long before another solution state can be found. And given MSCALE, starting from a transition matrix element whose square exhibits worse scaling than $2^{-N}$ would erase any chance of a quantum speedup, even when the oscillating fields of RFQA are included. However, all is not lost. Let us choose the transverse field strength $\kappa$ such that the paramagnet ground state $\ket{0'}$ lies in the band of solutions. Then, when oscillating fields are turned on, the initial solution state will mix with the paramagnet at a rate $O \of{2^{-0.75 N}}$, which can then mix with any of the $M-1$ other solutions also at a rate which scales as $O \of{2^{-0.75N}}$, regaining the quantum speedup. In any individual realization these two rates will have randomized $O \of{1}$ prefactors and thus differ from one another (though both will have the same scaling with $N$), and appealing to the well-known solution of multi-step chemical reactions, the short time behavior will not yield a linear increase in success probability with $N$ as it does in the simpler process of direct incoherent mixing of two states. Consequently, even ignoring the time costs of state preparation and measurement, the optimal per-trial runtime will scale as $2^{0.75N}/M$ to achieve constant success probability, providing the quantum speedup predicted earlier.

\section{Review of past work on environment assisted quantum algorithms}
\label{sec:EAss}
Before considering whether RFQA in flux qubits could qualify as fault-welcoming, it is important to review and acknowledge previous work on the topic of environment-assisted quantum algorithms, both in quantum annealing and closely related analog quantum walk models (which reduce to transverse field quantum annealing, with a pause in the schedule, if the walk adjacency matrix is taken to be the nearest neighbor $N$-dimensional hypercube). Interest in this topic perhaps began with the important early work of Amin et al \cite{aminlove2008}, who found that a superohmic bath could accelerate quantum annealing by thermal mixing ahead of the phase transition, though this speedup is removed by the presence of low-frequency potential noise, which is unfortunately ubiquitous in real systems. Subsequent work found that a cold bath helps find a solution more quickly in the physical d-Wave flux qubits \cite{dickson2013thermally,kadowaki2019experimental,marshall2019power}, and in theoretical constructions such as the p-spin model \cite{cattaneo2018quantum,passarelli2019may} or more general problems \cite{keck2017dissipation,smelyanskiy2017quantum,venuti2017relaxation,arceci2018optimal,suzuki2019quantum,roberts2019noise} including the frequently studied transverse field Ising chain. The MSCALE conjecture supports these results-- if we assume a constant system-bath coupling strength and ignore subtleties from the energy dependence of the bath density of states (which can be very important but may also be washed out by phase noise \cite{aminlove2008}), then relaxation after the phase transition should be faster than closed system, constant-rate annealing by a factor of $N$, since there are $N$ spins interacting with the bath and the scaling of the bath-assisted phase transition rate from single spin operations should be identical to the minimum gap scaling, up to a prefactor. This order $N$ enhancement was recently derived rigorously by Roberts \textit{et al} in a frustrated ring model \cite{roberts2019noise}; they also found good qualitative agreement with their theoretical prediction in experimental studies of the problem on a D-Wave 2X quantum annealing machine. Finally, a study of truly unstructured search \cite{wild2016adiabatic}, where both driver and problem Hamiltonians lack any bit structure, found that finite temperature baths with strictly longitudinal couplings erase the quantum speedup.

In these studies the benefit of a cold bath is clear, but the underlying problems are not ones where a quantum speedup exists, or at least has been shown. Further, theoretical studies proved that both finite temperature and imprecision in specifying interactions can both exponentially reduce the success probability for finding ground states of spin glasses at large $N$, \cite{albash2017temperature,slutskii2019analog}, though these issues can be mitigated by constructions based on quantum annealing correction \cite{vinci2018scalable,pearson2019analog}. Similar work in quantum walk models \cite{cattaneo2018quantum,novo2018environment,morley2019quantum} showed that there are regimes where a cold bath can increase the solution probability, but these schemes do not yield a provable quantum speedup when a realistic noise model is considered. In particular, Novo \textit{et al} \cite{novo2018environment} considered the Grover problem and found that dissipation via a bath can mitigate the effect of weak static disorder, and improve performance relative to constant schedule annealing, but were unable to recover a polynomial speedup in realistic limits. Note also that (non-oscillatory) analog implementations of the Grover problem are pathological in the sense that exponential precision in Hamiltonian terms and the annealing schedule is required to obtain the square root speedup, which is both unrealistic and is a sufficiently strong resource to produce to polynomial speedups in fully classical models \cite{hen2019quantum}. In population transfer protocols for the Grover problem and QREM \cite{baldwin2018quantum,faoro2018non,kechedzhi2018efficient,smelyanskiy2018non}, the algorithm is not exponentially sensitive to the ratio of the transverse field strength $\kappa$ to the problem Hamiltonian energy scale, as it is in the traditional continuous time formulation of the Grover problem \cite{rolandcerf2002}, but the ground states in question must be exponentially close in energy, with all corrections from the transverse fields taken into account, to obtain a quantum speedup in mixing them. Unitary longitudinal noise (which arises from both $1/f$ potential fluctuations and control imprecision) would broaden the band to inverse polynomial width and likely erase the speedup in a real analog implementation.

These results, while often encouraging, all stop short of establishing what we would consider fault welcoming behavior. 
First, no digital, gate-model quantum computer of any sort can be fault welcoming, since lacking any continuously applied Hamiltonian to set energy scales, all noise in those systems is effectively infinite temperature and induces a uniformly deleterious incoherent random walk through Hilbert space. They can, of course, be made fault tolerant though digital error correction, such as a surface code \cite{fowlersurface}. Similarly, the quantum annealers currently sold by D-Wave Systems, which use flux qubits arranged in a chimera graph, are not \textit{in current understanding} fault welcoming according to our definition, though the issue is considerably more complex. As cited above, D-Wave quantum annealers exhibit a helpful noise channel in bath-assisted relaxation, a necessary precondition for fault-welcoming quantum computing. But as mentioned earlier, to date no conclusive evidence of a quantum speedup over state of the art classical algorithms has emerged in these machines. A variety of proposals to improve performance, such as transverse couplers, counter-diabatic terms and inhomogenous driving \cite{crosson2014different,hormozibrown2017,sels2017minimizing,PhysRevA.98.042326,PhysRevLett.123.120501,hauke2019perspectives}, could prove effective but scalable quantum speedups have yet to be shown for realistic disordered problems. Further, a recent study by D-Wave \cite{dwave2019whitepaper} showed that reducing noise through improved fabrication techniques reduced the time to solution in a chosen trial problem, suggesting that the sum of noise effects is harmful in that machine. Finally, given that in many, though not all, cases quantum annealing with a stoquastic Hamiltonian can be efficiently simulated with quantum Monte Carlo \cite{isakov2016understanding,andriyash2017can,jiang2017scaling,jiang2017path,albash2018demonstration}, the prospects for demonstrating a broadly applicable quantum speedup in the simplest implementation of quantum annealing (near-uniform transverse field with local $z$ and $zz$ interactions) appear somewhat dim.

\bibliography{fullbib}

\begin{thebibliography}{108}%
\makeatletter
\providecommand \@ifxundefined [1]{%
 \@ifx{#1\undefined}
}%
\providecommand \@ifnum [1]{%
 \ifnum #1\expandafter \@firstoftwo
 \else \expandafter \@secondoftwo
 \fi
}%
\providecommand \@ifx [1]{%
 \ifx #1\expandafter \@firstoftwo
 \else \expandafter \@secondoftwo
 \fi
}%
\providecommand \natexlab [1]{#1}%
\providecommand \enquote  [1]{``#1''}%
\providecommand \bibnamefont  [1]{#1}%
\providecommand \bibfnamefont [1]{#1}%
\providecommand \citenamefont [1]{#1}%
\providecommand \href@noop [0]{\@secondoftwo}%
\providecommand \href [0]{\begingroup \@sanitize@url \@href}%
\providecommand \@href[1]{\@@startlink{#1}\@@href}%
\providecommand \@@href[1]{\endgroup#1\@@endlink}%
\providecommand \@sanitize@url [0]{\catcode `\\12\catcode `\$12\catcode
  `\&12\catcode `\#12\catcode `\^12\catcode `\_12\catcode `\%12\relax}%
\providecommand \@@startlink[1]{}%
\providecommand \@@endlink[0]{}%
\providecommand \url  [0]{\begingroup\@sanitize@url \@url }%
\providecommand \@url [1]{\endgroup\@href {#1}{\urlprefix }}%
\providecommand \urlprefix  [0]{URL }%
\providecommand \Eprint [0]{\href }%
\providecommand \doibase [0]{http://dx.doi.org/}%
\providecommand \selectlanguage [0]{\@gobble}%
\providecommand \bibinfo  [0]{\@secondoftwo}%
\providecommand \bibfield  [0]{\@secondoftwo}%
\providecommand \translation [1]{[#1]}%
\providecommand \BibitemOpen [0]{}%
\providecommand \bibitemStop [0]{}%
\providecommand \bibitemNoStop [0]{.\EOS\space}%
\providecommand \EOS [0]{\spacefactor3000\relax}%
\providecommand \BibitemShut  [1]{\csname bibitem#1\endcsname}%
\let\auto@bib@innerbib\@empty
%</preamble>
\bibitem [{\citenamefont {Aharonov}\ and\ \citenamefont
  {Ben-Or}(1999)}]{aharonov1999fault}%
  \BibitemOpen
  \bibfield  {author} {\bibinfo {author} {\bibfnamefont {D.}~\bibnamefont
  {Aharonov}}\ and\ \bibinfo {author} {\bibfnamefont {M.}~\bibnamefont
  {Ben-Or}},\ }\href@noop {} {\bibfield  {journal} {\bibinfo  {journal} {arXiv
  preprint quant-ph/9906129}\ } (\bibinfo {year} {1999})}\BibitemShut {NoStop}%
\bibitem [{\citenamefont {Knill}\ \emph {et~al.}(1998)\citenamefont {Knill},
  \citenamefont {Laflamme},\ and\ \citenamefont {Zurek}}]{knill1998resilient}%
  \BibitemOpen
  \bibfield  {author} {\bibinfo {author} {\bibfnamefont {E.}~\bibnamefont
  {Knill}}, \bibinfo {author} {\bibfnamefont {R.}~\bibnamefont {Laflamme}}, \
  and\ \bibinfo {author} {\bibfnamefont {W.~H.}\ \bibnamefont {Zurek}},\
  }\href@noop {} {\bibfield  {journal} {\bibinfo  {journal} {Science}\ }\textbf
  {\bibinfo {volume} {279}},\ \bibinfo {pages} {342} (\bibinfo {year}
  {1998})}\BibitemShut {NoStop}%
\bibitem [{\citenamefont {Terhal}(2015)}]{terhal2015}%
  \BibitemOpen
  \bibfield  {author} {\bibinfo {author} {\bibfnamefont {B.~M.}\ \bibnamefont
  {Terhal}},\ }\href {\doibase 10.1103/RevModPhys.87.307} {\bibfield  {journal}
  {\bibinfo  {journal} {Rev. Mod. Phys. \textbf{87}, 307}\ } (\bibinfo {year}
  {2015}),\ 10.1103/RevModPhys.87.307}\BibitemShut {NoStop}%
\bibitem [{\citenamefont {Fowler}\ \emph {et~al.}(2012)\citenamefont {Fowler},
  \citenamefont {Mariantoni}, \citenamefont {Martinis},\ and\ \citenamefont
  {Cleland}}]{fowlersurface}%
  \BibitemOpen
  \bibfield  {author} {\bibinfo {author} {\bibfnamefont {A.~G.}\ \bibnamefont
  {Fowler}}, \bibinfo {author} {\bibfnamefont {M.}~\bibnamefont {Mariantoni}},
  \bibinfo {author} {\bibfnamefont {J.~M.}\ \bibnamefont {Martinis}}, \ and\
  \bibinfo {author} {\bibfnamefont {A.~N.}\ \bibnamefont {Cleland}},\ }\href
  {\doibase 10.1103/PhysRevA.86.032324} {\bibfield  {journal} {\bibinfo
  {journal} {Phys. Rev. A \textbf{86}, 032324}\ } (\bibinfo {year} {2012}),\
  10.1103/PhysRevA.86.032324}\BibitemShut {NoStop}%
\bibitem [{\citenamefont {Peruzzo}\ \emph {et~al.}(2014)\citenamefont
  {Peruzzo}, \citenamefont {McClean}, \citenamefont {Shadbolt}, \citenamefont
  {Yung}, \citenamefont {Zhou}, \citenamefont {Love}, \citenamefont
  {Aspuru-Guzik},\ and\ \citenamefont {O?brien}}]{peruzzo2014variational}%
  \BibitemOpen
  \bibfield  {author} {\bibinfo {author} {\bibfnamefont {A.}~\bibnamefont
  {Peruzzo}}, \bibinfo {author} {\bibfnamefont {J.}~\bibnamefont {McClean}},
  \bibinfo {author} {\bibfnamefont {P.}~\bibnamefont {Shadbolt}}, \bibinfo
  {author} {\bibfnamefont {M.-H.}\ \bibnamefont {Yung}}, \bibinfo {author}
  {\bibfnamefont {X.-Q.}\ \bibnamefont {Zhou}}, \bibinfo {author}
  {\bibfnamefont {P.~J.}\ \bibnamefont {Love}}, \bibinfo {author}
  {\bibfnamefont {A.}~\bibnamefont {Aspuru-Guzik}}, \ and\ \bibinfo {author}
  {\bibfnamefont {J.~L.}\ \bibnamefont {O?brien}},\ }\href@noop {} {\bibfield
  {journal} {\bibinfo  {journal} {Nature communications}\ }\textbf {\bibinfo
  {volume} {5}},\ \bibinfo {pages} {4213} (\bibinfo {year} {2014})}\BibitemShut
  {NoStop}%
\bibitem [{\citenamefont {McClean}\ \emph {et~al.}(2016)\citenamefont
  {McClean}, \citenamefont {Romero}, \citenamefont {Babbush},\ and\
  \citenamefont {Aspuru-Guzik}}]{mcclean2016theory}%
  \BibitemOpen
  \bibfield  {author} {\bibinfo {author} {\bibfnamefont {J.~R.}\ \bibnamefont
  {McClean}}, \bibinfo {author} {\bibfnamefont {J.}~\bibnamefont {Romero}},
  \bibinfo {author} {\bibfnamefont {R.}~\bibnamefont {Babbush}}, \ and\
  \bibinfo {author} {\bibfnamefont {A.}~\bibnamefont {Aspuru-Guzik}},\
  }\href@noop {} {\bibfield  {journal} {\bibinfo  {journal} {New Journal of
  Physics}\ }\textbf {\bibinfo {volume} {18}},\ \bibinfo {pages} {023023}
  (\bibinfo {year} {2016})}\BibitemShut {NoStop}%
\bibitem [{\citenamefont {Farhi}\ \emph {et~al.}(2014)\citenamefont {Farhi},
  \citenamefont {Goldstone},\ and\ \citenamefont {Gutmann}}]{farhi2014quantum}%
  \BibitemOpen
  \bibfield  {author} {\bibinfo {author} {\bibfnamefont {E.}~\bibnamefont
  {Farhi}}, \bibinfo {author} {\bibfnamefont {J.}~\bibnamefont {Goldstone}}, \
  and\ \bibinfo {author} {\bibfnamefont {S.}~\bibnamefont {Gutmann}},\
  }\href@noop {} {\bibfield  {journal} {\bibinfo  {journal} {arXiv preprint
  arXiv:1411.4028}\ } (\bibinfo {year} {2014})}\BibitemShut {NoStop}%
\bibitem [{\citenamefont {Finnila}\ \emph {et~al.}(1994)\citenamefont
  {Finnila}, \citenamefont {Gomez}, \citenamefont {Sebenik}, \citenamefont
  {Stenson},\ and\ \citenamefont {Doll}}]{finnila1994quantum}%
  \BibitemOpen
  \bibfield  {author} {\bibinfo {author} {\bibfnamefont {A.}~\bibnamefont
  {Finnila}}, \bibinfo {author} {\bibfnamefont {M.}~\bibnamefont {Gomez}},
  \bibinfo {author} {\bibfnamefont {C.}~\bibnamefont {Sebenik}}, \bibinfo
  {author} {\bibfnamefont {C.}~\bibnamefont {Stenson}}, \ and\ \bibinfo
  {author} {\bibfnamefont {J.}~\bibnamefont {Doll}},\ }\href@noop {} {\bibfield
   {journal} {\bibinfo  {journal} {Chemical physics letters}\ }\textbf
  {\bibinfo {volume} {219}},\ \bibinfo {pages} {343} (\bibinfo {year}
  {1994})}\BibitemShut {NoStop}%
\bibitem [{\citenamefont {Kadowaki}\ and\ \citenamefont
  {Nishimori}(1998)}]{kadowakinishimori1998}%
  \BibitemOpen
  \bibfield  {author} {\bibinfo {author} {\bibfnamefont {T.}~\bibnamefont
  {Kadowaki}}\ and\ \bibinfo {author} {\bibfnamefont {H.}~\bibnamefont
  {Nishimori}},\ }\href@noop {} {\bibfield  {journal} {\bibinfo  {journal}
  {Physical Review E}\ }\textbf {\bibinfo {volume} {58}},\ \bibinfo {pages}
  {5355} (\bibinfo {year} {1998})}\BibitemShut {NoStop}%
\bibitem [{\citenamefont {Das}\ and\ \citenamefont
  {Chakrabarti}(2008)}]{das2008colloquium}%
  \BibitemOpen
  \bibfield  {author} {\bibinfo {author} {\bibfnamefont {A.}~\bibnamefont
  {Das}}\ and\ \bibinfo {author} {\bibfnamefont {B.~K.}\ \bibnamefont
  {Chakrabarti}},\ }\href@noop {} {\bibfield  {journal} {\bibinfo  {journal}
  {Reviews of Modern Physics}\ }\textbf {\bibinfo {volume} {80}},\ \bibinfo
  {pages} {1061} (\bibinfo {year} {2008})}\BibitemShut {NoStop}%
\bibitem [{\citenamefont {Johnson}\ \emph {et~al.}(2011)\citenamefont
  {Johnson}, \citenamefont {Amin}, \citenamefont {Gildert}, \citenamefont
  {Lanting}, \citenamefont {Hamze}, \citenamefont {Dickson}, \citenamefont
  {Harris}, \citenamefont {Berkley}, \citenamefont {Johansson}, \citenamefont
  {Bunyk} \emph {et~al.}}]{johnson2011quantum}%
  \BibitemOpen
  \bibfield  {author} {\bibinfo {author} {\bibfnamefont {M.~W.}\ \bibnamefont
  {Johnson}}, \bibinfo {author} {\bibfnamefont {M.~H.}\ \bibnamefont {Amin}},
  \bibinfo {author} {\bibfnamefont {S.}~\bibnamefont {Gildert}}, \bibinfo
  {author} {\bibfnamefont {T.}~\bibnamefont {Lanting}}, \bibinfo {author}
  {\bibfnamefont {F.}~\bibnamefont {Hamze}}, \bibinfo {author} {\bibfnamefont
  {N.}~\bibnamefont {Dickson}}, \bibinfo {author} {\bibfnamefont
  {R.}~\bibnamefont {Harris}}, \bibinfo {author} {\bibfnamefont {A.~J.}\
  \bibnamefont {Berkley}}, \bibinfo {author} {\bibfnamefont {J.}~\bibnamefont
  {Johansson}}, \bibinfo {author} {\bibfnamefont {P.}~\bibnamefont {Bunyk}},
  \emph {et~al.},\ }\href@noop {} {\bibfield  {journal} {\bibinfo  {journal}
  {Nature}\ }\textbf {\bibinfo {volume} {473}},\ \bibinfo {pages} {194}
  (\bibinfo {year} {2011})}\BibitemShut {NoStop}%
\bibitem [{\citenamefont {Boixo}\ \emph {et~al.}(2014)\citenamefont {Boixo},
  \citenamefont {R{\o}nnow}, \citenamefont {Isakov}, \citenamefont {Wang},
  \citenamefont {Wecker}, \citenamefont {Lidar}, \citenamefont {Martinis},\
  and\ \citenamefont {Troyer}}]{boixo2014evidence}%
  \BibitemOpen
  \bibfield  {author} {\bibinfo {author} {\bibfnamefont {S.}~\bibnamefont
  {Boixo}}, \bibinfo {author} {\bibfnamefont {T.~F.}\ \bibnamefont
  {R{\o}nnow}}, \bibinfo {author} {\bibfnamefont {S.~V.}\ \bibnamefont
  {Isakov}}, \bibinfo {author} {\bibfnamefont {Z.}~\bibnamefont {Wang}},
  \bibinfo {author} {\bibfnamefont {D.}~\bibnamefont {Wecker}}, \bibinfo
  {author} {\bibfnamefont {D.~A.}\ \bibnamefont {Lidar}}, \bibinfo {author}
  {\bibfnamefont {J.~M.}\ \bibnamefont {Martinis}}, \ and\ \bibinfo {author}
  {\bibfnamefont {M.}~\bibnamefont {Troyer}},\ }\href@noop {} {\bibfield
  {journal} {\bibinfo  {journal} {Nature Physics}\ }\textbf {\bibinfo {volume}
  {10}},\ \bibinfo {pages} {218} (\bibinfo {year} {2014})}\BibitemShut
  {NoStop}%
\bibitem [{\citenamefont {Albash}\ and\ \citenamefont
  {Lidar}(2017)}]{albashlidar2017}%
  \BibitemOpen
  \bibfield  {author} {\bibinfo {author} {\bibfnamefont {T.}~\bibnamefont
  {Albash}}\ and\ \bibinfo {author} {\bibfnamefont {D.~A.}\ \bibnamefont
  {Lidar}},\ }\href@noop {} {\bibfield  {journal} {\bibinfo  {journal}
  {arXiv:1611.04471}\ } (\bibinfo {year} {2017})}\BibitemShut {NoStop}%
\bibitem [{\citenamefont {Albash}\ and\ \citenamefont
  {Lidar}(2018)}]{albash2018demonstration}%
  \BibitemOpen
  \bibfield  {author} {\bibinfo {author} {\bibfnamefont {T.}~\bibnamefont
  {Albash}}\ and\ \bibinfo {author} {\bibfnamefont {D.~A.}\ \bibnamefont
  {Lidar}},\ }\href@noop {} {\bibfield  {journal} {\bibinfo  {journal}
  {Physical Review X}\ }\textbf {\bibinfo {volume} {8}},\ \bibinfo {pages}
  {031016} (\bibinfo {year} {2018})}\BibitemShut {NoStop}%
\bibitem [{\citenamefont {Mizel}\ \emph {et~al.}(2007)\citenamefont {Mizel},
  \citenamefont {Lidar},\ and\ \citenamefont
  {Mitchell}}]{PhysRevLett.99.070502}%
  \BibitemOpen
  \bibfield  {author} {\bibinfo {author} {\bibfnamefont {A.}~\bibnamefont
  {Mizel}}, \bibinfo {author} {\bibfnamefont {D.~A.}\ \bibnamefont {Lidar}}, \
  and\ \bibinfo {author} {\bibfnamefont {M.}~\bibnamefont {Mitchell}},\ }\href
  {\doibase 10.1103/PhysRevLett.99.070502} {\bibfield  {journal} {\bibinfo
  {journal} {Phys. Rev. Lett.}\ }\textbf {\bibinfo {volume} {99}},\ \bibinfo
  {pages} {070502} (\bibinfo {year} {2007})}\BibitemShut {NoStop}%
\bibitem [{\citenamefont {Aharonov}\ \emph {et~al.}(2008)\citenamefont
  {Aharonov}, \citenamefont {Van~Dam}, \citenamefont {Kempe}, \citenamefont
  {Landau}, \citenamefont {Lloyd},\ and\ \citenamefont
  {Regev}}]{aharonov2008adiabatic}%
  \BibitemOpen
  \bibfield  {author} {\bibinfo {author} {\bibfnamefont {D.}~\bibnamefont
  {Aharonov}}, \bibinfo {author} {\bibfnamefont {W.}~\bibnamefont {Van~Dam}},
  \bibinfo {author} {\bibfnamefont {J.}~\bibnamefont {Kempe}}, \bibinfo
  {author} {\bibfnamefont {Z.}~\bibnamefont {Landau}}, \bibinfo {author}
  {\bibfnamefont {S.}~\bibnamefont {Lloyd}}, \ and\ \bibinfo {author}
  {\bibfnamefont {O.}~\bibnamefont {Regev}},\ }\href@noop {} {\bibfield
  {journal} {\bibinfo  {journal} {SIAM review}\ }\textbf {\bibinfo {volume}
  {50}},\ \bibinfo {pages} {755} (\bibinfo {year} {2008})}\BibitemShut
  {NoStop}%
\bibitem [{\citenamefont {Grover}(1997)}]{grover1997}%
  \BibitemOpen
  \bibfield  {author} {\bibinfo {author} {\bibfnamefont {L.~K.}\ \bibnamefont
  {Grover}},\ }\href {\doibase 10.1103/PhysRevLett.79.325} {\bibfield
  {journal} {\bibinfo  {journal} {Phys. Rev. Lett.}\ }\textbf {\bibinfo
  {volume} {79}},\ \bibinfo {pages} {325} (\bibinfo {year} {1997})}\BibitemShut
  {NoStop}%
\bibitem [{\citenamefont {Zalka}(1999)}]{zalka1999}%
  \BibitemOpen
  \bibfield  {author} {\bibinfo {author} {\bibfnamefont {C.}~\bibnamefont
  {Zalka}},\ }\href@noop {} {\bibfield  {journal} {\bibinfo  {journal}
  {Physical Review A}\ }\textbf {\bibinfo {volume} {60}},\ \bibinfo {pages}
  {2746} (\bibinfo {year} {1999})}\BibitemShut {NoStop}%
\bibitem [{\citenamefont {Roland}\ and\ \citenamefont
  {Cerf}(2002)}]{rolandcerf2002}%
  \BibitemOpen
  \bibfield  {author} {\bibinfo {author} {\bibfnamefont {J.}~\bibnamefont
  {Roland}}\ and\ \bibinfo {author} {\bibfnamefont {N.~J.}\ \bibnamefont
  {Cerf}},\ }\href {\doibase 10.1103/PhysRevA.65.042308} {\bibfield  {journal}
  {\bibinfo  {journal} {Phys. Rev. A \textbf{65}, 042308}\ } (\bibinfo {year}
  {2002}),\ 10.1103/PhysRevA.65.042308}\BibitemShut {NoStop}%
\bibitem [{\citenamefont {Yoder}\ \emph {et~al.}(2014)\citenamefont {Yoder},
  \citenamefont {Low},\ and\ \citenamefont {Chuang}}]{yoderguang2014}%
  \BibitemOpen
  \bibfield  {author} {\bibinfo {author} {\bibfnamefont {T.~J.}\ \bibnamefont
  {Yoder}}, \bibinfo {author} {\bibfnamefont {G.~H.}\ \bibnamefont {Low}}, \
  and\ \bibinfo {author} {\bibfnamefont {I.~L.}\ \bibnamefont {Chuang}},\
  }\href {\doibase 10.1103/PhysRevLett.113.210501} {\bibfield  {journal}
  {\bibinfo  {journal} {Phys. Rev. Lett.}\ }\textbf {\bibinfo {volume} {113}},\
  \bibinfo {pages} {210501} (\bibinfo {year} {2014})}\BibitemShut {NoStop}%
\bibitem [{\citenamefont {Dalzell}\ \emph {et~al.}(2017)\citenamefont
  {Dalzell}, \citenamefont {Yoder},\ and\ \citenamefont
  {Chuang}}]{dalzellyoder2017}%
  \BibitemOpen
  \bibfield  {author} {\bibinfo {author} {\bibfnamefont {A.~M.}\ \bibnamefont
  {Dalzell}}, \bibinfo {author} {\bibfnamefont {T.~J.}\ \bibnamefont {Yoder}},
  \ and\ \bibinfo {author} {\bibfnamefont {I.~L.}\ \bibnamefont {Chuang}},\
  }\href@noop {} {\bibfield  {journal} {\bibinfo  {journal} {Physical Review
  A}\ }\textbf {\bibinfo {volume} {95}},\ \bibinfo {pages} {012311} (\bibinfo
  {year} {2017})}\BibitemShut {NoStop}%
\bibitem [{\citenamefont {Jiang}\ \emph
  {et~al.}(2017{\natexlab{a}})\citenamefont {Jiang}, \citenamefont {Rieffel},\
  and\ \citenamefont {Wang}}]{jiang2017near}%
  \BibitemOpen
  \bibfield  {author} {\bibinfo {author} {\bibfnamefont {Z.}~\bibnamefont
  {Jiang}}, \bibinfo {author} {\bibfnamefont {E.~G.}\ \bibnamefont {Rieffel}},
  \ and\ \bibinfo {author} {\bibfnamefont {Z.}~\bibnamefont {Wang}},\
  }\href@noop {} {\bibfield  {journal} {\bibinfo  {journal} {Physical Review
  A}\ }\textbf {\bibinfo {volume} {95}},\ \bibinfo {pages} {062317} (\bibinfo
  {year} {2017}{\natexlab{a}})}\BibitemShut {NoStop}%
\bibitem [{\citenamefont {Sarovar}\ and\ \citenamefont
  {Young}(2013)}]{sarovaryoung2013}%
  \BibitemOpen
  \bibfield  {author} {\bibinfo {author} {\bibfnamefont {M.}~\bibnamefont
  {Sarovar}}\ and\ \bibinfo {author} {\bibfnamefont {K.~C.}\ \bibnamefont
  {Young}},\ }\href {\doibase 10.1088/1367-2630/15/12/125032} {\bibfield
  {journal} {\bibinfo  {journal} {New J. Phys. \textbf{15}, 125032}\ }
  (\bibinfo {year} {2013}),\ 10.1088/1367-2630/15/12/125032}\BibitemShut
  {NoStop}%
\bibitem [{\citenamefont {Young}\ \emph {et~al.}(2013)\citenamefont {Young},
  \citenamefont {Sarovar},\ and\ \citenamefont
  {Blume-Kohout}}]{youngsarovar2013}%
  \BibitemOpen
  \bibfield  {author} {\bibinfo {author} {\bibfnamefont {K.~C.}\ \bibnamefont
  {Young}}, \bibinfo {author} {\bibfnamefont {M.}~\bibnamefont {Sarovar}}, \
  and\ \bibinfo {author} {\bibfnamefont {R.}~\bibnamefont {Blume-Kohout}},\
  }\href {\doibase 10.1103/PhysRevX.3.041013} {\bibfield  {journal} {\bibinfo
  {journal} {Phys. Rev. X \textbf{3}, 041013}\ } (\bibinfo {year} {2013}),\
  10.1103/PhysRevX.3.041013}\BibitemShut {NoStop}%
\bibitem [{\citenamefont {Pudenz}\ \emph {et~al.}(2014)\citenamefont {Pudenz},
  \citenamefont {Albash},\ and\ \citenamefont {Lidar}}]{pudenzalbash2014}%
  \BibitemOpen
  \bibfield  {author} {\bibinfo {author} {\bibfnamefont {K.~L.}\ \bibnamefont
  {Pudenz}}, \bibinfo {author} {\bibfnamefont {T.}~\bibnamefont {Albash}}, \
  and\ \bibinfo {author} {\bibfnamefont {D.~A.}\ \bibnamefont {Lidar}},\
  }\href@noop {} {\bibfield  {journal} {\bibinfo  {journal} {NATURE}\ }\textbf
  {\bibinfo {volume} {5}} (\bibinfo {year} {2014})}\BibitemShut {NoStop}%
\bibitem [{\citenamefont {Pudenz}\ \emph {et~al.}(2015)\citenamefont {Pudenz},
  \citenamefont {Albash},\ and\ \citenamefont {Lidar}}]{pudenzalbash2015}%
  \BibitemOpen
  \bibfield  {author} {\bibinfo {author} {\bibfnamefont {K.~L.}\ \bibnamefont
  {Pudenz}}, \bibinfo {author} {\bibfnamefont {T.}~\bibnamefont {Albash}}, \
  and\ \bibinfo {author} {\bibfnamefont {D.~A.}\ \bibnamefont {Lidar}},\
  }\href@noop {} {\bibfield  {journal} {\bibinfo  {journal} {Physical Review
  A}\ }\textbf {\bibinfo {volume} {91}},\ \bibinfo {pages} {042302} (\bibinfo
  {year} {2015})}\BibitemShut {NoStop}%
\bibitem [{\citenamefont {Vinci}\ \emph {et~al.}(2015)\citenamefont {Vinci},
  \citenamefont {Albash}, \citenamefont {Paz-Silva}, \citenamefont {Hen},\ and\
  \citenamefont {Lidar}}]{vincialbash2015}%
  \BibitemOpen
  \bibfield  {author} {\bibinfo {author} {\bibfnamefont {W.}~\bibnamefont
  {Vinci}}, \bibinfo {author} {\bibfnamefont {T.}~\bibnamefont {Albash}},
  \bibinfo {author} {\bibfnamefont {G.}~\bibnamefont {Paz-Silva}}, \bibinfo
  {author} {\bibfnamefont {I.}~\bibnamefont {Hen}}, \ and\ \bibinfo {author}
  {\bibfnamefont {D.~A.}\ \bibnamefont {Lidar}},\ }\href@noop {} {\bibfield
  {journal} {\bibinfo  {journal} {Physical Review A}\ }\textbf {\bibinfo
  {volume} {92}},\ \bibinfo {pages} {042310} (\bibinfo {year}
  {2015})}\BibitemShut {NoStop}%
\bibitem [{\citenamefont {Vinci}\ \emph {et~al.}(2016)\citenamefont {Vinci},
  \citenamefont {Albash},\ and\ \citenamefont {Lidar}}]{vinci2016nested}%
  \BibitemOpen
  \bibfield  {author} {\bibinfo {author} {\bibfnamefont {W.}~\bibnamefont
  {Vinci}}, \bibinfo {author} {\bibfnamefont {T.}~\bibnamefont {Albash}}, \
  and\ \bibinfo {author} {\bibfnamefont {D.~A.}\ \bibnamefont {Lidar}},\
  }\href@noop {} {\bibfield  {journal} {\bibinfo  {journal} {npj Quantum
  Information}\ }\textbf {\bibinfo {volume} {2}},\ \bibinfo {pages} {16017}
  (\bibinfo {year} {2016})}\BibitemShut {NoStop}%
\bibitem [{Note1()}]{Note1}%
  \BibitemOpen
  \bibinfo {note} {The acronym RFQA has the dual meaning of random field
  quantum annealing and radio frequency quantum annealing.}\BibitemShut {Stop}%
\bibitem [{\citenamefont {Lucas}(2014)}]{lucas2014ising}%
  \BibitemOpen
  \bibfield  {author} {\bibinfo {author} {\bibfnamefont {A.}~\bibnamefont
  {Lucas}},\ }\href@noop {} {\bibfield  {journal} {\bibinfo  {journal}
  {Frontiers in Physics}\ }\textbf {\bibinfo {volume} {2}},\ \bibinfo {pages}
  {5} (\bibinfo {year} {2014})}\BibitemShut {NoStop}%
\bibitem [{\citenamefont {Farhi}\ \emph {et~al.}(2008)\citenamefont {Farhi},
  \citenamefont {Goldstone}, \citenamefont {Gutmann},\ and\ \citenamefont
  {Nagaj}}]{farhigoldstone2008}%
  \BibitemOpen
  \bibfield  {author} {\bibinfo {author} {\bibfnamefont {E.}~\bibnamefont
  {Farhi}}, \bibinfo {author} {\bibfnamefont {J.}~\bibnamefont {Goldstone}},
  \bibinfo {author} {\bibfnamefont {S.}~\bibnamefont {Gutmann}}, \ and\
  \bibinfo {author} {\bibfnamefont {D.}~\bibnamefont {Nagaj}},\ }\href@noop {}
  {\bibfield  {journal} {\bibinfo  {journal} {International Journal of Quantum
  Information}\ }\textbf {\bibinfo {volume} {6}},\ \bibinfo {pages} {503}
  (\bibinfo {year} {2008})}\BibitemShut {NoStop}%
\bibitem [{\citenamefont {Baldwin}\ \emph {et~al.}(2016)\citenamefont
  {Baldwin}, \citenamefont {Laumann}, \citenamefont {Pal},\ and\ \citenamefont
  {Scardicchio}}]{baldwinlaumann2016}%
  \BibitemOpen
  \bibfield  {author} {\bibinfo {author} {\bibfnamefont {C.}~\bibnamefont
  {Baldwin}}, \bibinfo {author} {\bibfnamefont {C.}~\bibnamefont {Laumann}},
  \bibinfo {author} {\bibfnamefont {A.}~\bibnamefont {Pal}}, \ and\ \bibinfo
  {author} {\bibfnamefont {A.}~\bibnamefont {Scardicchio}},\ }\href@noop {}
  {\bibfield  {journal} {\bibinfo  {journal} {Physical Review B}\ }\textbf
  {\bibinfo {volume} {93}},\ \bibinfo {pages} {024202} (\bibinfo {year}
  {2016})}\BibitemShut {NoStop}%
\bibitem [{\citenamefont {Baldwin}\ \emph {et~al.}(2017)\citenamefont
  {Baldwin}, \citenamefont {Laumann}, \citenamefont {Pal},\ and\ \citenamefont
  {Scardicchio}}]{baldwinlaumann2017}%
  \BibitemOpen
  \bibfield  {author} {\bibinfo {author} {\bibfnamefont {C.}~\bibnamefont
  {Baldwin}}, \bibinfo {author} {\bibfnamefont {C.}~\bibnamefont {Laumann}},
  \bibinfo {author} {\bibfnamefont {A.}~\bibnamefont {Pal}}, \ and\ \bibinfo
  {author} {\bibfnamefont {A.}~\bibnamefont {Scardicchio}},\ }\href@noop {}
  {\bibfield  {journal} {\bibinfo  {journal} {Physical Review Letters}\
  }\textbf {\bibinfo {volume} {118}},\ \bibinfo {pages} {127201} (\bibinfo
  {year} {2017})}\BibitemShut {NoStop}%
\bibitem [{\citenamefont {Baldwin}\ and\ \citenamefont
  {Laumann}(2018)}]{baldwin2018quantum}%
  \BibitemOpen
  \bibfield  {author} {\bibinfo {author} {\bibfnamefont {C.}~\bibnamefont
  {Baldwin}}\ and\ \bibinfo {author} {\bibfnamefont {C.}~\bibnamefont
  {Laumann}},\ }\href@noop {} {\bibfield  {journal} {\bibinfo  {journal}
  {Physical Review B}\ }\textbf {\bibinfo {volume} {97}},\ \bibinfo {pages}
  {224201} (\bibinfo {year} {2018})}\BibitemShut {NoStop}%
\bibitem [{\citenamefont {Faoro}\ \emph {et~al.}(2018)\citenamefont {Faoro},
  \citenamefont {Feigel'man},\ and\ \citenamefont {Ioffe}}]{faoro2018non}%
  \BibitemOpen
  \bibfield  {author} {\bibinfo {author} {\bibfnamefont {L.}~\bibnamefont
  {Faoro}}, \bibinfo {author} {\bibfnamefont {M.}~\bibnamefont {Feigel'man}}, \
  and\ \bibinfo {author} {\bibfnamefont {L.}~\bibnamefont {Ioffe}},\
  }\href@noop {} {\bibfield  {journal} {\bibinfo  {journal} {arXiv preprint
  arXiv:1812.06016}\ } (\bibinfo {year} {2018})}\BibitemShut {NoStop}%
\bibitem [{\citenamefont {Smelyanskiy}\ \emph {et~al.}(2019)\citenamefont
  {Smelyanskiy}, \citenamefont {Kechedzhi}, \citenamefont {Boixo},
  \citenamefont {Neven},\ and\ \citenamefont
  {Altshuler}}]{smelyanskiy2019intermittency}%
  \BibitemOpen
  \bibfield  {author} {\bibinfo {author} {\bibfnamefont {V.~N.}\ \bibnamefont
  {Smelyanskiy}}, \bibinfo {author} {\bibfnamefont {K.}~\bibnamefont
  {Kechedzhi}}, \bibinfo {author} {\bibfnamefont {S.}~\bibnamefont {Boixo}},
  \bibinfo {author} {\bibfnamefont {H.}~\bibnamefont {Neven}}, \ and\ \bibinfo
  {author} {\bibfnamefont {B.}~\bibnamefont {Altshuler}},\ }\href@noop {}
  {\bibfield  {journal} {\bibinfo  {journal} {arXiv preprint arXiv:1907.01609}\
  } (\bibinfo {year} {2019})}\BibitemShut {NoStop}%
\bibitem [{\citenamefont {Isakov}\ \emph {et~al.}(2016)\citenamefont {Isakov},
  \citenamefont {Mazzola}, \citenamefont {Smelyanskiy}, \citenamefont {Jiang},
  \citenamefont {Boixo}, \citenamefont {Neven},\ and\ \citenamefont
  {Troyer}}]{isakov2016understanding}%
  \BibitemOpen
  \bibfield  {author} {\bibinfo {author} {\bibfnamefont {S.~V.}\ \bibnamefont
  {Isakov}}, \bibinfo {author} {\bibfnamefont {G.}~\bibnamefont {Mazzola}},
  \bibinfo {author} {\bibfnamefont {V.~N.}\ \bibnamefont {Smelyanskiy}},
  \bibinfo {author} {\bibfnamefont {Z.}~\bibnamefont {Jiang}}, \bibinfo
  {author} {\bibfnamefont {S.}~\bibnamefont {Boixo}}, \bibinfo {author}
  {\bibfnamefont {H.}~\bibnamefont {Neven}}, \ and\ \bibinfo {author}
  {\bibfnamefont {M.}~\bibnamefont {Troyer}},\ }\href@noop {} {\bibfield
  {journal} {\bibinfo  {journal} {Physical review letters}\ }\textbf {\bibinfo
  {volume} {117}},\ \bibinfo {pages} {180402} (\bibinfo {year}
  {2016})}\BibitemShut {NoStop}%
\bibitem [{\citenamefont {Andriyash}\ and\ \citenamefont
  {Amin}(2017)}]{andriyash2017can}%
  \BibitemOpen
  \bibfield  {author} {\bibinfo {author} {\bibfnamefont {E.}~\bibnamefont
  {Andriyash}}\ and\ \bibinfo {author} {\bibfnamefont {M.~H.}\ \bibnamefont
  {Amin}},\ }\href@noop {} {\bibfield  {journal} {\bibinfo  {journal} {arXiv
  preprint arXiv:1703.09277}\ } (\bibinfo {year} {2017})}\BibitemShut {NoStop}%
\bibitem [{\citenamefont {Jiang}\ \emph
  {et~al.}(2017{\natexlab{b}})\citenamefont {Jiang}, \citenamefont
  {Smelyanskiy}, \citenamefont {Isakov}, \citenamefont {Boixo}, \citenamefont
  {Mazzola}, \citenamefont {Troyer},\ and\ \citenamefont
  {Neven}}]{jiang2017scaling}%
  \BibitemOpen
  \bibfield  {author} {\bibinfo {author} {\bibfnamefont {Z.}~\bibnamefont
  {Jiang}}, \bibinfo {author} {\bibfnamefont {V.~N.}\ \bibnamefont
  {Smelyanskiy}}, \bibinfo {author} {\bibfnamefont {S.~V.}\ \bibnamefont
  {Isakov}}, \bibinfo {author} {\bibfnamefont {S.}~\bibnamefont {Boixo}},
  \bibinfo {author} {\bibfnamefont {G.}~\bibnamefont {Mazzola}}, \bibinfo
  {author} {\bibfnamefont {M.}~\bibnamefont {Troyer}}, \ and\ \bibinfo {author}
  {\bibfnamefont {H.}~\bibnamefont {Neven}},\ }\href@noop {} {\bibfield
  {journal} {\bibinfo  {journal} {Physical Review A}\ }\textbf {\bibinfo
  {volume} {95}},\ \bibinfo {pages} {012322} (\bibinfo {year}
  {2017}{\natexlab{b}})}\BibitemShut {NoStop}%
\bibitem [{\citenamefont {Jiang}\ \emph
  {et~al.}(2017{\natexlab{c}})\citenamefont {Jiang}, \citenamefont
  {Smelyanskiy}, \citenamefont {Boixo},\ and\ \citenamefont
  {Neven}}]{jiang2017path}%
  \BibitemOpen
  \bibfield  {author} {\bibinfo {author} {\bibfnamefont {Z.}~\bibnamefont
  {Jiang}}, \bibinfo {author} {\bibfnamefont {V.~N.}\ \bibnamefont
  {Smelyanskiy}}, \bibinfo {author} {\bibfnamefont {S.}~\bibnamefont {Boixo}},
  \ and\ \bibinfo {author} {\bibfnamefont {H.}~\bibnamefont {Neven}},\
  }\href@noop {} {\bibfield  {journal} {\bibinfo  {journal} {Physical Review
  A}\ }\textbf {\bibinfo {volume} {96}},\ \bibinfo {pages} {042330} (\bibinfo
  {year} {2017}{\natexlab{c}})}\BibitemShut {NoStop}%
\bibitem [{\citenamefont {King}\ \emph
  {et~al.}(2019{\natexlab{a}})\citenamefont {King}, \citenamefont {Raymond},
  \citenamefont {Lanting}, \citenamefont {Isakov}, \citenamefont {Mohseni},
  \citenamefont {Poulin-Lamarre}, \citenamefont {Ejtemaee}, \citenamefont
  {Bernoudy}, \citenamefont {Ozfidan}, \citenamefont {Smirnov} \emph
  {et~al.}}]{king2019scaling}%
  \BibitemOpen
  \bibfield  {author} {\bibinfo {author} {\bibfnamefont {A.~D.}\ \bibnamefont
  {King}}, \bibinfo {author} {\bibfnamefont {J.}~\bibnamefont {Raymond}},
  \bibinfo {author} {\bibfnamefont {T.}~\bibnamefont {Lanting}}, \bibinfo
  {author} {\bibfnamefont {S.~V.}\ \bibnamefont {Isakov}}, \bibinfo {author}
  {\bibfnamefont {M.}~\bibnamefont {Mohseni}}, \bibinfo {author} {\bibfnamefont
  {G.}~\bibnamefont {Poulin-Lamarre}}, \bibinfo {author} {\bibfnamefont
  {S.}~\bibnamefont {Ejtemaee}}, \bibinfo {author} {\bibfnamefont
  {W.}~\bibnamefont {Bernoudy}}, \bibinfo {author} {\bibfnamefont
  {I.}~\bibnamefont {Ozfidan}}, \bibinfo {author} {\bibfnamefont {A.~Y.}\
  \bibnamefont {Smirnov}},  \emph {et~al.},\ }\href@noop {} {\bibfield
  {journal} {\bibinfo  {journal} {arXiv preprint arXiv:1911.03446}\ } (\bibinfo
  {year} {2019}{\natexlab{a}})}\BibitemShut {NoStop}%
\bibitem [{\citenamefont {Harris}\ \emph {et~al.}(2010)\citenamefont {Harris},
  \citenamefont {Johansson}, \citenamefont {Berkley}, \citenamefont {Johnson},
  \citenamefont {Lanting}, \citenamefont {Han}, \citenamefont {Bunyk},
  \citenamefont {Ladizinsky}, \citenamefont {Oh}, \citenamefont {Perminov}
  \emph {et~al.}}]{harris2010experimental}%
  \BibitemOpen
  \bibfield  {author} {\bibinfo {author} {\bibfnamefont {R.}~\bibnamefont
  {Harris}}, \bibinfo {author} {\bibfnamefont {J.}~\bibnamefont {Johansson}},
  \bibinfo {author} {\bibfnamefont {A.}~\bibnamefont {Berkley}}, \bibinfo
  {author} {\bibfnamefont {M.}~\bibnamefont {Johnson}}, \bibinfo {author}
  {\bibfnamefont {T.}~\bibnamefont {Lanting}}, \bibinfo {author} {\bibfnamefont
  {S.}~\bibnamefont {Han}}, \bibinfo {author} {\bibfnamefont {P.}~\bibnamefont
  {Bunyk}}, \bibinfo {author} {\bibfnamefont {E.}~\bibnamefont {Ladizinsky}},
  \bibinfo {author} {\bibfnamefont {T.}~\bibnamefont {Oh}}, \bibinfo {author}
  {\bibfnamefont {I.}~\bibnamefont {Perminov}},  \emph {et~al.},\ }\href@noop
  {} {\bibfield  {journal} {\bibinfo  {journal} {Physical Review B}\ }\textbf
  {\bibinfo {volume} {81}},\ \bibinfo {pages} {134510} (\bibinfo {year}
  {2010})}\BibitemShut {NoStop}%
\bibitem [{\citenamefont {Bylander}\ \emph {et~al.}(2011)\citenamefont
  {Bylander}, \citenamefont {Gustavsson}, \citenamefont {Yan}, \citenamefont
  {Yoshihara}, \citenamefont {Harrabi}, \citenamefont {Fitch}, \citenamefont
  {Cory}, \citenamefont {Nakamura}, \citenamefont {Tsai},\ and\ \citenamefont
  {Oliver}}]{bylandergustavsson2011}%
  \BibitemOpen
  \bibfield  {author} {\bibinfo {author} {\bibfnamefont {J.}~\bibnamefont
  {Bylander}}, \bibinfo {author} {\bibfnamefont {S.}~\bibnamefont
  {Gustavsson}}, \bibinfo {author} {\bibfnamefont {F.}~\bibnamefont {Yan}},
  \bibinfo {author} {\bibfnamefont {F.}~\bibnamefont {Yoshihara}}, \bibinfo
  {author} {\bibfnamefont {K.}~\bibnamefont {Harrabi}}, \bibinfo {author}
  {\bibfnamefont {G.}~\bibnamefont {Fitch}}, \bibinfo {author} {\bibfnamefont
  {D.~G.}\ \bibnamefont {Cory}}, \bibinfo {author} {\bibfnamefont
  {Y.}~\bibnamefont {Nakamura}}, \bibinfo {author} {\bibfnamefont {J.-S.}\
  \bibnamefont {Tsai}}, \ and\ \bibinfo {author} {\bibfnamefont {W.~D.}\
  \bibnamefont {Oliver}},\ }\href {\doibase 10.1038/nphys1994} {\bibfield
  {journal} {\bibinfo  {journal} {Nature Physics \textbf{7}, 565}\ } (\bibinfo
  {year} {2011}),\ 10.1038/nphys1994}\BibitemShut {NoStop}%
\bibitem [{\citenamefont {Yan}\ \emph {et~al.}(2013)\citenamefont {Yan},
  \citenamefont {Gustavsson}, \citenamefont {Bylander}, \citenamefont {Jin},
  \citenamefont {Yoshihara}, \citenamefont {Cory}, \citenamefont {Nakamura},
  \citenamefont {Orlando},\ and\ \citenamefont {Oliver}}]{yangustavsson2013}%
  \BibitemOpen
  \bibfield  {author} {\bibinfo {author} {\bibfnamefont {F.}~\bibnamefont
  {Yan}}, \bibinfo {author} {\bibfnamefont {S.}~\bibnamefont {Gustavsson}},
  \bibinfo {author} {\bibfnamefont {J.}~\bibnamefont {Bylander}}, \bibinfo
  {author} {\bibfnamefont {X.}~\bibnamefont {Jin}}, \bibinfo {author}
  {\bibfnamefont {F.}~\bibnamefont {Yoshihara}}, \bibinfo {author}
  {\bibfnamefont {D.~G.}\ \bibnamefont {Cory}}, \bibinfo {author}
  {\bibfnamefont {Y.}~\bibnamefont {Nakamura}}, \bibinfo {author}
  {\bibfnamefont {T.~P.}\ \bibnamefont {Orlando}}, \ and\ \bibinfo {author}
  {\bibfnamefont {W.~D.}\ \bibnamefont {Oliver}},\ }\href {\doibase
  10.1038/ncomms3337} {\bibfield  {journal} {\bibinfo  {journal} {Nature
  Communications 4, 2337}\ } (\bibinfo {year} {2013}),\
  10.1038/ncomms3337}\BibitemShut {NoStop}%
\bibitem [{\citenamefont {Yan}\ \emph {et~al.}(2016)\citenamefont {Yan},
  \citenamefont {Gustavsson}, \citenamefont {Kamal}, \citenamefont {Birenbaum},
  \citenamefont {Sears}, \citenamefont {Hover}, \citenamefont {Gudmundsen},
  \citenamefont {Yoder}, \citenamefont {Orlando}, \citenamefont {Clarke},
  \citenamefont {Kerman},\ and\ \citenamefont {Oliver}}]{yangustavsson2015}%
  \BibitemOpen
  \bibfield  {author} {\bibinfo {author} {\bibfnamefont {F.}~\bibnamefont
  {Yan}}, \bibinfo {author} {\bibfnamefont {S.}~\bibnamefont {Gustavsson}},
  \bibinfo {author} {\bibfnamefont {A.}~\bibnamefont {Kamal}}, \bibinfo
  {author} {\bibfnamefont {J.}~\bibnamefont {Birenbaum}}, \bibinfo {author}
  {\bibfnamefont {A.}~\bibnamefont {Sears}}, \bibinfo {author} {\bibfnamefont
  {D.}~\bibnamefont {Hover}}, \bibinfo {author} {\bibfnamefont
  {T.}~\bibnamefont {Gudmundsen}}, \bibinfo {author} {\bibfnamefont
  {J.}~\bibnamefont {Yoder}}, \bibinfo {author} {\bibfnamefont
  {T.}~\bibnamefont {Orlando}}, \bibinfo {author} {\bibfnamefont
  {J.}~\bibnamefont {Clarke}}, \bibinfo {author} {\bibfnamefont
  {A.}~\bibnamefont {Kerman}}, \ and\ \bibinfo {author} {\bibfnamefont
  {W.}~\bibnamefont {Oliver}},\ }\href {\doibase 10.1038/ncomms12964}
  {\bibfield  {journal} {\bibinfo  {journal} {Nature Communications 7, 12964}\
  } (\bibinfo {year} {2016}),\ 10.1038/ncomms12964}\BibitemShut {NoStop}%
\bibitem [{\citenamefont {Weber}\ \emph {et~al.}(2017)\citenamefont {Weber},
  \citenamefont {Samach}, \citenamefont {Hover}, \citenamefont {Gustavsson},
  \citenamefont {Kim}, \citenamefont {Melville}, \citenamefont {Rosenberg},
  \citenamefont {Sears}, \citenamefont {Yan}, \citenamefont {Yoder},
  \citenamefont {Oliver},\ and\ \citenamefont {Kerman}}]{webersamach2017}%
  \BibitemOpen
  \bibfield  {author} {\bibinfo {author} {\bibfnamefont {S.~J.}\ \bibnamefont
  {Weber}}, \bibinfo {author} {\bibfnamefont {G.~O.}\ \bibnamefont {Samach}},
  \bibinfo {author} {\bibfnamefont {D.}~\bibnamefont {Hover}}, \bibinfo
  {author} {\bibfnamefont {S.}~\bibnamefont {Gustavsson}}, \bibinfo {author}
  {\bibfnamefont {D.~K.}\ \bibnamefont {Kim}}, \bibinfo {author} {\bibfnamefont
  {A.}~\bibnamefont {Melville}}, \bibinfo {author} {\bibfnamefont
  {D.}~\bibnamefont {Rosenberg}}, \bibinfo {author} {\bibfnamefont {A.~P.}\
  \bibnamefont {Sears}}, \bibinfo {author} {\bibfnamefont {F.}~\bibnamefont
  {Yan}}, \bibinfo {author} {\bibfnamefont {J.~L.}\ \bibnamefont {Yoder}},
  \bibinfo {author} {\bibfnamefont {W.~D.}\ \bibnamefont {Oliver}}, \ and\
  \bibinfo {author} {\bibfnamefont {A.~J.}\ \bibnamefont {Kerman}},\ }\href
  {\doibase 10.1103/PhysRevApplied.8.014004} {\bibfield  {journal} {\bibinfo
  {journal} {Phys. Rev. Applied}\ }\textbf {\bibinfo {volume} {8}},\ \bibinfo
  {pages} {014004} (\bibinfo {year} {2017})}\BibitemShut {NoStop}%
\bibitem [{\citenamefont {Prokof'ev}\ and\ \citenamefont
  {Stamp}(2000)}]{prokof2000theory}%
  \BibitemOpen
  \bibfield  {author} {\bibinfo {author} {\bibfnamefont {N.}~\bibnamefont
  {Prokof'ev}}\ and\ \bibinfo {author} {\bibfnamefont {P.}~\bibnamefont
  {Stamp}},\ }\href@noop {} {\bibfield  {journal} {\bibinfo  {journal} {Reports
  on Progress in Physics}\ }\textbf {\bibinfo {volume} {63}},\ \bibinfo {pages}
  {669} (\bibinfo {year} {2000})}\BibitemShut {NoStop}%
\bibitem [{Note2()}]{Note2}%
  \BibitemOpen
  \bibinfo {note} {Based on numerous conversations with other researchers in
  the field of quantum optimization, we feel that a loose version of this
  conjecture is tacitly assumed to varying degrees in the thinking of most
  scholars working on these problems. This is particularly true in studies
  which consider to how a many-body quantum spin glass interacts with a bath.
  However, we have been unable to find it stated in print, so we do so
  here.}\BibitemShut {Stop}%
\bibitem [{\citenamefont {Pietracaprina}\ \emph {et~al.}(2016)\citenamefont
  {Pietracaprina}, \citenamefont {Ros},\ and\ \citenamefont
  {Scardicchio}}]{pietracaprina2016forward}%
  \BibitemOpen
  \bibfield  {author} {\bibinfo {author} {\bibfnamefont {F.}~\bibnamefont
  {Pietracaprina}}, \bibinfo {author} {\bibfnamefont {V.}~\bibnamefont {Ros}},
  \ and\ \bibinfo {author} {\bibfnamefont {A.}~\bibnamefont {Scardicchio}},\
  }\href@noop {} {\bibfield  {journal} {\bibinfo  {journal} {Physical Review
  B}\ }\textbf {\bibinfo {volume} {93}},\ \bibinfo {pages} {054201} (\bibinfo
  {year} {2016})}\BibitemShut {NoStop}%
\bibitem [{\citenamefont {Scardicchio}\ and\ \citenamefont
  {Thiery}(2017)}]{scardicchio2017perturbation}%
  \BibitemOpen
  \bibfield  {author} {\bibinfo {author} {\bibfnamefont {A.}~\bibnamefont
  {Scardicchio}}\ and\ \bibinfo {author} {\bibfnamefont {T.}~\bibnamefont
  {Thiery}},\ }\href@noop {} {\bibfield  {journal} {\bibinfo  {journal} {arXiv
  preprint arXiv:1710.01234}\ } (\bibinfo {year} {2017})}\BibitemShut {NoStop}%
\bibitem [{Note3()}]{Note3}%
  \BibitemOpen
  \bibinfo {note} {In cases where $\Delta _{local}$ has inverse polynomial
  scaling in $N$, we expect that the bare drive amplitudes and/or frequencies
  $f_i$ would be similarly reduced if we were to remain in the vicinity of the
  phase transition, ensuring constant scaling; see our derivation of the
  quantum speedup in the Grover problem for an example of this
  effect.}\BibitemShut {Stop}%
\bibitem [{\citenamefont {Kadowaki}\ and\ \citenamefont
  {Ohzeki}(2019)}]{kadowaki2019experimental}%
  \BibitemOpen
  \bibfield  {author} {\bibinfo {author} {\bibfnamefont {T.}~\bibnamefont
  {Kadowaki}}\ and\ \bibinfo {author} {\bibfnamefont {M.}~\bibnamefont
  {Ohzeki}},\ }\href@noop {} {\bibfield  {journal} {\bibinfo  {journal}
  {Journal of the Physical Society of Japan}\ }\textbf {\bibinfo {volume}
  {88}},\ \bibinfo {pages} {061008} (\bibinfo {year} {2019})}\BibitemShut
  {NoStop}%
\bibitem [{\citenamefont {Marshall}\ \emph {et~al.}(2019)\citenamefont
  {Marshall}, \citenamefont {Venturelli}, \citenamefont {Hen},\ and\
  \citenamefont {Rieffel}}]{marshall2019power}%
  \BibitemOpen
  \bibfield  {author} {\bibinfo {author} {\bibfnamefont {J.}~\bibnamefont
  {Marshall}}, \bibinfo {author} {\bibfnamefont {D.}~\bibnamefont
  {Venturelli}}, \bibinfo {author} {\bibfnamefont {I.}~\bibnamefont {Hen}}, \
  and\ \bibinfo {author} {\bibfnamefont {E.~G.}\ \bibnamefont {Rieffel}},\
  }\href@noop {} {\bibfield  {journal} {\bibinfo  {journal} {Physical Review
  Applied}\ }\textbf {\bibinfo {volume} {11}},\ \bibinfo {pages} {044083}
  (\bibinfo {year} {2019})}\BibitemShut {NoStop}%
\bibitem [{\citenamefont {King}\ \emph
  {et~al.}(2019{\natexlab{b}})\citenamefont {King}, \citenamefont {Mohseni},
  \citenamefont {Bernoudy}, \citenamefont {Fr{\'e}chette}, \citenamefont
  {Sadeghi}, \citenamefont {Isakov}, \citenamefont {Neven},\ and\ \citenamefont
  {Amin}}]{king2019quantum}%
  \BibitemOpen
  \bibfield  {author} {\bibinfo {author} {\bibfnamefont {J.}~\bibnamefont
  {King}}, \bibinfo {author} {\bibfnamefont {M.}~\bibnamefont {Mohseni}},
  \bibinfo {author} {\bibfnamefont {W.}~\bibnamefont {Bernoudy}}, \bibinfo
  {author} {\bibfnamefont {A.}~\bibnamefont {Fr{\'e}chette}}, \bibinfo {author}
  {\bibfnamefont {H.}~\bibnamefont {Sadeghi}}, \bibinfo {author} {\bibfnamefont
  {S.~V.}\ \bibnamefont {Isakov}}, \bibinfo {author} {\bibfnamefont
  {H.}~\bibnamefont {Neven}}, \ and\ \bibinfo {author} {\bibfnamefont {M.~H.}\
  \bibnamefont {Amin}},\ }\href@noop {} {\bibfield  {journal} {\bibinfo
  {journal} {arXiv preprint arXiv:1907.00707}\ } (\bibinfo {year}
  {2019}{\natexlab{b}})}\BibitemShut {NoStop}%
\bibitem [{Note4()}]{Note4}%
  \BibitemOpen
  \bibinfo {note} {One could also oscillate the magnitudes of the transverse
  fields (RFQA-M), or the magnitudes and/or directions of transverse coupling
  elements (RFQA-C); more exotic variants of RFQA could be implemented using
  novel qubit designs \cite {kerman2019superconducting} or through simulation
  of RFQA evolution in a digital quantum computer.}\BibitemShut {Stop}%
\bibitem [{\citenamefont {Atia}\ and\ \citenamefont
  {Aharonov}(2019)}]{atia2019high}%
  \BibitemOpen
  \bibfield  {author} {\bibinfo {author} {\bibfnamefont {Y.}~\bibnamefont
  {Atia}}\ and\ \bibinfo {author} {\bibfnamefont {D.}~\bibnamefont
  {Aharonov}},\ }\href@noop {} {\bibfield  {journal} {\bibinfo  {journal}
  {arXiv preprint arXiv:1906.02581}\ } (\bibinfo {year} {2019})}\BibitemShut
  {NoStop}%
\bibitem [{Note5()}]{Note5}%
  \BibitemOpen
  \bibinfo {note} {Of course, for problems with more structure, the optimal
  $\protect \mathaccentV {bar}016{\alpha }$ may depend on the problem class or
  even on the details of a given instance. However, our argument is generic
  enough to provide a good starting point for a broad range of
  cases.}\BibitemShut {Stop}%
\bibitem [{\citenamefont {D?Alessio}\ and\ \citenamefont
  {Rigol}(2014)}]{d2014long}%
  \BibitemOpen
  \bibfield  {author} {\bibinfo {author} {\bibfnamefont {L.}~\bibnamefont
  {D?Alessio}}\ and\ \bibinfo {author} {\bibfnamefont {M.}~\bibnamefont
  {Rigol}},\ }\href@noop {} {\bibfield  {journal} {\bibinfo  {journal}
  {Physical Review X}\ }\textbf {\bibinfo {volume} {4}},\ \bibinfo {pages}
  {041048} (\bibinfo {year} {2014})}\BibitemShut {NoStop}%
\bibitem [{\citenamefont {Haldar}\ \emph
  {et~al.}(2018{\natexlab{a}})\citenamefont {Haldar}, \citenamefont
  {Moessner},\ and\ \citenamefont {Das}}]{PhysRevB.97.245122}%
  \BibitemOpen
  \bibfield  {author} {\bibinfo {author} {\bibfnamefont {A.}~\bibnamefont
  {Haldar}}, \bibinfo {author} {\bibfnamefont {R.}~\bibnamefont {Moessner}}, \
  and\ \bibinfo {author} {\bibfnamefont {A.}~\bibnamefont {Das}},\ }\href
  {\doibase 10.1103/PhysRevB.97.245122} {\bibfield  {journal} {\bibinfo
  {journal} {Phys. Rev. B}\ }\textbf {\bibinfo {volume} {97}},\ \bibinfo
  {pages} {245122} (\bibinfo {year} {2018}{\natexlab{a}})}\BibitemShut
  {NoStop}%
\bibitem [{\citenamefont {Nandkishore}\ and\ \citenamefont
  {Huse}(2015{\natexlab{a}})}]{nandkishore2015many}%
  \BibitemOpen
  \bibfield  {author} {\bibinfo {author} {\bibfnamefont {R.}~\bibnamefont
  {Nandkishore}}\ and\ \bibinfo {author} {\bibfnamefont {D.~A.}\ \bibnamefont
  {Huse}},\ }\href@noop {} {\bibfield  {journal} {\bibinfo  {journal} {Annu.
  Rev. Condens. Matter Phys.}\ }\textbf {\bibinfo {volume} {6}},\ \bibinfo
  {pages} {15} (\bibinfo {year} {2015}{\natexlab{a}})}\BibitemShut {NoStop}%
\bibitem [{\citenamefont {Bachmann}\ \emph {et~al.}(2017)\citenamefont
  {Bachmann}, \citenamefont {De~Roeck},\ and\ \citenamefont
  {Fraas}}]{PhysRevLett.119.060201}%
  \BibitemOpen
  \bibfield  {author} {\bibinfo {author} {\bibfnamefont {S.}~\bibnamefont
  {Bachmann}}, \bibinfo {author} {\bibfnamefont {W.}~\bibnamefont {De~Roeck}},
  \ and\ \bibinfo {author} {\bibfnamefont {M.}~\bibnamefont {Fraas}},\ }\href
  {\doibase 10.1103/PhysRevLett.119.060201} {\bibfield  {journal} {\bibinfo
  {journal} {Phys. Rev. Lett.}\ }\textbf {\bibinfo {volume} {119}},\ \bibinfo
  {pages} {060201} (\bibinfo {year} {2017})}\BibitemShut {NoStop}%
\bibitem [{Note6()}]{Note6}%
  \BibitemOpen
  \bibinfo {note} {This is not the case for proposals to engineer quantum
  annealing or similar continuous-time analog quantum optimization protocols in
  more general, driven quantum information systems, such as trapped ions or
  transmon qubits, where the states are generated and manipulated through
  oscillating fields. In general the ``bath'' in those systems can be well
  approximated by infinite temperature\cite {kapit2017review} and would pose
  severe challenges at large $N$ unless the system-bath coupling rates are
  extremely small.}\BibitemShut {Stop}%
\bibitem [{\citenamefont {Kechedzhi}\ \emph {et~al.}(2018)\citenamefont
  {Kechedzhi}, \citenamefont {Smelyanskiy}, \citenamefont {McClean},
  \citenamefont {Denchev}, \citenamefont {Mohseni}, \citenamefont {Isakov},
  \citenamefont {Boixo}, \citenamefont {Altshuler},\ and\ \citenamefont
  {Neven}}]{kechedzhi2018efficient}%
  \BibitemOpen
  \bibfield  {author} {\bibinfo {author} {\bibfnamefont {K.}~\bibnamefont
  {Kechedzhi}}, \bibinfo {author} {\bibfnamefont {V.}~\bibnamefont
  {Smelyanskiy}}, \bibinfo {author} {\bibfnamefont {J.~R.}\ \bibnamefont
  {McClean}}, \bibinfo {author} {\bibfnamefont {V.~S.}\ \bibnamefont
  {Denchev}}, \bibinfo {author} {\bibfnamefont {M.}~\bibnamefont {Mohseni}},
  \bibinfo {author} {\bibfnamefont {S.}~\bibnamefont {Isakov}}, \bibinfo
  {author} {\bibfnamefont {S.}~\bibnamefont {Boixo}}, \bibinfo {author}
  {\bibfnamefont {B.}~\bibnamefont {Altshuler}}, \ and\ \bibinfo {author}
  {\bibfnamefont {H.}~\bibnamefont {Neven}},\ }\href@noop {} {\bibfield
  {journal} {\bibinfo  {journal} {arXiv preprint arXiv:1807.04792}\ } (\bibinfo
  {year} {2018})}\BibitemShut {NoStop}%
\bibitem [{\citenamefont {Smelyanskiy}\ \emph {et~al.}(2018)\citenamefont
  {Smelyanskiy}, \citenamefont {Kechedzhi}, \citenamefont {Boixo},
  \citenamefont {Isakov}, \citenamefont {Neven},\ and\ \citenamefont
  {Altshuler}}]{smelyanskiy2018non}%
  \BibitemOpen
  \bibfield  {author} {\bibinfo {author} {\bibfnamefont {V.~N.}\ \bibnamefont
  {Smelyanskiy}}, \bibinfo {author} {\bibfnamefont {K.}~\bibnamefont
  {Kechedzhi}}, \bibinfo {author} {\bibfnamefont {S.}~\bibnamefont {Boixo}},
  \bibinfo {author} {\bibfnamefont {S.~V.}\ \bibnamefont {Isakov}}, \bibinfo
  {author} {\bibfnamefont {H.}~\bibnamefont {Neven}}, \ and\ \bibinfo {author}
  {\bibfnamefont {B.}~\bibnamefont {Altshuler}},\ }\href@noop {} {\bibfield
  {journal} {\bibinfo  {journal} {arXiv preprint arXiv:1802.09542}\ } (\bibinfo
  {year} {2018})}\BibitemShut {NoStop}%
\bibitem [{Note7()}]{Note7}%
  \BibitemOpen
  \bibinfo {note} {As a side note, it is worth comparing these estimates to the
  real experimental parameters observed in the d-Wave quantum annealers \cite
  {dwave2019whitepaper}, the only commercially available quantum hardware. The
  maximum transverse field strength $\Delta / h$ is around 5 GHz; we shall use
  2.5 GHz in our comparison since we are considering dynamics near the phase
  transition point. The Ramsey $T_2$ of these qubits varies somewhat from one
  report to the next, with a typical value of around 15 ns. In our simulations,
  at $N=10$ the transverse field strength is set to 0.05, so equating that with
  2.5 GHz results in a noise $T_{2R}$ of 750. In other words, our strong noise
  simulations ($T_{2R} \simeq 25$) correspond to an isolated coherence which is
  around a factor of thirty times \protect \textit {lower} than the d-Wave flux
  qubits.}\BibitemShut {Stop}%
\bibitem [{\citenamefont {Jaschke}\ \emph {et~al.}(2019)\citenamefont
  {Jaschke}, \citenamefont {Carr},\ and\ \citenamefont
  {de~Vega}}]{jaschke2019thermalization}%
  \BibitemOpen
  \bibfield  {author} {\bibinfo {author} {\bibfnamefont {D.}~\bibnamefont
  {Jaschke}}, \bibinfo {author} {\bibfnamefont {L.~D.}\ \bibnamefont {Carr}}, \
  and\ \bibinfo {author} {\bibfnamefont {I.}~\bibnamefont {de~Vega}},\
  }\href@noop {} {\bibfield  {journal} {\bibinfo  {journal} {Quantum Science
  and Technology}\ }\textbf {\bibinfo {volume} {4}},\ \bibinfo {pages} {034002}
  (\bibinfo {year} {2019})}\BibitemShut {NoStop}%
\bibitem [{\citenamefont {Albash}\ \emph {et~al.}(2017)\citenamefont {Albash},
  \citenamefont {Martin-Mayor},\ and\ \citenamefont
  {Hen}}]{albash2017temperature}%
  \BibitemOpen
  \bibfield  {author} {\bibinfo {author} {\bibfnamefont {T.}~\bibnamefont
  {Albash}}, \bibinfo {author} {\bibfnamefont {V.}~\bibnamefont
  {Martin-Mayor}}, \ and\ \bibinfo {author} {\bibfnamefont {I.}~\bibnamefont
  {Hen}},\ }\href@noop {} {\bibfield  {journal} {\bibinfo  {journal} {Physical
  review letters}\ }\textbf {\bibinfo {volume} {119}},\ \bibinfo {pages}
  {110502} (\bibinfo {year} {2017})}\BibitemShut {NoStop}%
\bibitem [{\citenamefont {Gardiner}\ and\ \citenamefont
  {Zoller}(2004)}]{gardinerzoller}%
  \BibitemOpen
  \bibfield  {author} {\bibinfo {author} {\bibfnamefont {C.}~\bibnamefont
  {Gardiner}}\ and\ \bibinfo {author} {\bibfnamefont {P.}~\bibnamefont
  {Zoller}},\ }\href@noop {} {\emph {\bibinfo {title} {Quantum Noise: A
  Handbook of Markovian and Non-Markovian Quantum Stochastic Methods with
  Applications to Quantum Optics}}}\ (\bibinfo  {publisher} {Springer},\
  \bibinfo {year} {2004})\BibitemShut {NoStop}%
\bibitem [{\citenamefont {Nandkishore}\ and\ \citenamefont
  {Huse}(2015{\natexlab{b}})}]{RNDAH2015}%
  \BibitemOpen
  \bibfield  {author} {\bibinfo {author} {\bibfnamefont {R.}~\bibnamefont
  {Nandkishore}}\ and\ \bibinfo {author} {\bibfnamefont {D.~A.}\ \bibnamefont
  {Huse}},\ }\href {\doibase 10.1146/annurev-conmatphys-031214-014726}
  {\bibfield  {journal} {\bibinfo  {journal} {Annual Review of Condensed Matter
  Physics}\ }\textbf {\bibinfo {volume} {6}},\ \bibinfo {pages} {15} (\bibinfo
  {year} {2015}{\natexlab{b}})},\ \Eprint
  {http://arxiv.org/abs/https://doi.org/10.1146/annurev-conmatphys-031214-014726}
  {https://doi.org/10.1146/annurev-conmatphys-031214-014726} \BibitemShut
  {NoStop}%
\bibitem [{\citenamefont {Kaufman}\ \emph {et~al.}(2016)\citenamefont
  {Kaufman}, \citenamefont {Tai}, \citenamefont {Lukin}, \citenamefont
  {Rispoli}, \citenamefont {Schittko}, \citenamefont {Preiss},\ and\
  \citenamefont {Greiner}}]{Kaufman794}%
  \BibitemOpen
  \bibfield  {author} {\bibinfo {author} {\bibfnamefont {A.~M.}\ \bibnamefont
  {Kaufman}}, \bibinfo {author} {\bibfnamefont {M.~E.}\ \bibnamefont {Tai}},
  \bibinfo {author} {\bibfnamefont {A.}~\bibnamefont {Lukin}}, \bibinfo
  {author} {\bibfnamefont {M.}~\bibnamefont {Rispoli}}, \bibinfo {author}
  {\bibfnamefont {R.}~\bibnamefont {Schittko}}, \bibinfo {author}
  {\bibfnamefont {P.~M.}\ \bibnamefont {Preiss}}, \ and\ \bibinfo {author}
  {\bibfnamefont {M.}~\bibnamefont {Greiner}},\ }\href {\doibase
  10.1126/science.aaf6725} {\bibfield  {journal} {\bibinfo  {journal}
  {Science}\ }\textbf {\bibinfo {volume} {353}},\ \bibinfo {pages} {794}
  (\bibinfo {year} {2016})},\ \Eprint
  {http://arxiv.org/abs/https://science.sciencemag.org/content/353/6301/794.full.pdf}
  {https://science.sciencemag.org/content/353/6301/794.full.pdf} \BibitemShut
  {NoStop}%
\bibitem [{\citenamefont {D'Alessio}\ and\ \citenamefont
  {Rigol}(2014)}]{Rigol2014}%
  \BibitemOpen
  \bibfield  {author} {\bibinfo {author} {\bibfnamefont {L.}~\bibnamefont
  {D'Alessio}}\ and\ \bibinfo {author} {\bibfnamefont {M.}~\bibnamefont
  {Rigol}},\ }\href {\doibase 10.1103/PhysRevX.4.041048} {\bibfield  {journal}
  {\bibinfo  {journal} {Phys. Rev. X}\ }\textbf {\bibinfo {volume} {4}},\
  \bibinfo {pages} {041048} (\bibinfo {year} {2014})}\BibitemShut {NoStop}%
\bibitem [{\citenamefont {Haldar}\ \emph
  {et~al.}(2018{\natexlab{b}})\citenamefont {Haldar}, \citenamefont
  {Moessner},\ and\ \citenamefont {Das}}]{RM2018}%
  \BibitemOpen
  \bibfield  {author} {\bibinfo {author} {\bibfnamefont {A.}~\bibnamefont
  {Haldar}}, \bibinfo {author} {\bibfnamefont {R.}~\bibnamefont {Moessner}}, \
  and\ \bibinfo {author} {\bibfnamefont {A.}~\bibnamefont {Das}},\ }\href
  {\doibase 10.1103/PhysRevB.97.245122} {\bibfield  {journal} {\bibinfo
  {journal} {Phys. Rev. B}\ }\textbf {\bibinfo {volume} {97}},\ \bibinfo
  {pages} {245122} (\bibinfo {year} {2018}{\natexlab{b}})}\BibitemShut
  {NoStop}%
\bibitem [{\citenamefont {Mallayya}\ and\ \citenamefont
  {Rigol}(2019)}]{Rigol2019}%
  \BibitemOpen
  \bibfield  {author} {\bibinfo {author} {\bibfnamefont {K.}~\bibnamefont
  {Mallayya}}\ and\ \bibinfo {author} {\bibfnamefont {M.}~\bibnamefont
  {Rigol}},\ }\href {\doibase 10.1103/PhysRevLett.123.240603} {\bibfield
  {journal} {\bibinfo  {journal} {Phys. Rev. Lett.}\ }\textbf {\bibinfo
  {volume} {123}},\ \bibinfo {pages} {240603} (\bibinfo {year}
  {2019})}\BibitemShut {NoStop}%
\bibitem [{\citenamefont {Sahni}\ and\ \citenamefont
  {Gonzalez}(1976)}]{sahni1976p}%
  \BibitemOpen
  \bibfield  {author} {\bibinfo {author} {\bibfnamefont {S.}~\bibnamefont
  {Sahni}}\ and\ \bibinfo {author} {\bibfnamefont {T.}~\bibnamefont
  {Gonzalez}},\ }\href@noop {} {\bibfield  {journal} {\bibinfo  {journal}
  {Journal of the ACM (JACM)}\ }\textbf {\bibinfo {volume} {23}},\ \bibinfo
  {pages} {555} (\bibinfo {year} {1976})}\BibitemShut {NoStop}%
\bibitem [{\citenamefont {Farhi}\ \emph {et~al.}(2010)\citenamefont {Farhi},
  \citenamefont {Goldstone}, \citenamefont {Gosset}, \citenamefont {Gutmann},\
  and\ \citenamefont {Shor}}]{farhi2010unstructured}%
  \BibitemOpen
  \bibfield  {author} {\bibinfo {author} {\bibfnamefont {E.}~\bibnamefont
  {Farhi}}, \bibinfo {author} {\bibfnamefont {J.}~\bibnamefont {Goldstone}},
  \bibinfo {author} {\bibfnamefont {D.}~\bibnamefont {Gosset}}, \bibinfo
  {author} {\bibfnamefont {S.}~\bibnamefont {Gutmann}}, \ and\ \bibinfo
  {author} {\bibfnamefont {P.}~\bibnamefont {Shor}},\ }\href@noop {} {\bibfield
   {journal} {\bibinfo  {journal} {arXiv preprint arXiv:1010.0009}\ } (\bibinfo
  {year} {2010})}\BibitemShut {NoStop}%
\bibitem [{Note8()}]{Note8}%
  \BibitemOpen
  \bibinfo {note} {We expect that the speed limit in this system arises from
  MSCALE. To exceed the $2^{N/2}$ bound the system-bath coupling likely must
  increase exponentially in system size, which would eventually break down the
  perturbative assumptions required to separate the system from its bath, and
  in turn freeze evolution through continuous measurement and/or exponentially
  dilute the probability of finding the solution state as an eigenstate of the
  combined system, depending on the type of coupling.}\BibitemShut {Stop}%
\bibitem [{\citenamefont {Preskill}(2011)}]{preskill2011}%
  \BibitemOpen
  \bibfield  {author} {\bibinfo {author} {\bibfnamefont {J.}~\bibnamefont
  {Preskill}},\ }\href@noop {} {\bibfield  {journal} {\bibinfo  {journal}
  {\textit{The Theory of the Quantum World}, (eds Gross, D., Henneaux, M. and
  Sevrin, A.) 63, 80}\ } (\bibinfo {year} {2011})}\BibitemShut {NoStop}%
\bibitem [{Note9()}]{Note9}%
  \BibitemOpen
  \bibinfo {note} {On physical grounds we expect that short runtimes $t/t_f\lll
  1$ should help further suppress the influence of other environmental
  effects.}\BibitemShut {Stop}%
\bibitem [{\citenamefont {Amin}\ \emph {et~al.}(2008)\citenamefont {Amin},
  \citenamefont {Love},\ and\ \citenamefont {Truncik}}]{aminlove2008}%
  \BibitemOpen
  \bibfield  {author} {\bibinfo {author} {\bibfnamefont {M.~H.~S.}\
  \bibnamefont {Amin}}, \bibinfo {author} {\bibfnamefont {P.~J.}\ \bibnamefont
  {Love}}, \ and\ \bibinfo {author} {\bibfnamefont {C.~J.~S.}\ \bibnamefont
  {Truncik}},\ }\href {\doibase 10.1103/PhysRevLett.100.060503} {\bibfield
  {journal} {\bibinfo  {journal} {Phys. Rev. Lett. \textbf{100}, 060503}\ }
  (\bibinfo {year} {2008}),\ 10.1103/PhysRevLett.100.060503}\BibitemShut
  {NoStop}%
\bibitem [{\citenamefont {Wild}\ \emph {et~al.}(2016)\citenamefont {Wild},
  \citenamefont {Gopalakrishnan}, \citenamefont {Knap}, \citenamefont {Yao},\
  and\ \citenamefont {Lukin}}]{wild2016adiabatic}%
  \BibitemOpen
  \bibfield  {author} {\bibinfo {author} {\bibfnamefont {D.~S.}\ \bibnamefont
  {Wild}}, \bibinfo {author} {\bibfnamefont {S.}~\bibnamefont
  {Gopalakrishnan}}, \bibinfo {author} {\bibfnamefont {M.}~\bibnamefont
  {Knap}}, \bibinfo {author} {\bibfnamefont {N.~Y.}\ \bibnamefont {Yao}}, \
  and\ \bibinfo {author} {\bibfnamefont {M.~D.}\ \bibnamefont {Lukin}},\
  }\href@noop {} {\bibfield  {journal} {\bibinfo  {journal} {Physical review
  letters}\ }\textbf {\bibinfo {volume} {117}},\ \bibinfo {pages} {150501}
  (\bibinfo {year} {2016})}\BibitemShut {NoStop}%
\bibitem [{\citenamefont {Yang}\ \emph {et~al.}(2017)\citenamefont {Yang},
  \citenamefont {Rahmani}, \citenamefont {Shabani}, \citenamefont {Neven},\
  and\ \citenamefont {Chamon}}]{yang2017optimizing}%
  \BibitemOpen
  \bibfield  {author} {\bibinfo {author} {\bibfnamefont {Z.-C.}\ \bibnamefont
  {Yang}}, \bibinfo {author} {\bibfnamefont {A.}~\bibnamefont {Rahmani}},
  \bibinfo {author} {\bibfnamefont {A.}~\bibnamefont {Shabani}}, \bibinfo
  {author} {\bibfnamefont {H.}~\bibnamefont {Neven}}, \ and\ \bibinfo {author}
  {\bibfnamefont {C.}~\bibnamefont {Chamon}},\ }\href@noop {} {\bibfield
  {journal} {\bibinfo  {journal} {Physical Review X}\ }\textbf {\bibinfo
  {volume} {7}},\ \bibinfo {pages} {021027} (\bibinfo {year}
  {2017})}\BibitemShut {NoStop}%
\bibitem [{\citenamefont {Hen}(2019)}]{hen2019quantum}%
  \BibitemOpen
  \bibfield  {author} {\bibinfo {author} {\bibfnamefont {I.}~\bibnamefont
  {Hen}},\ }\href@noop {} {\bibfield  {journal} {\bibinfo  {journal} {Quantum
  Information Processing}\ }\textbf {\bibinfo {volume} {18}},\ \bibinfo {pages}
  {162} (\bibinfo {year} {2019})}\BibitemShut {NoStop}%
\bibitem [{\citenamefont {Altshuler}\ \emph {et~al.}(2010)\citenamefont
  {Altshuler}, \citenamefont {Krovi},\ and\ \citenamefont
  {Roland}}]{altshulerkrovi2010}%
  \BibitemOpen
  \bibfield  {author} {\bibinfo {author} {\bibfnamefont {B.}~\bibnamefont
  {Altshuler}}, \bibinfo {author} {\bibfnamefont {H.}~\bibnamefont {Krovi}}, \
  and\ \bibinfo {author} {\bibfnamefont {J.}~\bibnamefont {Roland}},\
  }\href@noop {} {\bibfield  {journal} {\bibinfo  {journal} {Proceedings of the
  National Academy of Sciences}\ }\textbf {\bibinfo {volume} {107}},\ \bibinfo
  {pages} {12446} (\bibinfo {year} {2010})}\BibitemShut {NoStop}%
\bibitem [{\citenamefont {Knysh}(2016)}]{knysh2016}%
  \BibitemOpen
  \bibfield  {author} {\bibinfo {author} {\bibfnamefont {S.}~\bibnamefont
  {Knysh}},\ }\href@noop {} {\bibfield  {journal} {\bibinfo  {journal} {Nature
  communications}\ }\textbf {\bibinfo {volume} {7}} (\bibinfo {year}
  {2016})}\BibitemShut {NoStop}%
\bibitem [{\citenamefont {Laumann}\ \emph {et~al.}(2012)\citenamefont
  {Laumann}, \citenamefont {Moessner}, \citenamefont {Scardicchio},\ and\
  \citenamefont {Sondhi}}]{laumannmoessner2012}%
  \BibitemOpen
  \bibfield  {author} {\bibinfo {author} {\bibfnamefont {C.}~\bibnamefont
  {Laumann}}, \bibinfo {author} {\bibfnamefont {R.}~\bibnamefont {Moessner}},
  \bibinfo {author} {\bibfnamefont {A.}~\bibnamefont {Scardicchio}}, \ and\
  \bibinfo {author} {\bibfnamefont {S.~L.}\ \bibnamefont {Sondhi}},\ }\href
  {\doibase 10.1103/PhysRevLett.109.030502} {\bibfield  {journal} {\bibinfo
  {journal} {Phys. Rev. Lett. \textbf{109}, 030502}\ } (\bibinfo {year}
  {2012}),\ 10.1103/PhysRevLett.109.030502}\BibitemShut {NoStop}%
\bibitem [{Note10()}]{Note10}%
  \BibitemOpen
  \bibinfo {note} {It is important to distinguish changes in a given pair of
  ground states' \protect \textit {configuration} from their \protect \textit
  {energetic heirarchy} in parameter space. Near a first order phase
  transition, where the gap is exponentially small, small perturbations can
  change the energetic hierarchy of competing states. However, if the system is
  in either state, its qualitative character will not change, and one would
  need to wait an exponentially long time to detect the mixing. The tunneling
  matrix element between the two states, which is set by their structure and
  the many-body Hamiltonian, should not be exponentially sensitive
  (percentage-wise, since it is typically exponentially small to begin with at
  first order transitions) to changes in these local energies.}\BibitemShut
  {Stop}%
\bibitem [{Note11()}]{Note11}%
  \BibitemOpen
  \bibinfo {note} {The degree to which this self-interference effect will occur
  in more realistic problems, where no such optimality constraint exists, is
  not clear in general and likely depends on the details of the problem
  class.}\BibitemShut {Stop}%
\bibitem [{\citenamefont {Dickson}\ \emph {et~al.}(2013)\citenamefont
  {Dickson}, \citenamefont {Johnson}, \citenamefont {Amin}, \citenamefont
  {Harris}, \citenamefont {Altomare}, \citenamefont {Berkley}, \citenamefont
  {Bunyk}, \citenamefont {Cai}, \citenamefont {Chapple}, \citenamefont {Chavez}
  \emph {et~al.}}]{dickson2013thermally}%
  \BibitemOpen
  \bibfield  {author} {\bibinfo {author} {\bibfnamefont {N.~G.}\ \bibnamefont
  {Dickson}}, \bibinfo {author} {\bibfnamefont {M.}~\bibnamefont {Johnson}},
  \bibinfo {author} {\bibfnamefont {M.}~\bibnamefont {Amin}}, \bibinfo {author}
  {\bibfnamefont {R.}~\bibnamefont {Harris}}, \bibinfo {author} {\bibfnamefont
  {F.}~\bibnamefont {Altomare}}, \bibinfo {author} {\bibfnamefont
  {A.}~\bibnamefont {Berkley}}, \bibinfo {author} {\bibfnamefont
  {P.}~\bibnamefont {Bunyk}}, \bibinfo {author} {\bibfnamefont
  {J.}~\bibnamefont {Cai}}, \bibinfo {author} {\bibfnamefont {E.}~\bibnamefont
  {Chapple}}, \bibinfo {author} {\bibfnamefont {P.}~\bibnamefont {Chavez}},
  \emph {et~al.},\ }\href@noop {} {\bibfield  {journal} {\bibinfo  {journal}
  {Nature communications}\ }\textbf {\bibinfo {volume} {4}},\ \bibinfo {pages}
  {1903} (\bibinfo {year} {2013})}\BibitemShut {NoStop}%
\bibitem [{\citenamefont {Cattaneo}\ \emph {et~al.}(2018)\citenamefont
  {Cattaneo}, \citenamefont {Rossi}, \citenamefont {Paris},\ and\ \citenamefont
  {Maniscalco}}]{cattaneo2018quantum}%
  \BibitemOpen
  \bibfield  {author} {\bibinfo {author} {\bibfnamefont {M.}~\bibnamefont
  {Cattaneo}}, \bibinfo {author} {\bibfnamefont {M.~A.}\ \bibnamefont {Rossi}},
  \bibinfo {author} {\bibfnamefont {M.~G.}\ \bibnamefont {Paris}}, \ and\
  \bibinfo {author} {\bibfnamefont {S.}~\bibnamefont {Maniscalco}},\
  }\href@noop {} {\bibfield  {journal} {\bibinfo  {journal} {Physical Review
  A}\ }\textbf {\bibinfo {volume} {98}},\ \bibinfo {pages} {052347} (\bibinfo
  {year} {2018})}\BibitemShut {NoStop}%
\bibitem [{\citenamefont {Passarelli}\ \emph {et~al.}(2019)\citenamefont
  {Passarelli}, \citenamefont {De~Filippis}, \citenamefont {Cataudella},\ and\
  \citenamefont {Lucignano}}]{passarelli2019may}%
  \BibitemOpen
  \bibfield  {author} {\bibinfo {author} {\bibfnamefont {G.}~\bibnamefont
  {Passarelli}}, \bibinfo {author} {\bibfnamefont {G.}~\bibnamefont
  {De~Filippis}}, \bibinfo {author} {\bibfnamefont {V.}~\bibnamefont
  {Cataudella}}, \ and\ \bibinfo {author} {\bibfnamefont {P.}~\bibnamefont
  {Lucignano}},\ }\href@noop {} {\bibfield  {journal} {\bibinfo  {journal}
  {arXiv preprint arXiv:1901.07787}\ } (\bibinfo {year} {2019})}\BibitemShut
  {NoStop}%
\bibitem [{\citenamefont {Keck}\ \emph {et~al.}(2017)\citenamefont {Keck},
  \citenamefont {Montangero}, \citenamefont {Santoro}, \citenamefont {Fazio},\
  and\ \citenamefont {Rossini}}]{keck2017dissipation}%
  \BibitemOpen
  \bibfield  {author} {\bibinfo {author} {\bibfnamefont {M.}~\bibnamefont
  {Keck}}, \bibinfo {author} {\bibfnamefont {S.}~\bibnamefont {Montangero}},
  \bibinfo {author} {\bibfnamefont {G.~E.}\ \bibnamefont {Santoro}}, \bibinfo
  {author} {\bibfnamefont {R.}~\bibnamefont {Fazio}}, \ and\ \bibinfo {author}
  {\bibfnamefont {D.}~\bibnamefont {Rossini}},\ }\href@noop {} {\bibfield
  {journal} {\bibinfo  {journal} {New Journal of Physics}\ }\textbf {\bibinfo
  {volume} {19}},\ \bibinfo {pages} {113029} (\bibinfo {year}
  {2017})}\BibitemShut {NoStop}%
\bibitem [{\citenamefont {Smelyanskiy}\ \emph {et~al.}(2017)\citenamefont
  {Smelyanskiy}, \citenamefont {Venturelli}, \citenamefont {Perdomo-Ortiz},
  \citenamefont {Knysh},\ and\ \citenamefont
  {Dykman}}]{smelyanskiy2017quantum}%
  \BibitemOpen
  \bibfield  {author} {\bibinfo {author} {\bibfnamefont {V.~N.}\ \bibnamefont
  {Smelyanskiy}}, \bibinfo {author} {\bibfnamefont {D.}~\bibnamefont
  {Venturelli}}, \bibinfo {author} {\bibfnamefont {A.}~\bibnamefont
  {Perdomo-Ortiz}}, \bibinfo {author} {\bibfnamefont {S.}~\bibnamefont
  {Knysh}}, \ and\ \bibinfo {author} {\bibfnamefont {M.~I.}\ \bibnamefont
  {Dykman}},\ }\href@noop {} {\bibfield  {journal} {\bibinfo  {journal}
  {Physical review letters}\ }\textbf {\bibinfo {volume} {118}},\ \bibinfo
  {pages} {066802} (\bibinfo {year} {2017})}\BibitemShut {NoStop}%
\bibitem [{\citenamefont {Venuti}\ \emph {et~al.}(2017)\citenamefont {Venuti},
  \citenamefont {Albash}, \citenamefont {Marvian}, \citenamefont {Lidar},\ and\
  \citenamefont {Zanardi}}]{venuti2017relaxation}%
  \BibitemOpen
  \bibfield  {author} {\bibinfo {author} {\bibfnamefont {L.~C.}\ \bibnamefont
  {Venuti}}, \bibinfo {author} {\bibfnamefont {T.}~\bibnamefont {Albash}},
  \bibinfo {author} {\bibfnamefont {M.}~\bibnamefont {Marvian}}, \bibinfo
  {author} {\bibfnamefont {D.}~\bibnamefont {Lidar}}, \ and\ \bibinfo {author}
  {\bibfnamefont {P.}~\bibnamefont {Zanardi}},\ }\href@noop {} {\bibfield
  {journal} {\bibinfo  {journal} {Physical Review A}\ }\textbf {\bibinfo
  {volume} {95}},\ \bibinfo {pages} {042302} (\bibinfo {year}
  {2017})}\BibitemShut {NoStop}%
\bibitem [{\citenamefont {Arceci}\ \emph {et~al.}(2018)\citenamefont {Arceci},
  \citenamefont {Barbarino}, \citenamefont {Rossini},\ and\ \citenamefont
  {Santoro}}]{arceci2018optimal}%
  \BibitemOpen
  \bibfield  {author} {\bibinfo {author} {\bibfnamefont {L.}~\bibnamefont
  {Arceci}}, \bibinfo {author} {\bibfnamefont {S.}~\bibnamefont {Barbarino}},
  \bibinfo {author} {\bibfnamefont {D.}~\bibnamefont {Rossini}}, \ and\
  \bibinfo {author} {\bibfnamefont {G.~E.}\ \bibnamefont {Santoro}},\
  }\href@noop {} {\bibfield  {journal} {\bibinfo  {journal} {Physical Review
  B}\ }\textbf {\bibinfo {volume} {98}},\ \bibinfo {pages} {064307} (\bibinfo
  {year} {2018})}\BibitemShut {NoStop}%
\bibitem [{\citenamefont {Suzuki}\ \emph {et~al.}(2019)\citenamefont {Suzuki},
  \citenamefont {Oshiyama},\ and\ \citenamefont {Shibata}}]{suzuki2019quantum}%
  \BibitemOpen
  \bibfield  {author} {\bibinfo {author} {\bibfnamefont {S.}~\bibnamefont
  {Suzuki}}, \bibinfo {author} {\bibfnamefont {H.}~\bibnamefont {Oshiyama}}, \
  and\ \bibinfo {author} {\bibfnamefont {N.}~\bibnamefont {Shibata}},\
  }\href@noop {} {\bibfield  {journal} {\bibinfo  {journal} {Journal of the
  Physical Society of Japan}\ }\textbf {\bibinfo {volume} {88}},\ \bibinfo
  {pages} {061003} (\bibinfo {year} {2019})}\BibitemShut {NoStop}%
\bibitem [{\citenamefont {Roberts}\ \emph {et~al.}(2019)\citenamefont
  {Roberts}, \citenamefont {Cincio}, \citenamefont {Saxena}, \citenamefont
  {Petukhov},\ and\ \citenamefont {Knysh}}]{roberts2019noise}%
  \BibitemOpen
  \bibfield  {author} {\bibinfo {author} {\bibfnamefont {D.}~\bibnamefont
  {Roberts}}, \bibinfo {author} {\bibfnamefont {L.}~\bibnamefont {Cincio}},
  \bibinfo {author} {\bibfnamefont {A.}~\bibnamefont {Saxena}}, \bibinfo
  {author} {\bibfnamefont {A.}~\bibnamefont {Petukhov}}, \ and\ \bibinfo
  {author} {\bibfnamefont {S.}~\bibnamefont {Knysh}},\ }\href@noop {}
  {\bibfield  {journal} {\bibinfo  {journal} {arXiv preprint arXiv:1909.00322}\
  } (\bibinfo {year} {2019})}\BibitemShut {NoStop}%
\bibitem [{\citenamefont {Slutskii}\ \emph {et~al.}(2019)\citenamefont
  {Slutskii}, \citenamefont {Albash}, \citenamefont {Barash},\ and\
  \citenamefont {Hen}}]{slutskii2019analog}%
  \BibitemOpen
  \bibfield  {author} {\bibinfo {author} {\bibfnamefont {M.}~\bibnamefont
  {Slutskii}}, \bibinfo {author} {\bibfnamefont {T.}~\bibnamefont {Albash}},
  \bibinfo {author} {\bibfnamefont {L.}~\bibnamefont {Barash}}, \ and\ \bibinfo
  {author} {\bibfnamefont {I.}~\bibnamefont {Hen}},\ }\href@noop {} {\bibfield
  {journal} {\bibinfo  {journal} {arXiv preprint arXiv:1904.04420}\ } (\bibinfo
  {year} {2019})}\BibitemShut {NoStop}%
\bibitem [{\citenamefont {Vinci}\ and\ \citenamefont
  {Lidar}(2018)}]{vinci2018scalable}%
  \BibitemOpen
  \bibfield  {author} {\bibinfo {author} {\bibfnamefont {W.}~\bibnamefont
  {Vinci}}\ and\ \bibinfo {author} {\bibfnamefont {D.~A.}\ \bibnamefont
  {Lidar}},\ }\href@noop {} {\bibfield  {journal} {\bibinfo  {journal}
  {Physical Review A}\ }\textbf {\bibinfo {volume} {97}},\ \bibinfo {pages}
  {022308} (\bibinfo {year} {2018})}\BibitemShut {NoStop}%
\bibitem [{\citenamefont {Pearson}\ \emph {et~al.}(2019)\citenamefont
  {Pearson}, \citenamefont {Mishra}, \citenamefont {Hen},\ and\ \citenamefont
  {Lidar}}]{pearson2019analog}%
  \BibitemOpen
  \bibfield  {author} {\bibinfo {author} {\bibfnamefont {A.}~\bibnamefont
  {Pearson}}, \bibinfo {author} {\bibfnamefont {A.}~\bibnamefont {Mishra}},
  \bibinfo {author} {\bibfnamefont {I.}~\bibnamefont {Hen}}, \ and\ \bibinfo
  {author} {\bibfnamefont {D.}~\bibnamefont {Lidar}},\ }\href@noop {}
  {\bibfield  {journal} {\bibinfo  {journal} {arXiv preprint arXiv:1907.12678}\
  } (\bibinfo {year} {2019})}\BibitemShut {NoStop}%
\bibitem [{\citenamefont {Novo}\ \emph {et~al.}(2018)\citenamefont {Novo},
  \citenamefont {Chakraborty}, \citenamefont {Mohseni},\ and\ \citenamefont
  {Omar}}]{novo2018environment}%
  \BibitemOpen
  \bibfield  {author} {\bibinfo {author} {\bibfnamefont {L.}~\bibnamefont
  {Novo}}, \bibinfo {author} {\bibfnamefont {S.}~\bibnamefont {Chakraborty}},
  \bibinfo {author} {\bibfnamefont {M.}~\bibnamefont {Mohseni}}, \ and\
  \bibinfo {author} {\bibfnamefont {Y.}~\bibnamefont {Omar}},\ }\href@noop {}
  {\bibfield  {journal} {\bibinfo  {journal} {Physical Review A}\ }\textbf
  {\bibinfo {volume} {98}},\ \bibinfo {pages} {022316} (\bibinfo {year}
  {2018})}\BibitemShut {NoStop}%
\bibitem [{\citenamefont {Morley}\ \emph {et~al.}(2019)\citenamefont {Morley},
  \citenamefont {Chancellor}, \citenamefont {Bose},\ and\ \citenamefont
  {Kendon}}]{morley2019quantum}%
  \BibitemOpen
  \bibfield  {author} {\bibinfo {author} {\bibfnamefont {J.~G.}\ \bibnamefont
  {Morley}}, \bibinfo {author} {\bibfnamefont {N.}~\bibnamefont {Chancellor}},
  \bibinfo {author} {\bibfnamefont {S.}~\bibnamefont {Bose}}, \ and\ \bibinfo
  {author} {\bibfnamefont {V.}~\bibnamefont {Kendon}},\ }\href@noop {}
  {\bibfield  {journal} {\bibinfo  {journal} {Physical Review A}\ }\textbf
  {\bibinfo {volume} {99}},\ \bibinfo {pages} {022339} (\bibinfo {year}
  {2019})}\BibitemShut {NoStop}%
\bibitem [{\citenamefont {Crosson}\ \emph {et~al.}(2014)\citenamefont
  {Crosson}, \citenamefont {Farhi}, \citenamefont {Lin}, \citenamefont {Lin},\
  and\ \citenamefont {Shor}}]{crosson2014different}%
  \BibitemOpen
  \bibfield  {author} {\bibinfo {author} {\bibfnamefont {E.}~\bibnamefont
  {Crosson}}, \bibinfo {author} {\bibfnamefont {E.}~\bibnamefont {Farhi}},
  \bibinfo {author} {\bibfnamefont {C.~Y.-Y.}\ \bibnamefont {Lin}}, \bibinfo
  {author} {\bibfnamefont {H.-H.}\ \bibnamefont {Lin}}, \ and\ \bibinfo
  {author} {\bibfnamefont {P.}~\bibnamefont {Shor}},\ }\href@noop {} {\bibfield
   {journal} {\bibinfo  {journal} {arXiv preprint arXiv:1401.7320}\ } (\bibinfo
  {year} {2014})}\BibitemShut {NoStop}%
\bibitem [{\citenamefont {Hormozi}\ \emph {et~al.}(2017)\citenamefont
  {Hormozi}, \citenamefont {Brown}, \citenamefont {Carleo},\ and\ \citenamefont
  {Troyer}}]{hormozibrown2017}%
  \BibitemOpen
  \bibfield  {author} {\bibinfo {author} {\bibfnamefont {L.}~\bibnamefont
  {Hormozi}}, \bibinfo {author} {\bibfnamefont {E.~W.}\ \bibnamefont {Brown}},
  \bibinfo {author} {\bibfnamefont {G.}~\bibnamefont {Carleo}}, \ and\ \bibinfo
  {author} {\bibfnamefont {M.}~\bibnamefont {Troyer}},\ }\href {\doibase
  10.1103/PhysRevB.95.184416} {\bibfield  {journal} {\bibinfo  {journal} {Phys.
  Rev. B}\ }\textbf {\bibinfo {volume} {95}},\ \bibinfo {pages} {184416}
  (\bibinfo {year} {2017})}\BibitemShut {NoStop}%
\bibitem [{\citenamefont {Sels}\ and\ \citenamefont
  {Polkovnikov}(2017)}]{sels2017minimizing}%
  \BibitemOpen
  \bibfield  {author} {\bibinfo {author} {\bibfnamefont {D.}~\bibnamefont
  {Sels}}\ and\ \bibinfo {author} {\bibfnamefont {A.}~\bibnamefont
  {Polkovnikov}},\ }\href@noop {} {\bibfield  {journal} {\bibinfo  {journal}
  {Proceedings of the National Academy of Sciences}\ }\textbf {\bibinfo
  {volume} {114}},\ \bibinfo {pages} {E3909} (\bibinfo {year}
  {2017})}\BibitemShut {NoStop}%
\bibitem [{\citenamefont {Susa}\ \emph {et~al.}(2018)\citenamefont {Susa},
  \citenamefont {Yamashiro}, \citenamefont {Yamamoto}, \citenamefont {Hen},
  \citenamefont {Lidar},\ and\ \citenamefont {Nishimori}}]{PhysRevA.98.042326}%
  \BibitemOpen
  \bibfield  {author} {\bibinfo {author} {\bibfnamefont {Y.}~\bibnamefont
  {Susa}}, \bibinfo {author} {\bibfnamefont {Y.}~\bibnamefont {Yamashiro}},
  \bibinfo {author} {\bibfnamefont {M.}~\bibnamefont {Yamamoto}}, \bibinfo
  {author} {\bibfnamefont {I.}~\bibnamefont {Hen}}, \bibinfo {author}
  {\bibfnamefont {D.~A.}\ \bibnamefont {Lidar}}, \ and\ \bibinfo {author}
  {\bibfnamefont {H.}~\bibnamefont {Nishimori}},\ }\href {\doibase
  10.1103/PhysRevA.98.042326} {\bibfield  {journal} {\bibinfo  {journal} {Phys.
  Rev. A}\ }\textbf {\bibinfo {volume} {98}},\ \bibinfo {pages} {042326}
  (\bibinfo {year} {2018})}\BibitemShut {NoStop}%
\bibitem [{\citenamefont {Gra\ss{}}(2019)}]{PhysRevLett.123.120501}%
  \BibitemOpen
  \bibfield  {author} {\bibinfo {author} {\bibfnamefont {T.}~\bibnamefont
  {Gra\ss{}}},\ }\href {\doibase 10.1103/PhysRevLett.123.120501} {\bibfield
  {journal} {\bibinfo  {journal} {Phys. Rev. Lett.}\ }\textbf {\bibinfo
  {volume} {123}},\ \bibinfo {pages} {120501} (\bibinfo {year}
  {2019})}\BibitemShut {NoStop}%
\bibitem [{\citenamefont {Hauke}\ \emph {et~al.}(2019)\citenamefont {Hauke},
  \citenamefont {Katzgraber}, \citenamefont {Lechner}, \citenamefont
  {Nishimori},\ and\ \citenamefont {Oliver}}]{hauke2019perspectives}%
  \BibitemOpen
  \bibfield  {author} {\bibinfo {author} {\bibfnamefont {P.}~\bibnamefont
  {Hauke}}, \bibinfo {author} {\bibfnamefont {H.~G.}\ \bibnamefont
  {Katzgraber}}, \bibinfo {author} {\bibfnamefont {W.}~\bibnamefont {Lechner}},
  \bibinfo {author} {\bibfnamefont {H.}~\bibnamefont {Nishimori}}, \ and\
  \bibinfo {author} {\bibfnamefont {W.~D.}\ \bibnamefont {Oliver}},\
  }\href@noop {} {\bibfield  {journal} {\bibinfo  {journal} {arXiv preprint
  arXiv:1903.06559}\ } (\bibinfo {year} {2019})}\BibitemShut {NoStop}%
\bibitem [{dwa(2019)}]{dwave2019whitepaper}%
  \BibitemOpen
  \href@noop {} {\bibfield  {journal} {\bibinfo  {journal} {\textit{Improved
  coherence leads to gains in quantum annealing performance}, D-Wave Whitepaper
  Series}\ } (\bibinfo {year} {2019})}\BibitemShut {NoStop}%
\end{thebibliography}%

\end{document}